\documentclass[a4paper,11pt]{article}
\pdfoutput=1 % if your are submitting a pdflatex (i.e. if you have
\usepackage{jheppub} % for details on the use of the package, please
                     \usepackage[T1]{fontenc} % if needed
\usepackage[dvipsnames]{xcolor}
\usepackage{amsmath}
\usepackage{amssymb}
\usepackage{comment}
\usepackage{enumerate}
\usepackage{changes}
\usepackage{color}
\usepackage{tablefootnote}
\usepackage{slashed}
\usepackage{graphicx}

\def\be{\begin{equation}}
\def\ee{\end{equation}}
\def\ba{\begin{eqnarray}}
\def\ea{\end{eqnarray}}
\def\bi{\begin{itemize}}
\def\ei{\end{itemize}}	
\def\l{\left}
\def\r{\right}

\def\d{\partial}

\def\mpl{M_{\rm pl}}

\def\bea{\begin{eqnarray}}
\def\eea{\end{eqnarray}}

\title{An effective formalism for testing extensions to General Relativity with gravitational waves}

\author[a]{Solomon Endlich,}
\author[a]{Victor Gorbenko,}
\author[a]{Junwu Huang,}
\author[a]{Leonardo Senatore}

\affiliation[a]{Stanford Institute for Theoretical Physics, Stanford University, Stanford, CA 94305, USA}

\emailAdd{sendlich@stanford.edu}
\emailAdd{vitya@stanford.edu}
\emailAdd{curlyh@stanford.edu}
\emailAdd{senatore@stanford.edu}

\abstract{The recent direct observation of gravitational waves (GW) from merging black holes opens up the possibility of exploring the theory of gravity in the strong regime at an unprecedented level. It is therefore interesting to explore which extensions to General Relativity (GR) could be detected. We construct an Effective Field Theory (EFT) satisfying the following requirements. It is testable with GW observations; it is consistent with other experiments, including short distance tests of GR; it agrees with widely accepted principles of physics, such as locality, causality and unitarity; and it does not involve new light degrees of freedom. The most general theory satisfying these requirements  corresponds to adding to the GR Lagrangian operators constructed out of powers of the Riemann tensor, suppressed by a  scale comparable to the curvature of the observed merging binaries. The presence of these operators modifies the gravitational potential between the compact objects, as well as their effective mass and current quadrupoles, ultimately correcting the waveform of the emitted~GW.}

\begin{document}

\maketitle

\section{Introduction}

The importance of the recent direct detection by the LIGO-VIRGO collaboration~\cite{Abbott:2016blz,TheLIGOScientific:2016qqj,Abbott:2016nmj} of gravitational waves emitted from the merger of two black holes  can hardly be over-emphasized. It confirms one of the most beautiful predictions of General Relativity (GR), the existence of gravitational waves; it detects the presence and the coalescence of two black holes, another remarkable prediction of the same theory; and, finally, it provides mankind with a new set of eyes to look at the cosmos in a way that we had never done before. These detection events were a first taste of the incredible vista before us.

It is clear that the availability of gravitational wave observatories will enable us to learn a great deal about the physics of compact objects and, more generally, the whole of astrophysics. However, it is unclear to what extent they will allow us to increase our knowledge of the fundamental laws of physics~\footnote{Here, by fundamental laws, we mean those laws that describe the most basic phenomena, and from which, at least in principle and possibly at the cost of great complexity, all phenomena can be described. In this sense, usual astrophysical laws are not fundamental laws of physics, even though astrophysics is a fundamentally important discipline.}. One possible scenario where they could provide some insight is in the case of additional weakly interacting---to the standard model sector---light particles, such as the axion. In this case, the large gravitational field in the proximity of  black  holes and their rapid rotation can source the clustering of large numbers of these light particles, which in turn can have observational consequences on the dynamics of the black holes and their associated gravitational wave emission~\cite{Arvanitaki:2010sy,Arvanitaki:2014wva,Arvanitaki:2016qwi}. It is a fascinating possibility for a variety of reasons and has led to a flourishing body of work in the literature.

Here however, we would like to answer the following question: can the new observational window offered by gravitational wave astronomy teach us something about the nature of gravity? We will not be able to answer this question in full generality. However, we will be able to do so if we restrict ourselves to the case in which the modifications of GR is associated to states that are heavier than the curvature scale of the compact objects responsible for the emission of the gravitational waves.  Even with our assumptions this is quite a general scenario and consequently our statement might appear over-ambitious.  Our confidence stems from the fact that we will use a technique known as Effective Field Theory (EFT) that allows us to construct a Lagrangian and the associated equations of motion that encode the {\it most general} extension to GR of the kind we just described, {\it i.e.} where the new states are heavier than the curvature scale of the compact objects. We will construct this EFT in Sec.~\ref{sec:action}, and we will find that it takes the remarkably simple (given its generality) form

\begin{equation}
S _{\rm eff}=2\mpl^2\int { d}^4 x \sqrt{-g}\ \left(-R + \frac{\mathcal{C}^2}{\Lambda^6} + \frac{\tilde{\mathcal{C     }}^2}{\tilde{\Lambda}^6}+ \frac{\tilde{\mathcal{C     }}\mathcal{C}}{\Lambda_-^6}+\ldots\right)\label{eq:actionintro}
\end{equation}
where
\begin{equation}
\mathcal{C} \equiv  R_{\alpha \beta \gamma \delta} R^{\alpha \beta \gamma \delta},\quad \tilde{\mathcal{C     }} \equiv  R_{\alpha \beta \gamma \delta}\, \epsilon^{\alpha\beta}{}_{\mu\nu}\,{R}^{\mu \nu \gamma \delta}\ ,
\end{equation}
and $\ldots$ stands for terms that give subleading contributions~\footnote{We also consider the case in which the leading extension to GR is represented by three powers of the Riemann tensor, rather than by four. The most general action is given later in~(\ref{eq:sixderivative}). For reasons over which we elaborate later on, the case of the action~(\ref{eq:sixderivative}) appears to be disfavored from the UV point of view, and therefore it does not represent the main focus of our discussion. All the observational effects that we discuss to result from the action (\ref{eq:actionintro}) apply to the case of~(\ref{eq:sixderivative}) as well, with the replacement $(\Lambda r)^6\to (\Lambda r)^4$, where~$\Lambda$ is the typical scale suppressing the leading operators in the two effective actions.}. 

This EFT is very general. For example, it describes the extension to GR by string theory at energies below the string scale. Of course, in order for the effect to be measurable for experiments such as LIGO, we need to ensure that at least one of the scales $\Lambda,\;\tilde\Lambda$ and $\Lambda_-$ are not too much larger than the curvature scale of the compact objects themselves, which means that, crudely, we need to take the $\Lambda$'s $\sim {{\cal O}}\left({\rm km}^{-1}\right)$. This is a challenging scale for two reasons. 

First, on the theory side, we have a prejudice that we do not expect GR to be modified at such small energies. But, if the scales $\Lambda$, $\tilde\Lambda$ and $\Lambda_-$ were to be much greater than~${{\cal O}}\left({\rm km}^{-1}\right)$, there simply would be no possible signatures of UV modifications of GR expected by gravitational wave astronomy and similar astrophysical probes.  Therefore, with some apologies, we happily put aside our prejudices in favor of empirical verification, especially now that such measurements are actually possible.

Second, a more serious concern is that we have already probed gravity at scales much shorter than ${\rm km}$. How can we be sure such low values of the $\Lambda$'s are not already ruled out? The crucial difference between laboratory experiments and compact object observations is the size of the curvature tensor, which is much larger in the astrophysical setting. This allows us to argue in the main text that we can {\it assume} the following: there is a UV completion such that, {\it whenever} the Riemann tensor is sufficiently small, the modifications to GR are small even at length scales shorter than $\Lambda$, $\tilde\Lambda$ and $\Lambda_-$, as it happens in laboratory experiments.

Given our set of assumptions, we use the EFT in (\ref{eq:actionintro}) to compute observable consequences in the gravitational wave emission from compact objects from UV modifications of GR. We will perform our analysis in Sec.~\ref{sec:techsummary},\ref{sec:potential},\ref{sec:radiation}, where we will focus on describing the modifications of the signal in the post-Newtonian regime. We will find that the main effects of the operators $\mathcal{C}^2 $, $ \tilde{\mathcal{C     }}^2$ and $\mathcal{C}\tilde{\mathcal{C     }}$ are to rescale the amplitude and frequency of the emitted gravitational waves. We describe these findings from a phenomenological point of view in Sec.~\ref{sec:observeligo}. Explicitly, for the operator $\mathcal{C}^2$, for a binary of objects of equal mass $m$, in a quasi-circular orbit of radius $r$ and relative velocity $v$, we find
\begin{eqnarray}\label{result1}
&&\frac{\left[\Delta h^{TT}(t,\vec x)\right]_\Lambda}{h^{TT}}\sim\frac{\Delta\omega_{\Lambda}}{\omega_{PN}}\sim \frac{v^4}{(\Lambda r)^6} \; ,
\end{eqnarray}
where $h^{TT}$ is the strain (or amplitude) of the gravitational wave with frequency $\omega$ produced by the source incident upon the detector and $h^{TT}_\Lambda$ is the contribution to the strain generated by the $\mathcal{C}^2 $ term. Notice that the effect in~(\ref{result1}) depends on two independent parameters, $v$ and $1/(\Lambda r)$, both of which need to be smaller than one in the observable region. However, for $(\Lambda r)\sim 1$, the effect can be as large as $v^4$, i.e. second Post-Newtonian order (2PN), signaling that  the effect is potentially observable.

The effect of the operators $ \tilde{\mathcal{C     }}^2$ and ${\mathcal{C     }}\tilde{\mathcal{C     }}$ is similar, just differing by the suppression in the powers of the velocity or in the polarization of the emitted signal. 
We notice that, even after the inclusion of all the numerical factors, the signal can be rather large, and in fact probably the detection of the recent merger events can already put some interesting constraint on the scales $\Lambda$'s. In Sec.~\ref{sec:otherexp} we describe bounds from other experiments, which we find to be mild, apart from light X-ray binaries, that can potentially give interesting constraints.
In Sec.~\ref{sec:conclusions}, we conclude by summarizing our findings and discuss future directions.

Before we begin a deeper study of our EFT, it is worth spending some additional words elucidating on how our approach differs from others already present in the literature. Since the first observation of gravitational waves, there has been a vast number of publications related to tests of GR with compact object mergers and it is impossible to review them fairly in this short introduction. We would like, however, to compare our approach to that used by the LIGO-VIRGO collaboration in~\cite{TheLIGOScientific:2016src}. In this analysis the first few post-Newtonian~(PN) coefficients were allowed to deviate from the calculated GR values, thus changing the wave form of emitted gravitational waves. In this approach it is unfortunately very hard to see which variation of the parameters corresponds to  theories respecting physical principles like locality, Lorentz invariance or the equivalence principle and which do not. More concretely, it is difficult to track which physical principles we have to give up in order to produce one variation or the other of the PN coefficients.  Our approach, on the other hand, is automatically in agreement with the principle of locality, diffeomoprphism invariance, etc.. Even more strikingly, even though our proposed modification of GR has much fewer free parameters, it cannot be captured by the analysis of \cite{TheLIGOScientific:2016src}. The reason is that the prediction from our EFT corresponds to giving some very specific radial and time dependence of the PN parameters, in the form of factors of $1/(\Lambda r)^6$ in~(\ref{result1}). In order to be able to cover this signal with the phenomenological analysis of~\cite{TheLIGOScientific:2016src} one would need to give to the PN parameters some time and radial dependence whose form, without guidance from an EFT, would be arbitrary and therefore probably severely weakens the constraining power of the analysis.

An approach closer to ours was used in \cite{Yunes:2016jcc}. In that paper the authors studied how theoretically motivated modifications to GR can be constrained by observed BH mergers. However, the focus was on theories containing extra light degrees of freedom while none of the UV modifications captured by our EFT were discussed. In this sense our results are complimentary to those of \cite{Yunes:2016jcc}. Theories with extra light particles predict significant changes in the waveforms due to the existence of new emission channels. However, it turns out to be especially difficult to compute predictions for theories that contain additional light particles in the regime of strong gravity~\cite{Yunes:2016jcc}. Furthermore, in the presence of additional light degrees of freedom, one would need to work out a different prediction for each corresponding theory. One advantage of our EFT approach is that with just a couple of parameters all corrections to the merger process are well defined (to a given precision) and even though in the present paper we restrict to the perturbative phase of the merger, in principle the full numerical study of the corrections in the non-linear regime can be performed as we outline in Section~\ref{sec:numerical}.

\section{General Construction of the Effective Field Theory\label{sec:action}}

It is a common lore that, given a fixed set of light degrees of freedom, at low energies any possible effect of ultraviolet physics can be parametrized by a set of local interactions involving these light degrees of freedom only~\footnote{Recently, some investigations in the context of string theory have given indications that this theory, in the presence of backgrounds with horizon, might induce non-local effects at scales much longer than the string scale~\cite{Dodelson:2015toa}. Such phenomena would require a different description than the one we develop here, which crucially assumes locality. It would be interesting to extend our analysis to include these non-local effects. It is tempting to say that the leading effects will come from modifying the effective black-hole finite size operators, that we discuss later on at the end of Sec.~\ref{sec:observeligo}. However, we leave a study of this to future work.}. If one is interested in computing a physical observable to a given precision, this set is always finite. This approach to parametrizing physical observables is called Effective Field Theory (see for example~\cite{Weinberg:1995mt}). We are interested here in the most general theory that can be tested by precision experiments measuring gravitational waves produced by mergers of compact astrophysical objects that involves  a single light degree of freedom - the graviton~\footnote{Additional light degrees of freedom can be included in a straightforward way, however, we will restrain from doing so here and leave it to future work.} - and that is not already excluded by any other experiments. The most convenient way of classifying such theories is the EFT approach. A crucial ingredient of all EFTs is an energy scale $\Lambda_c$, usually referred to as a "cutoff". The cutoff defines what was meant by "low energies" above. At energies higher than the cutoff, the EFT loses its validity and the knowledge of ultraviolet physics, that often involves some new degrees of freedom, is necessary to do the computation. On the other hand, at energies below the cutoff, the EFT is not only absolutely universal but also is always under perturbative control with expansion parameter $E/\Lambda_c$. For this reason it is convenient to organize the terms in the effective actions in order of increasing number of derivatives and fields.

First of all, let us note that in spite of several intriguing mysteries associated with gravity, at low energies it is nothing more than another EFT and importantly its cutoff scale does not have to be parametrically close to $\mpl$. Some new physical effects can in principle appear at a much lower scale. In order to produce observable and calculable consequences for mergers of compact objects, this scale has to be close to or below the characteristic curvatures of the space-time nearby these objects, which, for stellar mass black holes and neutron stars, is of order of a few inverse kilometers. One may immediately object that we are already running into a contradiction: gravity has been tested to high precision at distances much smaller than a few kilometers and hence any modifications that we are discussing are by far excluded. We will show, however, that this objection is too quick. Under some broad assumptions about the UV completion at the scale $\Lambda_c$, modifications of gravity can be unobservably small unless the scale of the space-time curvature happens to be close to $\Lambda_c$. Since compact astrophysical objects like black holes and neutron start are the only known sources of large curvatures, it is consistent (though not necessary) for new physics to affect significantly a merger process while keeping all "weak field" processes practically intact.

Effective field theories are subject to a set of consistency conditions. A very important one is radiative stability. It implies that if one can construct a Feynman diagram containing UV divergence proportional to some local operator this operator has generically to be included in the action with a coefficient at least as large as given by the diagram with loop momentum integrals cut off at $\Lambda_c$. There are other, more subtle though very reasonable, constraints on effective field theories that usually involve additional, even more essential, assumptions, such as locality and causality~\cite{Adams:2006sv, Gruzinov:2006ie, Camanho:2014apa}. We will review the relevant ones below and will be specific about which extra assumptions are involved. In our case there will be further constraints imposed by the "testability" requirement, of which we will discuss later.

In order to optimize the set of terms present at each order in the effective action it is convenient to include only those operators that do not vanish on the equations of motion produced by the lower order action. For a reader not familiar with the EFT approach it could be instructive to consider a simple example. Consider the following action for a single scalar field:
\be\label{eq:example1}
\int d^4x\;\left( \frac{1}{2}\phi \square \phi+\frac{\square\phi \phi^2}{\Lambda_c}\right).
\ee
Naively there will be an interaction at the linear order in $1/\Lambda_c$, however, for the type of observable we are interested in, instead of using the field $\phi$, one can equivalently use the field $\phi'$ given by $\phi=\phi'-\phi'^2/\Lambda_c$, in terms of which the action only contains operators suppressed by $\Lambda_c^2$. This means that all physical effects in this theory will be suppressed at least by the second power of our cutoff scale\footnote{In fact in this simple example these interactions can be further redefined away and the leading physical interaction appears even at higher orders.} and we could have started classifying operators beginning from dimension 6. In case of (pure) gravity the leading equations of motion are the vacuum Einstein equations
\be
R_{\mu\nu}-\frac{1}{2}R g_{\mu\nu}=0\ ,
\ee
from which it follows that both $R$ and $R_{\mu\nu}$ vanish on the leading equations of motion, and hence one can consider only  operators constructed  from the Riemann tensor $R_{\mu\nu\rho\sigma}$.  

We are now ready to start to construct the effective action. This amounts to writing down, in a power law expansion, all the terms that are allowed by the symmetries of the problem, which in our case are operators built out of the Riemann tensor. Let us start classifying them. We will begin with operators that only involve the gravitational field and discuss possible mixed gravity-matter operators later on.  At the level of two derivatives there is a single term allowed by diffeomorphism invariance, the usual Einstein-Hilbert term. At the level of four derivatives there is also a single term one can write:
\be
\int d^4 x\,\sqrt{-g}\; R_{\mu\nu\rho\sigma}R^{\mu\nu\rho\sigma}\ ,
\ee
on top of the usual $\int d^4x\; \sqrt{-g} \epsilon_{\mu\nu\rho\sigma} R^{\mu\nu\alpha\beta} R_{\alpha\beta}{}^{\rho\sigma}$, which is a total derivative.
However, in four dimensions, after integration by parts, this operator can be reduced to terms involving $R$ and $R_{\mu\nu}$ and hence can be ignored according to the discussion above. In fact, the Euler density $E_4$: 
\be
E_4=R_{\mu\nu\rho\sigma}R^{\mu\nu\rho\sigma}-4R_{\mu\nu}R^{\mu\nu}+R^2\ ,
\ee
is a total derivative. 

As we show in section~\ref{sec:classification} there are two independent terms involving six derivatives, one parity even and one parity odd, that can be chosen in the following form:~\footnote{To define the second invariant we used $\tilde{R}^{\alpha \beta \gamma \delta}=\epsilon^{\alpha\beta}{}_{\mu\nu}R^{\mu\nu\gamma\delta}$. We define $\epsilon$ in a way that $\epsilon^{0123}=1/\sqrt{-g}$.}
\be\label{eq:sixderivative}
S_{\rm eff}=2 \mpl^2\int d^4 x\; \sqrt{-g}\;\left[- R+c_3 \frac{ R_{\mu\nu\rho\sigma} R^{\mu\nu}\,_{\alpha\beta}R^{\alpha\beta\rho\sigma}}{\Lambda^4}+\tilde{c}_3 \frac{ \tilde{R}_{\mu\nu\rho\sigma} R^{\mu\nu}\,_{\alpha\beta}R^{\alpha\beta\rho\sigma}}{\Lambda^4}\right]\ .
\ee

As we discussed, we are interested only in theories that can be tested with gravitational wave astronomy. We call this requirement the ``Testability" requirement. In order to satisfy this requirement, $\Lambda$ has to be picked of order a few km inverse, if $c_3$ and $\tilde c_3$ are taken to be order one. If radiative stability was the only constraint the two six-derivative operators would be perfect candidates for the leading corrections to General Relativity in four dimensions. However, a recent argument~\cite{Camanho:2014apa} that we briefly review in Sec.~\ref{sec:causality} shows that if $c_3$ or $\tilde{c}_3$ are non vanishing, and under certain assumptions about the UV complition, causality would require an infinite tower of higher spin particles coupled to standard model fields with gravitational strength. The mass of the lightest of those particles has to be of order $\Lambda$ and the couplings such as to allow mediation of long range forces of gravitational strength between any matter fields. Obviously on sub-kilometer distances we have not observed any additional long range forces and hence the $c_3$ and $\tilde c_3$ terms must be suppressed by a much higher scale. {We however warn the reader than the argument of~\cite{Camanho:2014apa} appears to assume that the UV completion of (\ref{eq:sixderivative}) enters at tree level. It could be that the argument of~\cite{Camanho:2014apa} can be extended to include all possible UV complitions. However at the moment we do not have such a proof, and we cautiously conclude that the theory in~(\ref{eq:sixderivative}) can still be considered as a viable theory, and we briefly discuss its phenomenological consequences in Sec.~\ref{sec:numerical} and~\ref{sec:techsummary}.}

In the rest of the paper we therefore concentrate mostly on the eight derivative terms. In four dimensions, as shown in Sec.~\ref{sec:classification}, there are three possible terms that we can add to the action:
\begin{equation}
S _{\rm eff}=\int { d}^4 x \sqrt{-g} 2 \mpl^2 \left(-R + \frac{\mathcal{C}^2}{\Lambda^6} + \frac{\tilde{\mathcal{C     }}^2}{\tilde{\Lambda}^6}+ \frac{\tilde{\mathcal{C     }}\mathcal{C}}{\Lambda_-^6}+\ldots\right)\label{eq:action}
\end{equation}
where
\begin{equation}
\mathcal{C} \equiv  R_{\alpha \beta \gamma \delta} R^{\alpha \beta \gamma \delta},\quad \tilde{\mathcal{C     }} \equiv  R_{\alpha \beta \gamma \delta} \tilde{R}^{\alpha \beta \gamma \delta}\ ,
\end{equation}
and $\ldots$ refers to higher order terms.

The coefficients of the parity even terms have to be positive due to causality~\cite{Gruzinov:2006ie} and analyticity~\cite{Bellazzini:2015cra} constraints. We keep in mind that these arguments can be subtle for gravity, but taking negative coefficients would not change significantly any part of our results. Furthermore, as we show in the Sec.~\ref{sec:causality}, the argument by~\cite{Gruzinov:2006ie} can be easily extended to incorporate the parity odd term which results in the following constraint:
\be
\Lambda_-^2\gtrsim\Lambda \tilde{\Lambda}\ .
\ee
Together with power-counting and symmetry arguments presented in Sec.~\ref{sec:observeligo}, this implies that the parity odd term can give the leading contribution to the physical observables only for some rather small region in parameter space~\footnote{Saturating the subluminality bound, the parity odd term can give the leading signal in the parametric window $v^2\lesssim \frac{\tilde\Lambda^6}{\Lambda^6}\lesssim v$.}. However, we still keep this term for generality. We also keep in mind that extensions of GR are very strongly constrained, and it could well be that our extension in (\ref{eq:action}) is incompatible with some general physical principle even within the parameter range that we identified. Still, to the best of our knowledge, such an argument has not yet been presented. We therefore proceed.

In the rest of the paper we are going to analyze the implications of the theory given by the action~(\ref{eq:action}) with all higher terms neglected (as they give a subleading effect) for the phenomenology of astrophysical compact objects. Before we go on, however, let us see a little more in detail how the higher order terms in~(\ref{eq:action}) look like. Schematically we can write
\be
S _{\rm eff}=\int { d}^4 x \sqrt{-g} \, 2 \mpl^2 \left(-R + \frac{\mathcal{C}^2}{\Lambda^6} + \frac{\tilde{\mathcal{C     }}^2}{\tilde{\Lambda}^6}+ \frac{\tilde{\mathcal{C     }}\mathcal{C}}{\Lambda_-^6}+c_{mn} \Lambda_R^2\sum_m\sum_n \l(\frac{\nabla_\gamma}{\Lambda_c}\r)^m \l(\frac{R_{\mu\nu\rho\sigma}}{\Lambda_R^2}\r)^n\right), \label{eq:actionfull}
\ee
where the $c_{mn}$'s are dimensionless constants. Notice that $\Lambda_c$ is the cutoff of the theory, {\it i.e.} when new states are expected to become relevant. Therefore, when computing loop corrections, we can Taylor expand in external momenta much less than $\Lambda_c$,  which implies that, in the effective theory, the scale suppressing the derivates is indeed the cutoff. Instead, we have suppressed powers of the Riemann tensor by a so-far arbitrary scale $\Lambda_R$. In the schematic writing of (\ref{eq:actionfull}), it is implied that, at each order, there are a few terms for a given $m$ and $n$,  suppressed by the same combination of scales. The scales $\Lambda$, $\tilde\Lambda$ and $\Lambda_-$ are in principle independent but for now it is convenient to keep them parametrically the same and of order a few kilometers inverse so that the corresponding operators give sizable yet perturbative corrections to Einstein equations in the black hole backgrounds. Obviously, in order to be able to use our effective theory for describing black hole mergers when distances and hence gradients of fields are of the order of the curvatures it is necessary to keep
\be
\Lambda_c\gtrsim\Lambda\ .
\label{ineq1}
\ee
If we focus on terms with $m=0$ and $n$ large it also becomes clear that in order for the theory to remain perturbative for $R\sim \Lambda^2$ it is necessary to keep
 \be
\Lambda_R\gtrsim\Lambda\ ,
\label{ineq2}
\ee
because otherwise terms with infinitely many powers of Riemann tensor will start to dominate before our quartic terms can produce a detectable correction. 

\begin{figure}[!ht]
  \centering
      \includegraphics[trim = 6cm 14cm 3cm 3cm, clip, width=0.3\textwidth]{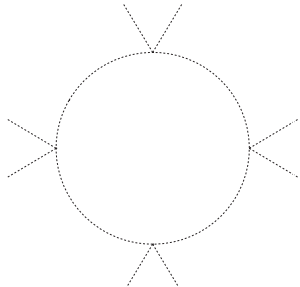}      
  \caption{Radiative generation of $R_{\mu\nu\rho\sigma}^{2n}$ operator, in this case $R_{\mu\nu\rho\sigma}{}^{8}$, through ${\mathcal{C}}^2$ and $\tilde{\mathcal{C}}^2$.  }\label{fig:operator1}
\end{figure}

Let us now see which constraints are imposed by the criteria of radiatiative stability. To do so we calculate the one-loop diagram drawn on Fig.~\ref{fig:operator1} for large number of vertices $n$ and cut off the internal loop momentum at the cutoff $\Lambda_c$. For large $n$ we generate the following correction to the effective action:
\be
  \frac{(R_{\mu\nu\rho\sigma})^{2n}\Lambda_c^{2n}}{\Lambda^{4n} }\ .
\ee
Radiative stability of the action then requires
\be
\frac{\Lambda_c}{\Lambda^{2}} \lesssim \frac{1}{\Lambda_R}\ .
\ee
Combining this requirement with (\ref{ineq1}) and (\ref{ineq2}), we are forced to set all the three scales approximately equal to each other:
\be
 \Lambda_c\approx\Lambda_R\approx\Lambda\ .
 \label{scales}
\ee
Let us note that this result, albeit natural, was not immediately obvious. For example the unitarity bound associated with the growth of the tree-level scattering around flat space produced by our quartic vertices would require $\Lambda_c^4<\Lambda^3\mpl$, while the suppression scale of power of fields in EFTs stemming from weakly coupled UV completions is usually $\Lambda_c/g$ where $g$ is some combination of coupling constants. 

Eq.~(\ref{scales}) corresponds to suppressing canonically normalized perturbations of the metric, $h^c_{\mu\nu}$, by $\mpl$: 
\ba\label{eq:radiative2}
&&S _{\rm eff}\supset \int { d}^4 x\; \sqrt{-g} \mpl^2\;  \l(\frac{\nabla_\gamma}{\Lambda}\r)^m  \l(\left(1+\tfrac{h_c}{\mpl}+\left(\tfrac{h_c}{\mpl}\right)^2+\ldots\right) \frac{\nabla_\alpha\nabla_\beta } {\Lambda^2}\frac{h_c}{ \mpl}\r)^n \\ \nonumber
&&\qquad \sim\int { d}^4 x\; \sqrt{-g}\; \mpl^2\,\Lambda^2  \l(\frac{\nabla_\gamma}{\Lambda}\r)^{2n+m}  \l(\frac{h^c}{ \mpl}\r)^n \ .
\ea

Above we noticed that the six derivative operator in $c_3$ in (\ref{eq:sixderivative}) has to be suppressed by a scale much higher than $\Lambda$. We can check with which scale this could possibly get generated by one-loop diagrams. Since $\Lambda=\Lambda_c$ we get $c_3=\Lambda^2/\mpl^2$, consequently even if we have to introduce new higher spin particles coupled gravitationally to the Standard Model following the argument of~\cite{Camanho:2014apa}, their mass should be of order $\sqrt{\Lambda \mpl}\gtrsim{\rm Gev} \sim \frac{10^{15}}{\rm meter}$, which is experimentally perfectly safe.

In principle, one could have suppressed the quartic terms in the action as well and start from even higher terms. Such a theory would be radiatively stable, however, we do not pursue it for the following reasons: first, there are no reasons that we aware of that would forbid the quartic terms; second, in all known UV completions of gravity (i.e. string theory) quartic terms do get generated; finally any modifications of General Relativity coming from such theories would be even harder to observe.

At this point, let us come back to our claim that the theory (\ref{eq:action}) with $\Lambda$ of order a few inverse ${\rm km}$ is not excluded by any flat-space or approximately flat-space experiments. Of course experiments of interest are performed at energies much larger than $\Lambda$, hence this claim necessarily depends on the UV completion. Our assumption will be that the UV completion is "soft" in the following sense: at energies above $\Lambda$, the vertices suppressed by $\Lambda$ get resolved and become renormalizable, that is they stop growing with energy. In ordinary quantum field theories such behavior is not at all exotic. In Appendix~\ref{app:example}, we present an example of a UV-complete quantum field theory where non-renormalizable operators present in low-energy effective theory get resolved in the "soft" way we just specified. We check explicitly that while the leading higher derivative operators give order one corrections to solutions with large values of the background fields, which is what makes them testable with compact objects, corrections to any experiments performed in a near-vacuum state are parametrically small. Of course it is notoriously hard to provide any example of such a UV completion for gravitational theories. In fact there is only one known (this happens to have the name of string theory). Indeed quartic vertices like those in (\ref{eq:action}) are present in low energy string actions in which the string scale plays the role of $\Lambda$. In that case, the growth of the four-graviton amplitude saturates at this scale and even goes to zero at very high energies~\footnote{As we will mention later, the UV completion of our EFT is not the normal string theory for the way the couplings to matter are affected.}. This is within the class of behaviors we require for the UV completion of our theory. Not surprisingly, similar behavior is also present in large-$N$ QCD. 

More explicitly, if we expand the metric around a flat background and introduce canonically normalized field for the metric perturbations, $g_{\mu\nu}=\eta_{\mu\nu}+h^c_{\mu\nu}/\mpl$, then the leading interaction vertex will schematically read 
\be
\mpl^{-2}\frac{(\partial^2 h^c)^4}{\Lambda^6}.
\ee
If we now consider some process that includes energy-momentum transfer of order $E\gg\Lambda$, all $
\Lambda$'s that naively stay in the denominator will get cancelled by powers of the cutoff that is also $\Lambda$ as a consequence of our "softness" assumption, while powers of $\mpl$ in the denominator will not get compensated by anything bigger than $E$. As a result all processes involving our vertex, or more precisely, whatever replaces it at energies above $\Lambda$, will have extra powers of $E/\mpl, h^c/\mpl\sim E/\mpl$ or $\Lambda/\mpl$, as compared to the leading contribution coming from GR and consequently will be practically unobservable. It is only when the metric $g_{\mu\nu}$ deviates from flat by order one that we have a chance to have sizable (order-one) corrections, as we can replace the fluctuating $h$ with its vacuum expectation value, which, for black holes, is of order $h^c\sim \mpl$.

We are  now ready to comment on operators that contain not only the metric, but also matter fields. These couplings are strongly constrained by various lab experiments, and so they better be small in order for our theory not to be ruled out. If one focuses within the regime of validity of the EFT, one finds that naively matter-matter interactions suppressed by a scale as low $\sqrt{\mpl\Lambda}$ are generated. Taken at face value, these operators are ruled out at collider experiments. However, the use of these operators at scales above $\Lambda$ is ill-defined, a fact that is made manifest by the fact that there is a series of higher derivative operators suppressed by the same scale and additional powers of $\d/\Lambda$, similarly to what we wrote in~(\ref{eq:radiative2}). Our `UV-softness' assumption is indeed crucial to forbid these operators to keep growing at scales above $\Lambda$, beyond their value at energies of order $\Lambda$, which is negligibly small, of order $(\Lambda/\mpl)^2$. In a sense, this discussion is indeed almost a repetition of the one we had a couple of paragraphs above. 

Finally, we need to discuss the special operators that are quadratic in the graviton fields, such as $R^2$. If we do not include them in the action, they are not radiatively generated because they vanish on shell. However, one might wonder what happens if we were to include them directly in the action, suppressed by a scale $\Lambda$. The fact that they vanish on-shell implies that we can perform several field redefinitions in such a way that these operators get replaced by operators of the form $T^2/(\mpl\Lambda)^2$ (similarly to the example in~(\ref{eq:example1})), where $T^2$ contains scalar functions of the energy momentum tensor $T_{\mu\nu}$, i.e., $(T^{\mu}_{\mu})^2$ and $T^{\mu \nu} T_{\mu \nu}$. At energies of order $\Lambda$, these operators give an order one correction to gravity, which is ruled out. In the case of the higher order operators, such as $\mathcal{C}^2$, we obtained smaller corrections at scales of order $\Lambda$, because these operators were interactions, and so we paid powers of the coupling constant $\mpl$. Instead, in this case, $R^2$ is just a kinetic term.
This discussion makes it clear therefore that adding these $R^2$-like operators corresponds to simply adding a new degree of freedom with mass of order $\Lambda$ directly coupled to matter with gravitational strength.  It is clear that this kind of theories will be better tested by lab experiments rather than by gravitational wave observations, as they do not require strong fields. Our purpose, instead, is to write theories that indeed can be tested best by strong gravity experiments, which we called `testability requirement' earlier on. This justifies us neglecting to include these operators in the action.

\subsection{Classification of operators made of $R_{\mu\nu\rho\sigma}$\label{sec:classification}}

In this section, we classify all operators containing up to eight derivatives that do not vanish for $R_{\mu\nu}=0$. An uninterested reader can skip this subsection.

First, it is useful to remember the second Bianchi identity:
\be\label{eq:bianchiII}
\nabla_\mu R^{\nu\rho\sigma\gamma}+\nabla_\gamma R^{\nu\rho\mu\sigma}+\nabla_\sigma R^{\nu\rho\gamma\mu}=0\ ,
\ee
which, upon contraction of $\mu$ and $\nu$ indices and the use of $R^{\nu\mu}=0$, gives
\be
\nabla_\mu R^{\mu\rho\sigma\gamma}=0\ .
\label{divR}
\ee
Since the commutator of two derivatives gives another Riemann tensor it is a straightforward exercise in integration by parts to rewrite all terms containing two Riemann tensors with extra covariant derivatives acting on them through terms with three Riemanns or more plus terms vanishing due to (\ref{divR})~\footnote{Let us sketch the proof. The only non-trivial terms are those where the covariant derivatives are contracted among themselves, as otherwise, after integration by parts and commutators we can use~(\ref{divR}). In this remaining case, one can use (\ref{eq:bianchiII}) to shuffle the derivative as being contracted with an index of the Riemann tensor, reducing therefore to the simple case. }.

It is a significantly more complicated task to classify all terms containing three Riemann tensors. Intuitively we expect two independent terms with three Riemanns and no extra derivatives and we expect to be able to rewrite all terms with extra derivatives through terms containing four Riemann tensors or more. The reason is that there are only two independent on-shell cubic vertices for gravitons in four dimensions and this vertices correspond to $c_3$ and $\tilde c_3$ terms in (\ref{eq:sixderivative}). Indeed, the results of \cite{doi:10.1063/1.529470} and \cite{Fulling:1992vm} give exactly these terms in case of six derivatives. Moving to eight derivatives (and of course three Riemanns), parity even terms were also classified in \cite{Fulling:1992vm}. In four dimensions there is only one independent term that can be chosen in the following form:
\be
R^{\mu\nu\rho\sigma}\nabla_\rho R^{\gamma\delta\beta}\,_\mu \nabla_\sigma R_{\gamma\delta\beta\nu}.
\ee
To simplify this term we can integrate $\nabla_\sigma$ by parts, because of (\ref{divR}) the derivative has to go on the second Riemann but then the two derivatives are anti-symmetrized so the term is reduced to four Riemann tensors. 

For three Riemanns, in order to classify parity odd eight-derivative terms we used {\it{Invar}} tensor package~\cite{MartinGarcia:2008qz}. In four dimensions there is again only one independent term with three Riemanns that can be chosen in the following form:
\be
\epsilon_{\mu\nu\rho\sigma} R^{\alpha\beta\mu\nu}\nabla^\rho R_\alpha\,^{\delta\gamma\kappa}\nabla^\sigma R_{\beta\gamma\delta\kappa}.
\ee
Now we integrate $\nabla^\sigma$ by parts, if it acts on the second Riemann the derivatives are again anti-symmetrized, while if it acts on the first Riemann we can use the second Bianchi identity (\ref{eq:bianchiII}) on indices $\sigma$, $\alpha$ and $\beta$ to get a structure proportional to $\epsilon_{\mu\nu\rho\sigma} R^{\sigma\alpha\mu\nu}$, which is zero due to the first Bianchi identity~\footnote{One can contract three indices in (\ref{eq:bianchiI}) with the epsilon tensor, and show that each one of the three terms of the Bianchi identity is proportional to $\epsilon_{\mu\nu\rho\sigma} R^{\sigma\alpha\mu\nu}$.}:
\be\label{eq:bianchiI}
 R^{\sigma\alpha\mu\nu}+ R^{\mu\sigma\alpha\nu}+ R^{\alpha\mu\sigma\nu}=0.
\ee

Parity even four-Riemann terms were classified in \cite{Fulling:1992vm}, while for parity odd terms we can use reference \cite{doi:10.1063/1.529470}. The results are that the terms present in the action (\ref{eq:actionfull}) are the only independent ones, with no extra derivatives~\footnote{The latter reference, which applies to parity odd operators, classified terms up to algebraic equivalence (which means that they consider as dependent different operators that are built out of products of operators that appeared at lower orders), consequently our $\Lambda_-$ term is not presented as an independent one. In fact, there is no single algebraic-independent parity-odd four-Riemann operator, which implies that, if there is a linearly independent one, it must be a product of lower order scalar operators, which have been classified. Therefore, the results of \cite{doi:10.1063/1.529470} are enough to argue that the term we include is the only possible term.}.

\subsection{Review of causality constraints on coefficients in the effective action\label{sec:causality}}

\subsubsection*{Quartic operators}

We begin with a brief summary of the argument of~\cite{Gruzinov:2006ie}. The authors considered the effect of quartic Riemann operators (\ref{eq:action}) on the dispersion relation of a graviton propagating in a background with $R_{\mu\nu}=0$. This takes the following form:{~\footnote{\cite{Gruzinov:2006ie} did not consider the parity odd term, however here we present the easily-generalized results.}}
\be\label{eq:superluminarR4}
k^2=\frac{64}{\Lambda^6}\l( S^{\alpha\beta}e_{\alpha\beta}\r)^2+\frac{64}{\tilde\Lambda^6}\l( \tilde S^{\alpha\beta}e_{\alpha\beta}\r)^2+\frac{64}{\Lambda_-^6}S^{\alpha\beta}e_{\alpha\beta} \tilde S^{\mu\nu}e_{\mu\nu},
\ee
where $k^\mu$ is the gravitons four-momentum, $e_{\alpha\beta}$ the polarization tensor and
\be
S_{\mu\nu}=k^\alpha k^\beta R_{\mu\alpha\beta\nu}, \quad \tilde S_{\mu\nu}=k^\alpha k^\beta \tilde R_{\mu\alpha\beta\nu}.
\ee
By picking different graviton polarizations and backgrounds we first conclude that indeed in order to avoid superluminal graviton propagation the first two coefficients must be positive and also that the coefficient of the parity odd term cannot be very large, namely,
\be\label{eq:superluminality1}
\Lambda^6>0\,,\qquad \tilde\Lambda^6>0\,, \qquad\frac{1}{\Lambda_{-}^{12}}\leq\frac{2}{\Lambda^6\tilde\Lambda^6}.
\ee
Constraints on the positivity of the parity even terms were obtained from independent arguments involving analyticity of the graviton amplitudes in~\cite{Bellazzini:2015cra}. 

To ensure that there is an obvious inconsistency, one should be sure that the time advance resulting from the change in the dispersion relation is larger than the time-delay that is present in normal GR.  Clearly, in order to trust the equations of motion, the curvature scale $\rho$ cannot be larger than $\Lambda$. In GR, the time delay will generically be proportional to $1/\rho$. For example for a black hole the time delay is of order of the Swartzschild radius. For $k\lesssim \Lambda$, the time advanced does not appear to be generically larger than the GR time delay. Naively, one can try to go to high $k$'s for the graviton, as the right hand side of~(\ref{eq:superluminarR4}) grows as $k^4$. However, for $k\gtrsim \Lambda$, other operators present in our EFT and containing more powers of Riemann can possibly give a contribution that grows faster than $k^4$. Consequently, we do not seem to be able to guarantee that the overall time advancement can beat the GR time delay within the validity of the computation. One can in principle be worried even by graviton propagation faster than in GR, however this does not seem to be a necessary requirement, as such a phenomenon already happens in QED~\cite{Drummond:1979pp}. We therefore consider the constraint given in~(\ref{eq:superluminality1}) simply as indicating a somewhat preferred  region, but we cautiously suggest that whole of the parameter space should be explored. 

Of course, while the set of inequalities in (\ref{eq:superluminality1}) means that, when they are violated, there is faster-than-GR propagation, we cannot state for sure that, by performing some additional analysis, one cannot find that even in the parametric regime allowed by (\ref{eq:superluminality1}), superluminality is present~\footnote{For example, the analysis of~\cite{Gruzinov:2006ie} is insensitive to the superluminal propagation induced by the  cubic operators, as the different analysis of~\cite{Camanho:2014apa} reveals.}. In such a case, the sensible parameter space should be further reduced.

\subsubsection*{Cubic operators}

Let us now turn to reviewing the constraints on the cubic couplings $c_3$ and $\tilde c_3$ derived in~\cite{Camanho:2014apa}~\footnote{We thank Sasha Zhiboedov for discussions about technical aspects of~\cite{Camanho:2014apa}.}. The authors  assume that the new UV physical states that, as we explained in our case, must be present at the scale $\Lambda$, couple at tree level (that is the scattering amplitude remains a meromorphic function of the kinematic invariants). The idea is to consider small-angle scattering of a gravitational wave off some arbitrary (Standard Model) particle.  This process is dominated by the eikonal approximation, or equivalently by ladder diagrams. The latter exponentiate in the impact-parameter representation and the leading order answer depends exclusively on the on-shell cubic vertices present in the theory. The result in the GR limit is the phase shift associated with the Shapiro time delay for gravitational waves. Of course this is always positive. At the impact parameter $b\sim c_3^{1/4}/\Lambda$ corrections from the $R^3$ vertices will become significant. Crucially, within the stated assumptions, these corrections can become observable while the calculation is under control. The result of \cite{Camanho:2014apa} is that independently of the signs of $c_3$ and $\tilde c_3$ there will be a polarization that instead of a time delay acquires time advance. This violates causality because the time-advancement becomes larger than the time-delay in GR. 

Ref.~\cite{Camanho:2014apa} also showed that the only way to cure superluminality is by introducing (an infinite tower of) higher spin particles with mass of order $\Lambda/c_3^{1/4}$ coupled both to the graviton and to the matter particle on which the graviton is scattering, Standard Model particles in our case. However, these new particles would mediate a new force between all Standard Model particles, basically through the same set of diagrams but with graviton replaced with the second matter field. The range of such a force will be of order $r\sim c_3^{1/4}/\Lambda$ and the strength will be parametrically equal to gravitational. Recently an independent argument that uses AdS/CFT correspondence~\cite{Afkhami-Jeddi:2016ntf} was given that leads to conclusions equivalent to those in~\cite{Camanho:2014apa}.  This argument puts a strong constraints on $c_3$ and $\tilde c_3$ for the values of $\Lambda$ of interest to us. We therefore conclude that, within the set of assumptions about the UV completion made in~\cite{Camanho:2014apa}, the six-derivatives operators made with three Riemanns must be suppressed by a scale smaller than the one probed in laboratory experiments, and therefore are completely negligible in the context of compact objects. 
 
Similarly to the case of quartic operators with negative coefficients, that we discussed just above, the cubic operators always lead to faster-than-GR propagation, independently of any assumption about the UV completion. However, this is not obviously enough to violate causality, and therefore we conclude that we cautiously should consider also the cubic operators.

\section{Classical equations of motion and Numerical Simulations\label{sec:numerical}}

The equations of motion resulting from action (\ref{eq:action}) in the $R_{\mu\nu}=0$ background are 
\begin{align}
R^{\mu\alpha} - \frac{1}{2}g^{\mu\alpha} R  = \frac{1}{\Lambda^6}\left(8  R^{\mu \nu \alpha \beta}  \nabla_{\nu} \nabla_{\beta}\mathcal{C} +  \frac{1}{2}  g^{\mu \alpha}  \mathcal{C}^2\right) + \frac{1}{\tilde\Lambda^6} \left( 8\tilde{R}^{\mu\rho\alpha\nu}\nabla_{\rho}\nabla_{\nu}\tilde{\mathcal{C     }} +  \frac{1}{2} g^{\mu\alpha} \tilde{\mathcal{C     }}^2\right)
\nonumber
\\
+\frac{1}{\Lambda_-^6} \left( 4\tilde{R}^{\mu\rho\alpha\nu}\nabla_{\rho}\nabla_{\nu}\mathcal{C} + 4 R^{\mu\rho\alpha\nu}\nabla_{\rho}\nabla_{\nu}\tilde{\mathcal{C     }} +  \frac{1}{2} g^{\mu\alpha} \tilde{\mathcal{C     }}\mathcal{C}\right).
\label{eq}
\end{align}
where we used $R_{\mu\nu}=0$ to simplify the right hand side. 

Higher derivative terms are sometimes feared because of potential instabilities of the equations of motion. EFTs always contain higher derivative operators, however, as long as one works at energies below the cutoff, instabilities never occur. Indeed, by definition, the solutions are small, perturbative deformations of the leading term in the action, in our case Einstein-Hilbert term, that only has healthy well-behaved solutions. 

 As we mentioned, one can consider also the case of an effective theory where the leading operators are cubic, which is given in~(\ref{eq:sixderivative}). In this case, the equations of motion are given by
\bea\label{eq:sixderivativeequations}
&&R_{\mu \nu} -\frac{1}{2}g_{\mu\nu}R = \frac{6 c_3}{\Lambda^4} \left(\nabla_{\alpha} R_{\mu \beta \delta \gamma}\right)\left(\nabla^{\beta} R_{\nu} {}^{\alpha \delta \gamma}\right)\\ \nonumber
&&\qquad\qquad\qquad+ \frac{2 \tilde c_3}{{\Lambda}^4}  \left\{ 
\epsilon_{\mu \delta \rho \sigma} R_{\nu}{}^{\alpha\beta\gamma} \left( \nabla^{\sigma}\nabla_{\alpha} R_{\beta\gamma}{}^{\delta\rho}\right) + \epsilon_{\nu \delta \rho \sigma} R_{\mu}{}^{\alpha\beta\gamma} \left( \nabla^{\sigma}\nabla_{\alpha} R_{\beta\gamma}{}^{\delta\rho}\right) \right. \nonumber\\
&&\qquad\qquad\qquad+ \left. \epsilon_{\mu \delta \rho \sigma} \left(\nabla_{\alpha} R_{\beta\gamma}{}^{\rho\sigma}\right) \left(\nabla^{\delta} R_{\nu}{}^{\alpha\beta\gamma}\right) + \epsilon_{\beta \gamma \rho \sigma} \left(\nabla_{\alpha} R_{\mu\delta}{}^{\rho\sigma}\right)\left(\nabla^{\delta} R_{\nu}{}^{\alpha\beta\gamma}\right) \right. \nonumber\\
&&\qquad\qquad\qquad+ \left. \epsilon_{\nu \delta \rho \sigma} \left(\nabla_{\alpha} R_{\beta\gamma}{}^{\rho\sigma}\right)\left(\nabla^{\delta} R_{\mu}{}^{\alpha\beta\gamma}\right) \right\}\nonumber\ .
\eea

In this paper, we will study the phenomenology of our effective field theory confining ourselves to the inspiralling phase where the velocity is non relativistic. This is the so-called post-Newtonian regime. Another regime that is prone to a perturbative treatment is the study of the quasi-normal modes. We leave this study to a subsequent publication. Of course, it would be interesting to study the merging phase, where the velocity is relativistic. In standard GR, this is done numerically using the renowned codes that can handle horizons and singularities~\cite{Pretorius:2005gq}. In this section, we wish to briefly highlight how it appears to us a potential adaptation of the same GR codes can be used to simulate the coalescence of black holes within our effective field theory. For simplicity, we will refer only to eq.~(\ref{eq}), but everything we say in this section applies equally also to (\ref{eq:sixderivativeequations}).

The equations of motion we wish to solve are given in~(\ref{eq}). However, they cannot be solved numerically as is. In fact, the terms on the right-hand side contain more than two time derivatives. If solved as is, these terms will induce  exponentially growing unstable modes that would destroy the ordinary GR solution. Why this statement does not rule out completely from the get-go the modifications of the Einstein-Hilbert action we are proposing? The answer is that we should not overinterpret the meaning of effective field theories. The new terms that we are inserting represent a consistent theory only in the limit in which these terms provide small perturbations. When the correction becomes of order one, the whole series of terms that we neglected to write down under the assumption that the correction is small will become important, and so, an effect of order one originating from the right-hand side of (\ref{eq}) simply cannot be trusted.

How therefore can  we numerically solve this equation~(\ref{eq}) assuming that the effect of the terms on the right-hand side are small? Here we outline how we can adapt a standard perturbative method to the study of our specific effective operators that requires rather possibly minimal modifications of the existing codes, but that we do not implement for lack of technical knowledge of the relevant codes (and not because the problem is mathematically-ill posed or there are physical instabilities in this theory). Our proposed approach is valid for the entire merger event for $\Lambda's\gtrsim 1/r_s$, where $r_s$ is the Schwarzschild radius of the compact object, while for $\Lambda's\lesssim 1/r_s$ it should be trusted, with a slight modification, until the distance $r$ between the compact objects is $r\gtrsim 1/\Lambda$, as we discussed next. Suppose we have solved the ordinary, $\Lambda,\tilde\Lambda,\Lambda_{-}\to \infty$, equations to obtain the solution of the inspiralling, merging, and ring down phases of GR, which is what is normally done to obtain the templates for experiments such as LIGO. Let us denote the obtained metric, and resulting Riemann tensor with the subscript $_{(0)}$: $g_{(0),\mu\nu},\, R_{(0),\mu\nu\rho\sigma}$. Let us suppose this solution is stored in our computer. Let us denote the leading correction from the right-hand side of~(\ref{eq}) with the subscript $_{(1)}$: $g_{(1),\mu\nu},\, R_{(1),\mu\nu\rho\sigma}$, so that the full solution is $g_{\mu\nu}=g_{(0),\mu\nu}+g_{(1),\mu\nu}$. To obtain the leading correction $g_{(1),\mu\nu}$, we then have just to solve~\footnote{We point out the following. The recipe we are going to describe will give the correct prediction of the EFT at order $1/\Lambda^6$. Iteration of this procedure using the same eq.~(\ref{eq}) would naively give the corrections to order $1/\Lambda^{12}$. However, at order $1/\Lambda^{12}$ one expects many new operators to appear in the EFT, so that the equation of motion, at this order, would need to be modified. Still, a similar procedure to what we describe here can be implemented. }

\begin{eqnarray}\label{eq_temp_source}
&&  R^{\mu\alpha} - \frac{1}{2}g^{\mu\alpha} R  =\\  \nonumber
&& \qquad\qquad \left[\frac{1}{\Lambda^6}\left(8  R^{\mu \nu \alpha \beta}  \nabla_{\nu} \nabla_{\beta}\mathcal{C} +  \frac{1}{2}  g^{\mu \alpha}  \mathcal{C}^2\right) + \frac{1}{\tilde\Lambda^6} \left( 8\tilde{R}^{\mu\rho\alpha\nu}\nabla_{\rho}\nabla_{\nu}\tilde{\mathcal{C     }} +  \frac{1}{2} g^{\mu\alpha} \tilde{\mathcal{C     }}^2\right)\right.\\ \nonumber
&&\qquad\qquad
\left.+\frac{1}{\Lambda_-^6} \left( 4\tilde{R}^{\mu\rho\alpha\nu}\nabla_{\rho}\nabla_{\nu}\mathcal{C} + 4 R^{\mu\rho\alpha\nu}\nabla_{\rho}\nabla_{\nu}\tilde{\mathcal{C     }} +  \frac{1}{2} g^{\mu\alpha} \tilde{\mathcal{C     }}\mathcal{C}\right)\right]_{g_{\mu\nu}=g_{(0),\mu\nu}} \ ,
\end{eqnarray}
with appropriate initial and boundary conditions.

The difference between (\ref{eq_temp_source}) and (\ref{eq}) is that in (\ref{eq_temp_source}) the right hand side is a known source, {\it i.e.} it does not contain any term in the unknown $g_{(1),\mu\nu}$. The differential operator acting on $g_{(1),\mu\nu}$ is the same as in the standard  Einstein equations, so it is second order in derivatives, and it {\it does not} lead to any unstable solutions or ill-posed mathematical problems. Notice, furthermore, with large enough $\Lambda$'s, {\it i.e} $\Lambda\gtrsim 1/r_s$, where $r_s$ is the Schwarzschild radius of the black hole, the right-hand side of (\ref{eq_temp_source}) is small over the whole spacetime of interest for the simulation, i.e. even at the horizon, so the perturbative expansion will apply.  Notice that solving this problem is very similar to solving the usual Einstein equations: the only difference is that there are two iterations: in the first iteration, ones saves $g_{(0),\mu\nu}$ to compute the sources, and in the second iteration one adds those to the right hand side and solve again using the standard Einstein solver. Additionally, if this were to be simpler, one could also linearize the left-hand side of (\ref{eq_temp_source}) in $g_{(1),\mu\nu}$ and solve the linearized Einstein equations with the known source provided by the right hand side of~(\ref{eq_temp_source}).

Let us comment to what extent this same simulations can be trusted in the regime $\Lambda's\lesssim 1/r_s$. The effective field theory breaks down for distances shorter than $1/\Lambda$'s. However, our assumptions about the UV complition tell us that the effects of new physics are suppressed away from the horizon. Therefore, even though one cannot trust the numerical solution inside the region $r<1/\Lambda$, one can still perform the simulation with exactly the same algorithm we just described by adding the following modiification: one should smoothly damp the right-hand-side source for distances $r\lesssim 1/\Lambda$'s, so that this never becomes non-perturbative, mimicking in this way the softening  of the assumed UV complition. While the black holes themselves are at distances longer than $1/\Lambda$'s, the emitted radiation, whose frequency is slower than $1/\Lambda$, can be trusted as being universal (and in particular independent of the softening procedure)~\footnote{There is a slightly technical issue associated with this setup that we would like to highlight. Strictly speaking, for $\Lambda's\lesssim 1/r_s$ this algorithm simulates a particular UV completion, given by the specifically-chosen softening of the source term. The only effect of this short-distance procedure at long distances is that it rescales (renormalizes) the coefficients, $\Lambda$'s, of the cubic or quartic operators. Therefore, in interpreting the size of the effect in terms of $\Lambda$'s, one should adjust (i.e. renormalize) the values of the simulated $\Lambda$'s  with the result of the PN calculations that we perform later on.}. In this regime parametric regime $ \Lambda's\lesssim 1/r_s$ simulations can be trusted only in the regime $v\lesssim 1$, where PN calculations are also reliable. However, there are clear different advantages in both approaches.

Let us make some comments on possible technical issues. In order to evaluate the right-hand side of (\ref{eq_temp_source}), one needs to evaluate four derivatives of the metric $g_{(0),\mu\nu}$. This probably will require to store the metric for a few time steps, which we are unable to judge if it is a technical challenge. 

Let us also comment more on the initial conditions for~(\ref{eq_temp_source}). Once the black holes are enough far apart in the past, we are tempted to argue that possibly the initial conditions are well approximated by  the perturbed metric of two isolated black holes, where the perturbed metric is obtained by solving perturbatively~(\ref{eq_temp_source}) again, but this time in a static and isolated configuration (where the boundary conditions are vanishing) and the right hand side of~(\ref{eq_temp_source}) is this time known analytically: it is the one given by a Kerr metric.

Finally, let us add a discussion about another possible technical issue~\footnote{We thank Frans Pretorius for pointing out to us the existence of such a potential difficulty and of the two possible solutions mentioned here. We also thank him for mentioning the possibility of using the standard Einstein solver in~(\ref{eq_temp_source}) instead of linearizing the left hand side, which makes the required modifications to existing code smaller.}. Notwithstanding the smallness of the corrections from our EFT, the accumulation of the phase difference with respect to the GR solution could, after enough orbits, become so large that, at a given instance of time, the actual solution is out of phase with respect to the GR one, and therefore naively one should not be able to solve perturbatively in $g_{(1),\mu\nu}\ll g_{(0),\mu\nu}$. If present, this issue can potentially be addressed with one of the two following approaches. In the first approach, one could solve (\ref{eq_temp_source}) in multiple iterations with smaller effective coupling parameters that slowly
build up to the desired values. In a second approach, one could break the full evolution down into tiny segments where in each one we evolve the system for a small time such that the phase accumulation is small and the solution given by  (\ref{eq_temp_source}) is reliable, and then we start the next segment
with the corrected solution as the initial conditions. In reality, we might need to use both method in some extreme cases, and run the simulation multiple times. For large enough values of $\Lambda$ numerical procedure outlined above should converge and give the precise results, however, the signal for those values of parameters will also be small and potentially high signal to noise events will be required to detect it. On the other hand, for smaller values of $\Lambda$ close to $1/r_s$, there is a possibility that significant systematic errors will be present due to the sensitivity to the UV completion of the theory. Indeed, it is plausible that locally in some near-horizon regions excitations of the UV modes become important rendering our approximate equations insufficient. A further, most likely numerical, study is required to determine for which values of the parameters this can happen and what is the largest value of the signal that can be obtained while staying in the regime of validity of the EFT.\footnote{We thank the referee for bringing to our attention two papers~\cite{Okounkova:2017yby,Cayuso:2017iqc} %that appeared after our preprint and 
that study numerical techniques which can be used for simulating merger events in extended gravitational theories containing higher-derivative operators.}

None of the authors of this paper have the expertise to tackle the numerical solution of (\ref{eq_temp_source}). However, we do hope that the strategy described in this section might encourage the experts of the field to attempt to solve this numerical problem, so that we will be able to study the effect of the UV-extension of GR in the relativistic regime as well. Instead, we will now move on to study the post-Newtonian regime.

\section{Outline of the post-Newtonian calculation}\label{sec:techsummary}

We are interested in studying the effects of adding the higher derivative terms
\ba
&&\frac{\mpl^2}{\Lambda^6}(R_{\alpha \beta \gamma \delta} R^{\alpha \beta \gamma \delta})^2 \ , \quad \quad \frac{\mpl^2}{\tilde\Lambda^6}(\epsilon^{\alpha \beta}\,_{\mu \nu}R_{\alpha \beta \gamma \delta} R^{\mu \nu  \gamma \delta})^2 \\ \nonumber
&&\quad \text{and} \quad \frac{\mpl^2}{\Lambda_-^6}(R_{\alpha \beta \gamma \delta} R^{\alpha \beta \gamma \delta})(\epsilon^{\alpha \beta}\,_{\mu \nu}R_{\alpha \beta \gamma \delta} R^{\mu \nu  \gamma \delta})\ ,
\ea
to the canonical Einstein-Hilbert action in the post-Newtonian regime. We talk about the EFT with cubic operators at the end of this section. The energy scales $\Lambda$, $\tilde \Lambda$ and $\Lambda_-$ control at what scales these terms become relevant. When treated perturbatively we can compute the effects of these terms. This leads to predictions that can be in principle measured, leading to a discovery of modification of GR, or, in absence of detection, can be used  put lower bounds on the size of $\Lambda$, $\tilde \Lambda$ and $\Lambda_-$. There are many different physical effects these terms can modify, but here we focus on the inspiral problem. We therefore compute the corrections that these terms generate both to the instantaneous potential as well as to the form of the radiation coupling (i.e. corrections to the quadrupole formula). This is sufficient to compute the modification to the gravitational wave signal from the inspiralling regime. To perform these calculations we utilize the EFT framework developed by Goldberger and Rothstein~\cite{Goldberger:2004jt}. 

The EFT framework is proposed for systematically calculating to any order in the PN expansion. Such an EFT is the result of integrating out gravitational and matter perturbations with wavelength shorter than the size of the extended object involved. The properties of the compact sources are encapsulated by a particular series of operators with coefficients respecting the symmetries of the extended object~\footnote{This is similar in spirit to the operators that we add to the GR action in Sec.~\ref{sec:action}: there we include all the operators compatible with diffeomorphism  invariance and built out of the graviton, now we include also operators build out of the world-line of the extended objects.}. Given a model of stellar structure, the precise value of the coefficients in front of these operators can be found from UV matching conditions. In the case where the extended object is a black hole, the properties of the source are captured by the mass and spin of the black hole. An EFT also has the advantage of having manifest power counting in the expansion parameters of the theory. In the case of this EFT for compact objects, such an expansion parameter is the relative velocity of the extended objects, $v$. In the non-relativistic limit ($v\ll 1$), the size of various post-Newtonian corrections compared to the leading newtonian potential can often be estimated by some simple power counting rules.

The gravitons in the problem can be generally separated into two categories according to the energy and momentum they carry. For gravitons that are responsible for mediating long range interactions between two extended objects, the typical energy and momentum carried by these gravitons is $(p^0 \sim v/r,{\bf p}\sim 1/r)$, while for gravitons that are emitted by the system, the typical energy and momentum carried by these gravitons is $(p^0 \sim v/r,{\bf p}\sim v/r)$ since they should be on shell (i.e. gravitational wave satisfy the relativistic dispersion relation~$p^0=p$).~\footnote{\label{footnote:omega}It is important to point out that in the non-relativistic regime of the inspiral, the velocity $v$ and the distance between the extended objects are related by the virial theorem as $v^2\sim G M/r$. The rotational frequency ($\omega \sim v/r$), as a result, is at order $v^3$. It should be noted that this power counting rule can be extended to a much more complicated potential $V$ as long as the potential respects a rotational symmetry. The rotational frequency can be found from $\omega \sim \sqrt{\frac{1}{r}\frac{{\rm d} V}{{\rm d} r}}$ and the order of PN-corrections can be read off accordingly.} With this in mind, we can work out the Feynman rules from the Einstein-Hilbert action, and in particular the lowest order vertices that captures the interaction between the source (mass and spin of a black hole) and the graviton. These are summarized in Appendix~\ref{app:A}.

Such a framework can be extend to include new interactions between the gravitons in the form of our leading higher dimensional operators. The new quartic interaction vertices from the new EFT operators are summarized in Appendix.~\ref{App:new_quartics}. In addition to~$v$, the post-Newtonian expansion parameter, there is one more expansion parameter of the new EFT, which is ${p}/\Lambda$. When this expansion parameter becomes $O(1)$, additional higher-dimensional operators or new resonances lead to significant deviations from our leading order predictions~\footnote{As we anticipated in the introduction and in Sec.~\ref{sec:action}, some observations/experiments are sensitive to the regime where $p\gg \Lambda$. We discuss more this regime in Sec.~\ref{sec:otherexp}.}.

The broad strategy of the calculation is the following. From far away, the inspiralling binary can be thought as a single compact object endowed with a small extension in space. It emits gravitational waves by the oscillations of its multipoles. The effective action (which is equivalent to its effective equations of motion) takes therefore the form of the one of a particle that is characterized by its multipole moments. In the center of mass frame, this takes the form
\begin{equation}\label{eq:PNeffectiveaction}
S_{\rm ext.\, obj.}=\int { d} t \,\left\{ \left[m_1+m_2+\frac{1}{2} \mu(t) {\bf v}_{\rm rel}^2 - V(r(t)) \right] + \frac{1}{2}Q_{ij}(t) R^{i0j0} -  \frac{1}{3}J_{ij}(t) \epsilon_{jkl}R^{kli0}+\ldots \right\},
\end{equation}
where $\mu$ is the reduced mass of the system, and ${\bf v}_{\rm rel}$ is the relative velocity between the inspiralling binary. We also included in the effective one-body action the kinetic energy and the potential energy $V(r(t))$, because, even though they describes internal degrees of freedom from the point of view of the single-body system, they allow us to compute the time dependence of the multipoles. Thinking of the binary as an extended object allows us to compute the gravitational wave emission directly using the standard multipole formulas. Therefore, the problem of GW emission is reduced to computing the effective action (\ref{eq:PNeffectiveaction}) of the binary system starting from the action of two point-like particles
\ba\label{eq:point}\nonumber
&& S_{\rm EH+p.p.}=S _{\rm eff}+\\ \nonumber
&&\qquad+ \int d^4x \,\left\{ \delta^{(3)}(\vec x-\vec x_1) \left( m_1(1+ {\bf v}_1^2/2)+ d_{2}^{(1)} \;\sqrt{g_{\alpha\beta}\dot x_1^\alpha\dot x_1^\beta}R^{\mu\nu\rho\sigma}R_{\mu\nu\rho\sigma}+\ldots\right)\right.\\ 
&&\qquad\qquad\qquad\left.+\delta^{(3)}(\vec x-\vec x_2)\left( m_2(1+{\bf v}_2^2/2)+\ldots \right)\right\}, 
\ea
where $S _{\rm eff}$ is our extension of GR of Sec.~\ref{sec:radiation}, and $\ldots$ represent the coupling between gravity and particles, given explicitly in (\ref{eq:pointverteces}) and other higher order terms associated to the finite size of the point particle, out of which we just wrote a representative one, $\sqrt{g_{\alpha\beta}\dot x^\alpha\dot x^\beta} R^{\mu\nu\rho\sigma}R_{\mu\nu\rho\sigma}$, which is proportional to an unknown coupling constant $d_2^{(1)}$.

While the effective quadrupole is trivially given by $Q^0_{ij}(t)=\sum_a m_a \left(x_a(t)^i x_a(t)^j-\frac{1}{3} x_a(t)^2\delta^{ij}\right)$ at leading order, the expressions for the multipoles and their time-dependence become more subtle at post-Newtonian level. In particular, the derivation of the time dependence requires knowledge of the potential between the two compact objects. To compute all of this, the EFT of gravity for extended objects~\cite{Goldberger:2004jt} provides the aforementioned Feynman rules (see appendix~\ref{app:A}), that we extend here to include our new vertices.
The computation of the effective multipoles and of the potential is presented in Sec.~\ref{sec:potential} and~\ref{sec:radiation}, and it has the following schematic structure:
\begin{enumerate}
\item We identify the leading contributing diagrams using the scaling arguments of  \cite{Goldberger:2004jt}. 
\item In order to facilitate the computations, we identify recurring computable subsections of the graphs.

\item We use these to evaluate the graphs in question.
\end{enumerate}

Notice that evaluation of these Feynman diagrams involves integration over momenta, represented as loop diagrams. This should not mislead us to think that we are computing quantum corrections. All the effects we are computing here are classical ones, higher loops are suppressed by powers of $v$.

In the following, we summarize the main results of Sec.~\ref{sec:potential} and~\ref{sec:radiation}, which contain only the technical details (and therefore can be skipped by an uninterested reader). The corrections to the potentials due to $\mathcal{C}^2$ and $\tilde{\mathcal{C}}^2$ are 
\be\label{Lambda_pot}
\Delta V_{\Lambda}= \frac{2}{\pi^6}\frac{G m_1 m_2}{r} \left(\frac{2\pi}{\Lambda r}\right)^6\frac{4 G^2 (m_1^2+m_2^2)}{r^2}
\ee
and
\be\label{Lambda_tilde_pot}
\Delta V_{\tilde \Lambda} = \frac{216}{11 \pi^6}\frac{G m_1 m_2}{r} \left(\frac{2\pi}{\tilde \Lambda r}\right)^6 \frac{4 G^2 \left(m_1S_1^{i}+m_2 S_2^{i}  \right)  \epsilon_{inm}v_{12}^n r_{12}^m}{r^{4}},
\ee
where $S^i = \epsilon^i{}_{jk} S^{j k}$, and $S_{ij}$, carefully defined in App.~\ref{app:A}, parametrizes the spin of the a compact object. For a maximally rotating black hole with spin along the $z$-direction, $S_z = G m^2$. From these we extract the correction to the orbital frequency $\omega$. For the~$\tilde{\mathcal{C}}^2$, the modification to gravitational wave emission from the potential is subleading with respect to the modification of the multipole moments that we discuss next. The same is true also for the operator~$\mathcal{C}\tilde{\mathcal{C}}$, for which we therefore neglect to compute the correction to the potential.

We have organized our result in (\ref{Lambda_pot}) and (\ref{Lambda_tilde_pot}) by factoring out the combination $\left(2\pi/(\tilde \Lambda r)\right)$, with the idea that our EFT is reliable only for $r\gtrsim (2\pi)/\Lambda$. Estimating the value when an EFT breaks down to order one numbers is not universal, {\it i.e.} it depends on the UV completion. The factor of $(2\pi)$ we have chosen  is expected to be an upper bound to the smallest value of $r$ at which our EFT is under control. This means that the estimate of the maximum sizes of the effect that one can exact from (\ref{Lambda_pot}), and (\ref{Lambda_tilde_pot}), and from the similar equations for the multipoles (\ref{eq:quad_one}) and (\ref{eq:quadtwo}) that follow, by setting $r=(2\pi)/\Lambda$ and $v=1$, should be meant only at a conservative level. In particular, the fact that the dependence on $(\Lambda r)$ is raised to the sixth power, makes the ambiguity in the estimate (not in the calculation) rather large. 

The $\mathcal{C}^2$ term can also lead to corrections to the quadrupole moment of a binary system, which will change the amplitude of the radiation. The modification to the quadrupole can be expressed as a renormalization of the initial quadrupole moment as
\ba\label{eq:quad_one}
Q_{ij} = \left(1+ \frac{21}{2\pi^6} \left(\frac{2\pi}{\Lambda r}\right)^6 \left(\frac{2G (m_1+m_2)}{r}\right)^2\right)Q^{(N)}_{ij},
\ea
where $Q^{(N)}_{ij}$ is the quadrupole of a binary system in GR at leading order. Similarly, the $\tilde{\mathcal{C}}^2$ and $\mathcal{C}{\tilde{\mathcal{C}}}$ terms can lead to corrections to the current quadrupole of a binary system. We express this modification as
\ba\label{eq:quadtwo}
J_{ij} &\rightarrow&\left(1-\frac{36}{\pi^6}\left(\frac{2\pi}{\tilde \Lambda r}\right)^6 \left(\frac{2G (m_1+m_2)}{r}\right)^2\right) J^{(N)}_{ij}\\ \nonumber 
&& +\frac{63}{8\pi^6} \left(\frac{2\pi}{\Lambda_- r}\right)^6 \left(\frac{2G (m_1+m_2)}{r}\right)^2 Q^{(N)}_{ij} \; ,
\ea
where $J^{(N)}_{ij}$ is the current quadrupole of a binary system in GR at leading order  (see Sec.~\ref{sec:radiationsummary} for the definition of $J^{(N)}_{ij}$). The corrected frequency and the corrected multipoles allow us to compute the GW emission using the standard formulas. 

Notice that the parameters associated to the finite size effects in the point-particle action in (\ref{eq:point}), such as $d_2^{(1)}$, did not appear in the former expressions. In reality, they do contribute both to the potential and the multipoles of the effective one-body system, but, as we argue in Sec.~\ref{sec:observeligo}, their effect is negligible for the regime of interest of post-Newtonian calculations.

In table~\ref{tab:vcount} in Sec.~\ref{sec:observeligo}, we summarize the post-Newtonian order of each contribution. Readers who are mainly interested in the effect of these terms in LIGO observables can skip to Sec.~\ref{sec:observeligo}.

Let us comment briefly also on the EFT with cubic operators~(\ref{eq:sixderivative}). In this case, one could expect that the leading corrections to the potential and multipoles appear at $v^2$ order. However, we find that these, as well as the order $v^3$, contributions cancel, and the leading corrections are expected to arise not earlier than order $v^4$. The computation of these effects becomes quite challenging (at least to us) given the proliferation of diagrams at this order. We  leave this problem to  future work, possibly attacking it with the help of on-shell techniques as discussed in~\cite{Neill:2013wsa}. In summary, at parametric level, the physical consequences of the theory with cubic operators are expected to be quite similar to the ones we discuss in greater detail for the theory with quartic operators, with just the replacement $(\Lambda r)^6\to (\Lambda r)^4$ in every formula in the rest of the paper, as, as we said, we expect the leading effect to be 2PN in this case as well. The factor of $(\Lambda r)^4$ makes the observational signatures of this EFT somewhat more promising.

\section{Corrections to the potential}\label{sec:potential}

In this section, we will show the various diagrams that will lead to corrections to the gravitational potential of a binary system. As outlined in section~\ref{sec:techsummary}, this is important to compute the time-dependence of the multipoles. We will not compute this correction to the potential for the $\mathcal{C}{\tilde{\mathcal{C}}}$ operator, as this is subleading.

\subsection{Corrections to potential: $\mathcal{C}^2$ term}

The diagrams that correct the gravitational potential are those which do not involve external gravitational waves. Given our quartic vertices, there are only two different topologies:
one is obtained by inserting one of our quartic vertices and contracting a pair of legs with each particle trajectory, and the other is when three legs are contracted on a single particle trajectory.  

\subsubsection{Cross Diagram}
\begin{figure}[!ht]
  \centering
      \includegraphics[width=0.3\textwidth]{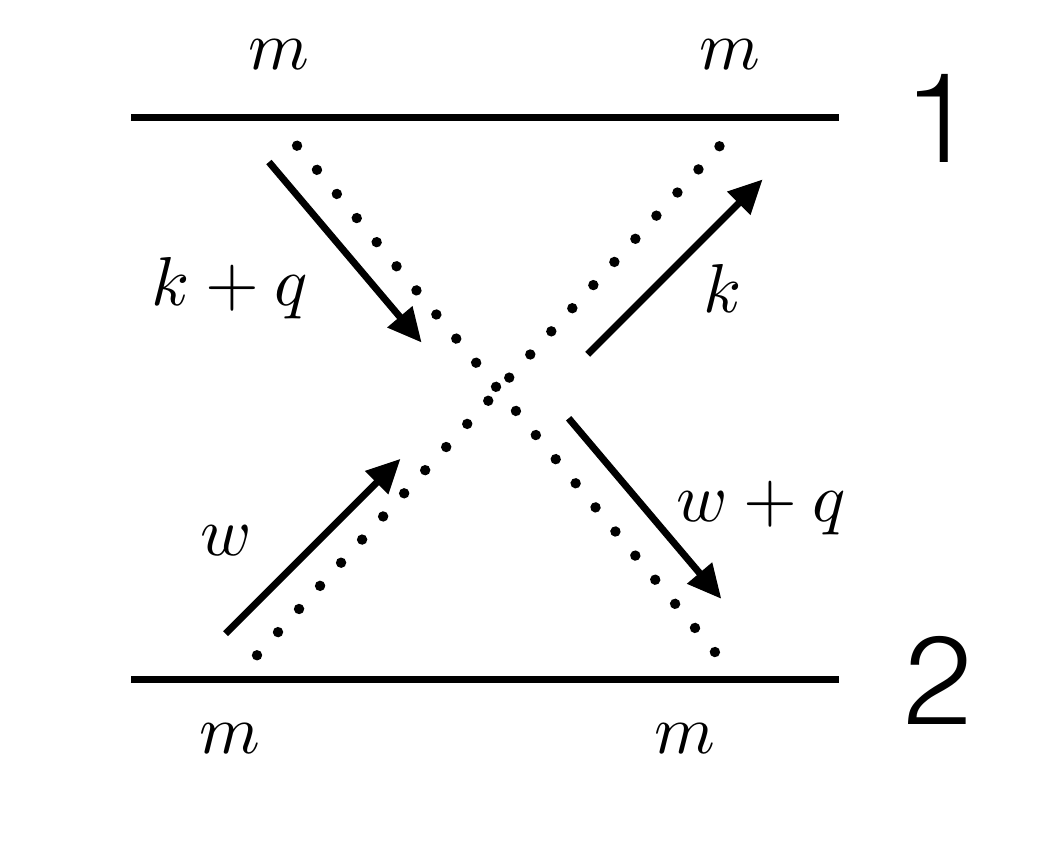}
  \caption{The two loop diagram with the ``cross" topology. Arrows indicate the direction of the momentum flow. The $m$'s in the vertex specifies that we used the leading source-gravity vertex.}
   \label{R4_cross}
\end{figure}

Let us begin by analyzing the diagram of the form in Fig. \ref{R4_cross}.
Such a diagram has the structure
\be
\text{Figure 1 } \sim \int_{\vec q} \int_{\vec k} \int_{\vec w} \frac{kk(q+k)(q+k)ww (q+w)(q+w)}{k^2(\vec q +\vec k)^2 w^2(\vec q +\vec w)^2}e^{i\vec q \cdot \vec {x}_{12}(t)}
\ee
where the momenta in the numerator are contracted with each other in some manner (which we will see is not important). In this expression there are no factors of $v$, as they give a  subleading contribution with respect to the main non-vanishing one. The crucial property to notice is that the resulting integrals factorize into two separate tensor loop integrals in $k$ and $w$ of the form
\be
\label{integral_to_solve_tem}
\int_{\vec k} \frac{k^{i_1} \ldots k^{i_N}}{k^2(k+q)^2}\ .
\ee
As we can see explicitly in the Appendix~\ref{app:loops}, in three-dimensions there are no divergences for a single loop integral. {{The intuitive reason is that all divergences should correspond to local counter-terms that are polynomial in momenta, however by dimensional analysis a one-loop counter-term would be proportional to $\sqrt{q^2}$, which corresponds to a non-local term.}} Consequently, the resulting integral over $q$ can only have the following form with a finite prefactor:
\be
\text{Figure \ref{R4_cross} } \sim \int_{\vec q}  ( q^2)^3 e^{i\vec q \cdot \vec {x}_{12}(t)} \; .
\ee
Note however, that this integral is zero in our dimensional regularization scheme as seen from~(\ref{qInt}),
and so, such a diagram vanishes for the $\mathcal{C}^2$ interaction vertex at lowest order in the PN expansion.
Physically, i.e. independent of regularization, this means that this diagram does not induce a long range force, but rather only a contact, $\delta$-function supported, force, which is inconsequential for the prediction of the time dependence of the system.

\subsubsection{Peace/Log Diagram}

\begin{figure}[!ht]
  \centering
      \includegraphics[width=0.3\textwidth]{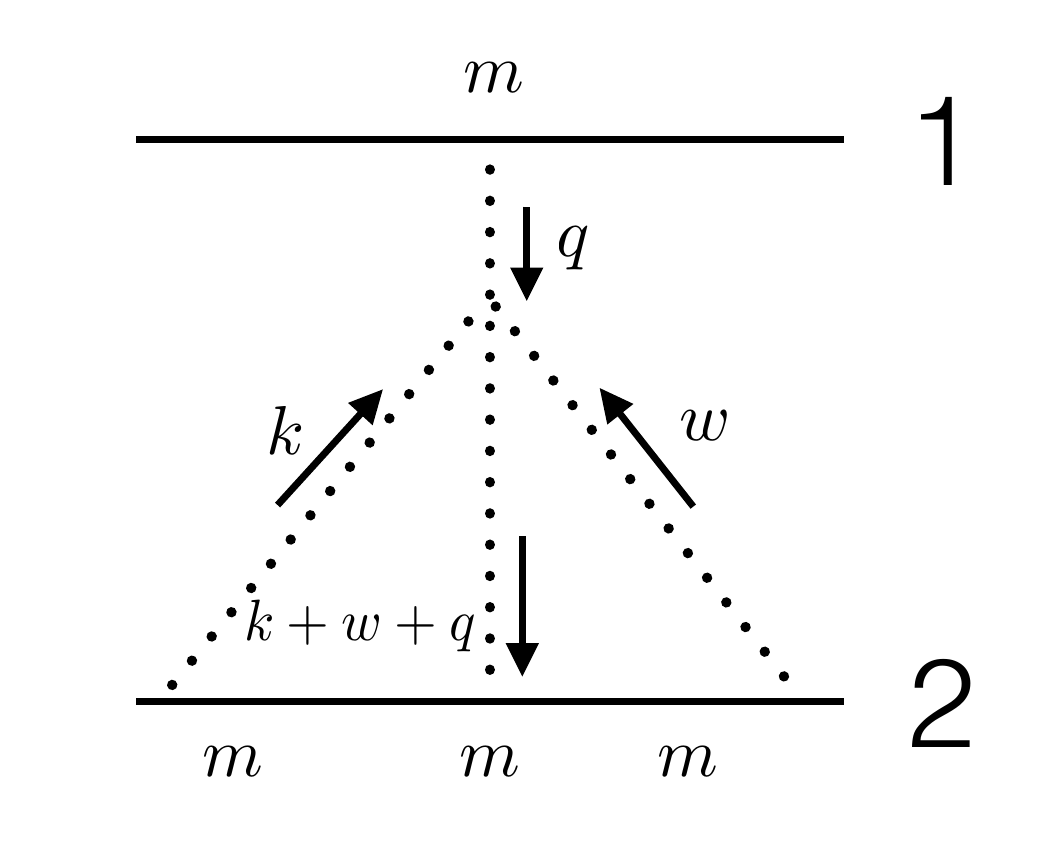}
  \caption{The two loop diagram with the ``peace/log" topology.}
  \label{R4_log}
\end{figure}

There is another topology we can investigate, one where three of the legs are associated with one source and one leg with the other. This leads to diagrams of the form in Fig.~\ref{R4_log}~(\footnote{The name Peace/Log clearly shows that all the authors of the paper are currently living in the San Francisco Area. Some things never die.}).
As we can redefine the loop momentum as we choose, the diagram in Fig. \ref{R4_log} is given by just a single momentum structure contraction. Using the result of (\ref{RR_H00_H00}), and after accounting for combinatorics,  we can write 
\be
\text{Figure \ref{R4_log} } = \frac{i}{8} \frac{m_1 m_2^3}{\Lambda^6 \mpl^6} \int dt \, \int_{\vec q}\int_{\vec w} \int_{\vec k}  \frac{(\vec q \cdot \vec w)^2(\vec k \cdot (\vec k+\vec q+\vec w))^2}{q^2 w^2 k^2 (\vec k+\vec q+\vec w)^2 } e^{i\vec q \cdot \vec {x}_{12}(t)} \; .
\ee 
The first simplification we can make is to drop all terms that are proportional to $k^2$ in the numerator (as they will vanish in dim. reg. as demonstrated in the App.~\ref{app:loops}). We begin by computing the integral over $k$ (the first loop). We do not need to keep the subleading in $\epsilon$ terms, as, as we will show momentarily, the $q$ integral gives a term proportional to $\epsilon$, so we need a $1/\epsilon$ term from the $k\&w$ integrals (there are no $1/\epsilon^2$ terms). Following the general results in  App.~\ref{app:loops} given by (\ref{final_loop_int_form}) we have that 
\be
\text{Figure \ref{R4_log} } = \frac{i}{8}\frac{m_1 m_2^3}{\Lambda^6 \mpl^6} \int dt \, \int_{\vec q}\int_{\vec w}   \frac{e^{i\vec q \cdot \vec {x}_{12}(t)}(\vec q \cdot \vec w)^2\bar{w}^i \bar{w}^j}{q^2 w^2}  \times \frac{\bar w}{32}T^{ij}(\hat{\bar{w}}) 
\ee 
where $\bar w \equiv w+q$ and the traceless symmetric tensor $T^{ij}$ is defined in the  App.~\ref{app:loops}. As $\bar{w}^i \bar{w}^jT^{ij}(\hat{\bar{w}})=\bar{w}^2$, the integral over $w$ (the second loop) can be calculated using our master formulas (\ref{eq:maseq1}), giving
\ba
\text{Figure \ref{R4_log} } = -\frac{1}{\epsilon}\frac{1}{2}\frac{i}{128}\frac{m_1 m_2^3}{\Lambda^6 \mpl^6} \left(-\frac{1}{315 \pi^2} \right) \int dt \, \int_{\vec q} q^6 e^{i\vec q \cdot \vec {x}_{12}(t)} \; ,
\ea
where we kept only the divergent peace of the two-loop integral to get a non-vanishing result. Notice in particular there are no $1/\epsilon^2$ divergencies.
Using (\ref{qInt}), the integral over~$q$ is readily evaluated:
\be
\int_{\vec q} q^6 e^{i\vec q \cdot \vec {x}_{12}(t)}=-\epsilon\frac{1260}{\pi r^9} \; .
\ee
Collecting everything we have, finally,  we obtain
\ba
\text{Figure \ref{R4_log} } =  i  \int dt \, \left(\frac{-m_1 m_2^3}{64 \pi^3 \Lambda^6 \mpl^6} \right) \frac{1}{r^9(t)} \; .
\ea
There is also the diagram with $1 \leftrightarrow 2$. Combining these together, we obtain a correction to the potential of the form
\be
\Delta V_{\text{$\Lambda$}}=\Delta V_{\text{$\Lambda$, Peace/Log}}= \frac{2}{\pi^6}\frac{G m_1 m_2}{r} \left(\frac{2\pi}{\Lambda r}\right)^6\frac{4 G^2 (m_1^2+m_2^2)}{r^2}\ .
\ee

\subsection{Correction to the potential: $\tilde{\mathcal{C}}^2$ term}

The correction to the potential from the ${\tilde{\mathcal{C}}}^2$ operator is subleading to the effect arising from the modification of the multipoles. However, we present the result as an illustration of the Feynman rules involving the $\tilde{\mathcal{C}}$ operator, and also because, if one were to compute the waveform, even a subleading effect can accumulate with time over many orbits and become sizable.

\subsubsection{Cross Diagram}

As mentioned in Appendix~\ref{App:new_quartics}, the $\delta R(h\rightarrow H_{00}) \delta \tilde R (h\rightarrow H_{00})$ vanishes. Therefore we need to have higher order vertices, hence the leading diagram from $\tilde{\mathcal{C}}^2$ will be proportional at least to a total of two powers of $v^i$ or $\partial_i S^{ij}$, where $S_{ij}$ parametrizes the spin of the black hole, carefully defined in~App.~\ref{app:A}. However, analogously to the cross diagram with the $\mathcal{C}^2$ vertex, the integrals over loop momenta factorize and after they are carried out the $q$ integral will be again proportional to 
\be
\int_{\vec q} q^m e^{-i\vec q \cdot \vec {x}_{12}(t)}
\ee
with $m$ even and positive. Consequently at this order, the contribution vanishes for non-zero $r$ as follows from~(\ref{qInt}).

\subsubsection{Peace/Log Diagram}

\begin{figure}[!ht]
  \centering
      \includegraphics[width=0.3\textwidth]{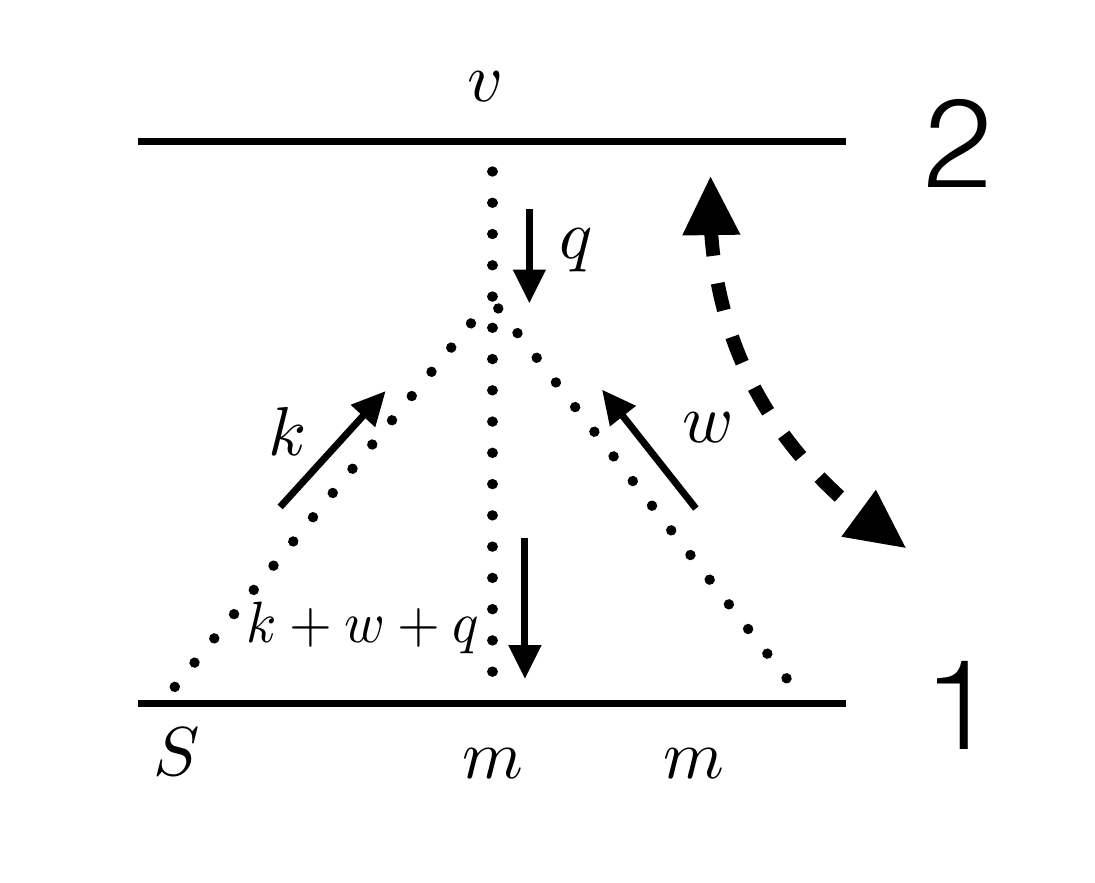} 
   \caption{One of the leading contributions to the effective potential from $\tilde{\mathcal{C     }}^2$ with the ``peace/log'' topology. The dashed arrow indicates the contraction of tensor indices within one of the two $\tilde{\mathcal{C}}$, the other  becoming at this point automatic.}
   \label{RtildeR_loop_1}
\end{figure}

\begin{figure}[!ht]
  \centering
      \includegraphics[width=0.3\textwidth]{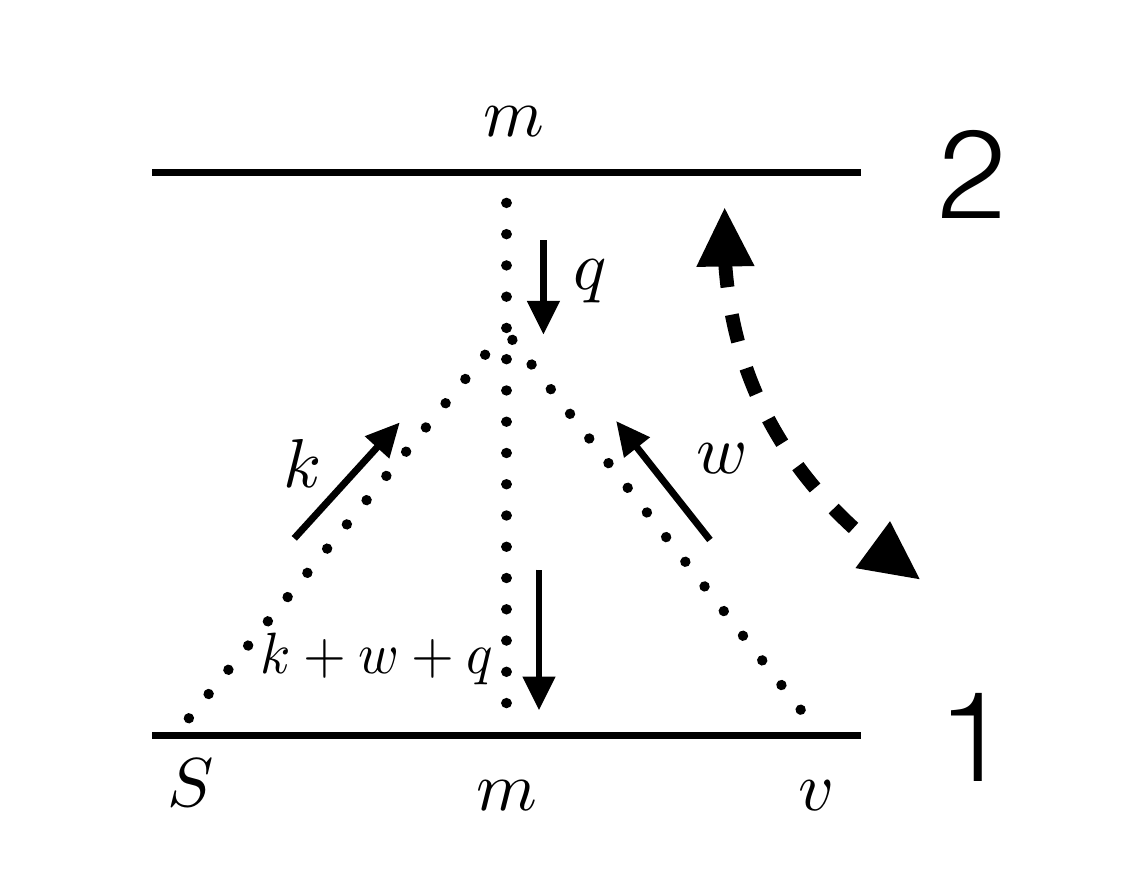}
   \caption{One of the other leading contributions to the effective potential from $\tilde{\mathcal{C     }}^2$ with the ``peace/log'' topology. The dashed arrow indicates contraction of indices within $\tilde{\mathcal{C     }}.$}
   \label{RtildeR_loop_2}
\end{figure}

For the same reasons as above, higher order source vertices are needed to produce non-vanishing contribution. It turns out that we need one velocity vertex and one spin vertex (the diagram with two velocity vertices vanishes). 
Just as in the $\mathcal{C}^2$ case the result will be proportional a pole in $1/\epsilon$ produced by the overlapping two loop integrals which them multiply the linear-in-$\epsilon$ $q$ integral. There are two choices of how to locate the spin and velocity vertices that give non-vanishing contributions that we show on Fig. \ref{RtildeR_loop_1} and~\ref{RtildeR_loop_2}. First, consider the diagram on Fig. \ref{RtildeR_loop_1}:

Performing the calculation, and taking into account of combinatorial factors, we have that
\ba
\text{Figure \ref{RtildeR_loop_1} } &=& 2 \frac{m_2 m_1^2 }{\tilde \Lambda^6 \mpl^6} \int dt \, \int_{\vec q}\int_{\vec w} \int_{\vec k} e^{-i\vec q \cdot \vec {x}_{12}(t)}  \nonumber \\ 
&&\frac{\epsilon_{nmi}(k+\bar w)_i (k+\bar w)_j k_j k_n k_p  \epsilon_{rsl}w_l w_k q_k q_r  S_1^{pm}(t) v_2^s(t)}{ q^2 w^2 k^2 ( k+ \bar w)^2 }  \; ,
\ea
where $\bar w=w+q$. Doing the loop integral over $k$ first we have that
\ba\nonumber
\int_{\vec k}\frac{\epsilon_{nmi}(k+\bar w)_i (k+\bar w)_j k_j k_n k_p S_1^{pm}(t)}{k^2 (k+\bar w)^2} \rightarrow -\frac{\epsilon_{nmi}}{2^6}\bar w^i T_2^{np}{(\hat {\bar w})}S_1^{pm} \bar w ^3 \rightarrow \frac{1}{2^7}\epsilon_{nmi}\bar w^i S_1^{nm} \bar w ^3 \; .\\
\end{eqnarray}
Now we need to do the loop integral over $w$ which takes the form
\ba
\int_{\vec w}\frac{\epsilon_{nmi}\epsilon_{rsl}q_k q_r\left(q_i w_l w_k+w_i w_l w_k\right)(w+q)^2 (w+q)^2 S_1^{nm}v_2^s}{w^2 \left((w+q)^{2}\right)^{1/2}}  \; .
\ea
Using the  formula~(\ref{eq:maseq2}), we can integrate over $\omega$. The remaining $q$ integral will be proportional to
\be
\int_{\vec q} q^6 (i q_n) e^{-i\vec q \cdot \vec {x}_{12}(t)}=-\epsilon \d_n \frac{1260}{\pi r^9} \; ,
\ee
where again (\ref{qInt}) was used. 
So, in totality, we obtain: 
\ba
\text{Figure \ref{RtildeR_loop_1} } &=& -i\left(\frac{27 }{176 \pi^3}\right) \frac{m_2 m_1^2 }{\tilde \Lambda^6 \mpl^6} \int dt \, \frac{ x_{12}^n S_1^{nm} v_2^m}{r^{11}} \; .
\ea

The other diagram configuration that we can construct has both the spin vertex and the velocity vertex associated with the same source. That diagram is given by Fig. \ref{RtildeR_loop_2}.
When we compute this diagram we find that it is structurally the same as that of Fig. \ref{RtildeR_loop_1} but with $\epsilon_{rsl}w_l w_k q_k q_r   v_2^s(t) \rightarrow\epsilon_{rsl}q_l q_k w_k w_r   v_1^s(t)$ which tells us that
\be
\text{Figure \ref{RtildeR_loop_2}}=-\text{Figure \ref{RtildeR_loop_1} (with $v_2 \rightarrow v_1$)}
\ee
and consequently 
\ba
\text{Figure \ref{RtildeR_loop_1}}+\text{Figure \ref{RtildeR_loop_2}}=i\left(\frac{27 }{176 \pi^3}\right)  \frac{m_2 m_1^2 }{\tilde \Lambda^6 \mpl^6} \int dt \, \frac{ x_{12}^n S_1^{nm} \Delta v_{12}^m}{r^{11}} \; .
\ea
Including the diagrams with $(m_1\leftrightarrow m_2)$, we have the correction to the potential in the form
\begin{align}
\Delta V_{\text{ $\tilde\Lambda$}}=\Delta V_{\text{ $\tilde\Lambda$ Peace/Log}} = \frac{216}{11 \pi^6}\frac{G m_1 m_2}{r} \left(\frac{2\pi}{\tilde \Lambda r}\right)^6 \frac{4 G^2 \left(m_1S_1^{i}+m_2 S_2^{i}  \right)  \epsilon_{inm}v_{12}^n r_{12}^m}{r^{4}}\ .
\end{align}

\section{Correction to radiation}\label{sec:radiation}

In this section, we will show the various diagrams that will lead to corrections to the quadrupole $Q_{ij}$ and the current quadrupole $J_{ij}$ of a binary system as outlined in section~\ref{sec:techsummary}. 

\subsection{Corrections of quadrupole: $\mathcal{C}^2$ \label{sec:radiativeonesec}}

\begin{figure}[!ht]
  \centering
      \includegraphics[width=0.3\textwidth]{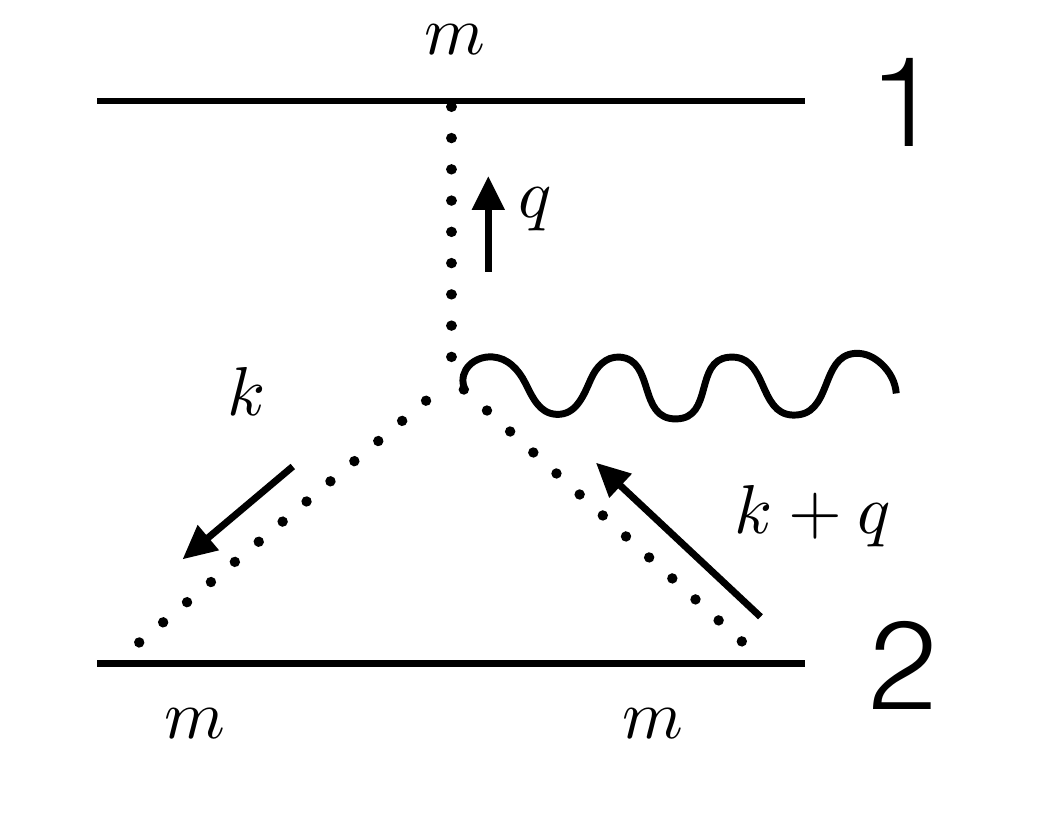}
   \caption{The leading order radiative correction diagram from the $\mathcal{C}^2$ operator. The wiggly line represents the outgoing gravitational wave. }
      \label{R4_rad}
\end{figure}

 Here the leading order diagram has the structure of Fig. \ref{R4_rad}.
Using (\ref{RR_H00_hbar}) and (\ref{RR_H00_H00}) there are two possible tensor structures of this digram coming from the two distinct places to contract the $\delta R (H\rightarrow H_{00}) \delta R (\bar h) $. After accounting for combinatorial factors, we have
\ba
\text{Figure \ref{R4_rad} } =i \frac{m_1 m_2^2}{\Lambda^6 \mpl^4} \int dt \, \int_{\vec q}\int_{\vec k}  \frac{2(\vec q \cdot ( \vec q+\vec k))^2k^i k^j+(\vec k \cdot ( \vec q+\vec k))^2q^i q^j}{q^2 k^2 (\vec k+\vec q)^2 } e^{-i\vec q \cdot \vec {x}_{12}(t)} \times R^{0i0j}(\bar h) \nonumber .\\ 
\ea
The factor of $2$ inside the integral comes from the fact that there are two contraction configurations that are identical upon a shift in the loop momentum. Dropping the $k^2$ terms in the numerator, which, as usual, gives a vanishing contribution, we have four structures when we expand the numerator. These are
\be
\text{Numerator}=2q^4k^i k^j+4q^2 q^n k^n k^i k^j+2q^n q^m k^n k^m k^i k^j +q^i q^j q^n q^m k^n k^m  \  .
\ee
Each of these can then be computed using formula~(\ref{general_structure}) and~(\ref{master_prefactor}) for the loop integral over momentum $k$. Performing these loop integrals we will obtain a result whose tensorial structure will be of the form 
\be
\int_{\vec k}\frac{\text{Numerator}}{k^2(\vec k+\vec q)^2}=\frac{q^3}{32}\left(\frac{q^2}{2}T^{ij}_2+q^i q^j \right) \; .
\ee
As this is contracted with $R^{0i0j}(\bar h)$ we can simplify further by explicitly removing the trace, and writing $q^iq^j\rightarrow \frac{2 }{3}q^2T_2^{ij}$. And so finally, we may write:
\ba
\text{Figure \ref{R4_rad} } = i \frac{m_1 m_2^2}{\Lambda^6 \mpl^4}\left(\frac{7}{192}\right) \int dt \, \int_{\vec q}  \frac{q^5T_2^{ij}(\hat q)}{q^2 } e^{-i\vec q \cdot \vec {x}_{12}(t)} \times R^{0i0j}(\bar h) \; .
\ea
We can compute the integral over $q$ using (\ref{eq:trensotial}) and (\ref{qInt}), obtaining 
\ba
\text{Figure \ref{R4_rad} } = i \frac{m_1 m_2^2}{\Lambda^6 \mpl^4}\left(\frac{7}{16 \pi^2}\right) \int dt \, \left(\frac{3r^ir^j}{r^8}-\frac{\delta^{ij}}{r^6} \right) \times R^{0i0j}(\bar h) \; .
\ea
When we include the same diagram but exchanging $1 \leftrightarrow 2$, we find that this diagram corresponds to adding to the effective action of a single object (which, as we explained, corresponds to the action of the two compact objects seen from far away) a term which has the functional form of a quadrupole (see eq.~(\ref{eq:PNeffectiveaction})):
\ba\label{eq:radiativeone}
S_{\Lambda,rad}=\int dt \,  \frac{21}{4\pi^6} \left(\frac{2\pi}{\Lambda r}\right)^6 \left(\frac{2G (m_1+m_2)}{r}\right)^2 \frac{m_1 m_2}{m_1+m_2} \left(r^ir^j-\frac{r^2 \delta^{ij}}{3} \right) \times R^{0i0j}(\bar h).
\ea 
We can therefore think of this term as a correction to the quadrupole of the binary.

\subsection{Corrections to current quadrupole: $\tilde{\mathcal{C}}^2$}

Similarly to the calculation of the potential presented in the previous Section, the lowest order diagrams are not as simple as the $\mathcal{C}^2$ case. Due to the epsilon structure, $\delta R(h\rightarrow H_{00}) \delta \tilde R (h\rightarrow H_{00})$ vanishes. This means that we must compute diagrams with higher order couplings to the world lines (graviton-source couplings), as explained in more detailed in~App.~\ref{App:new_quartics}. Consequently we will need the leading order contribution to structures like $\delta R(h\rightarrow H_{00}) \delta \tilde R (h\rightarrow v^i H_{0i})$. We will also need structures like $\delta R(h\rightarrow H_{00}) \delta \tilde R (\bar h )$, which are given in App.~\ref{App:new_quartics}. Putting these all together with the right pre-factors given by the Feynman rules, we have two types of diagrams: one where the velocity vertex is paired with a mass vertex acting on the same source, and the other where it is the only vertex acting on a source. More explicitly we have that the first diagram is given in Fig. \ref{RtildeR_radiative_1} while the other is give by Fig. \ref{RtildeR_radiative_2}.

\subsubsection{First Diagram: paired $v$}
\begin{figure}[!ht]
  \centering
      \includegraphics[width=0.3\textwidth]{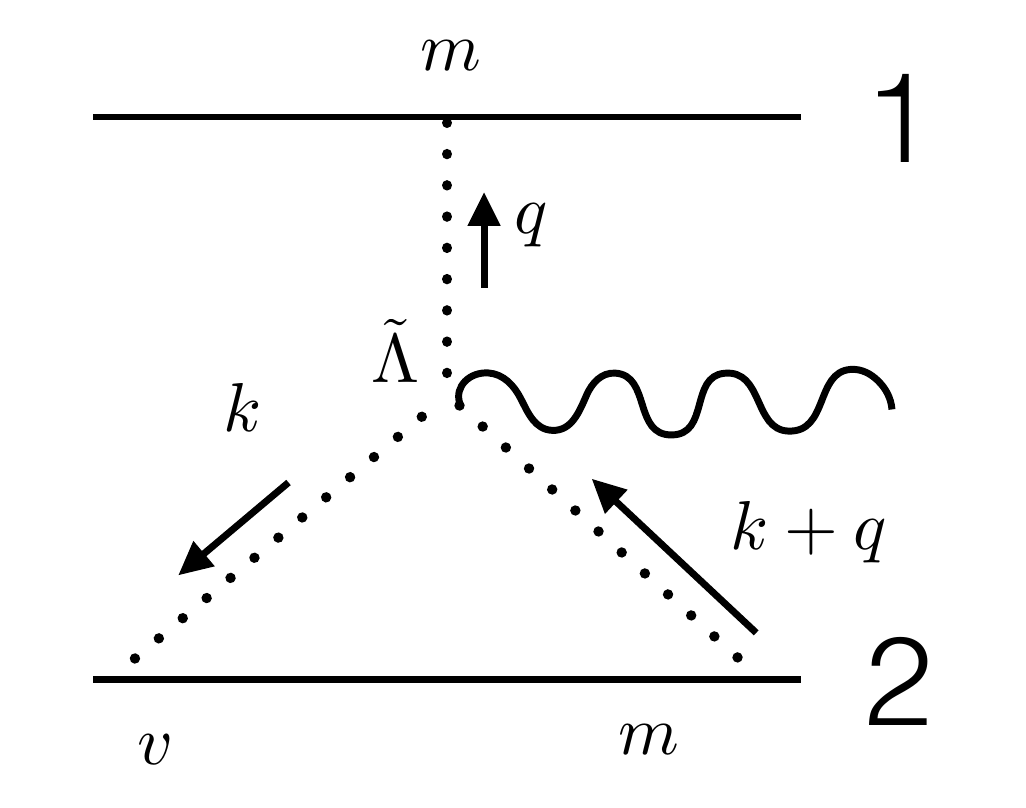}
   \caption{A diagram with the first ``topology"  for the leading order radiative correction diagram for $\tilde{\mathcal{C     }}^2$.}
      \label{RtildeR_radiative_1}
\end{figure}
This diagram has contributions from two likes of structures: one where the $\delta R(h\rightarrow H_{00}) \delta \tilde R (\bar h )$ is on source $1$ (which is contracted only with one leg); and one where it is on source $2$ (which is contracted with two legs). Let us examine the first case. The numerator will be proportional to
\be
\epsilon_{nmi}(k+q)^i (k+q)^j k^j k^n \rightarrow \epsilon_{nmi}q^i q^j k^j k^n
\ee
as anything proportional to $k^2$ vanishes in the loop (for the usual argument) and where we have taken advantage of the structure of the epsilon tensor. When we compute the loop integral over $\vec k$ we get
\be
 \epsilon_{nmi}q^i q^j T_2^{jn}\sim 3 \epsilon_{nmi}q^i q^j q^j q^n-\epsilon_{nmi}q^i q^n\ ,
\ee
each of which vanishes by anti-symmetry. Consequently, the only (possible) non-zero piece is when the contraction is on source $2$. Therefore, after a shift in the loop integral and after including combinatorics, we can write 
\ba
\text{Figure \ref{RtildeR_radiative_1} } &=& 4i\frac{m_1 m_2^2}{\tilde \Lambda^6 \mpl^4}  \int dt \, \int_{\vec q}\int_{\vec k}  \frac{\epsilon_{nmr}q^r q^s (k+q)^s k^n \epsilon_{ijk}k^k k^l}{q^2 k^2 (\vec k+\vec q)^2 } e^{-i\vec q \cdot \vec {x}_{12}(t)}  \nonumber \\
&&\times  v_2(t)^m  \left(R^{ijl0}(\bar h)+2R^{ilj0}(\bar h)\right) \; .
\ea
Performing the loop integral over $\vec k$, the momentum dependent tensor structure of the diagram becomes
\be
\frac{1}{8}\frac{\epsilon_{nmr}\epsilon_{ijk}q^r \left(q^s T_4^{snkl}q^3-2q^2 T_3^{nkl}q^2\right)}{q^2}\quad\rightarrow\quad -\frac{1}{8}\frac{\epsilon_{nmr}\epsilon_{ijk}q^r T_3^{nkl}q^4}{q^2} \; .
\ee
Examining the structure of $T_3$ we see that the term $\propto \hat q^n \hat q^k \hat q^l$ will vanish as will the term with $\hat q^n \delta^{kl}$, leaving us with just 
\ba
\text{Figure \ref{RtildeR_radiative_1} } &=& i\frac{m_1 m_2^2}{\tilde \Lambda^6 \mpl^4}\frac{1}{64}  \int dt \, \int_{\vec q} \frac{\epsilon_{nmr}\epsilon_{ijk}q^r \left(q^k \delta^{nl}+q^l \delta^{nk} \right)q^3}{q^2  } e^{-i\vec q \cdot \vec {x}_{12}(t)}  \nonumber \\
&&\times  v_2(t)^m  \left(R^{ijl0}(\bar h)+2R^{ilj0}(\bar h)\right) \nonumber \\
&\rightarrow&\quad i\frac{m_1 m_2^2}{\tilde \Lambda^6 \mpl^4}\frac{1}{64}  \int dt \, \int_{\vec q} \frac{8 \,q^j q^l q^3}{q^2  } e^{-i\vec q \cdot \vec {x}_{12}(t)} \times v_2(t)^i  R^{ijl0}(\bar h)\; ,
\ea
where we have utilized the trace free condition of the on-shell radiation graviton in our final manipulations. Performing the final integrals over $\vec q$ as illustrated in App.~\ref{app:loops} we arrive at 
\ba
\text{Figure \ref{RtildeR_radiative_1} } &=& i\int dt\, \frac{m_1 m_2^2}{\tilde \Lambda^6 \mpl^4}\frac{1}{2\pi^2} \frac{1}{r^8}\left(6v_2(t)^i r^j r^l-v_2(t)^i r^2 \delta^{jl}  \right)  \times R^{ijl0}(\bar h)\; .
\ea

\subsubsection{Second Diagram: alone $v$}
\begin{figure}[!ht]
  \centering
      \includegraphics[width=0.3\textwidth]{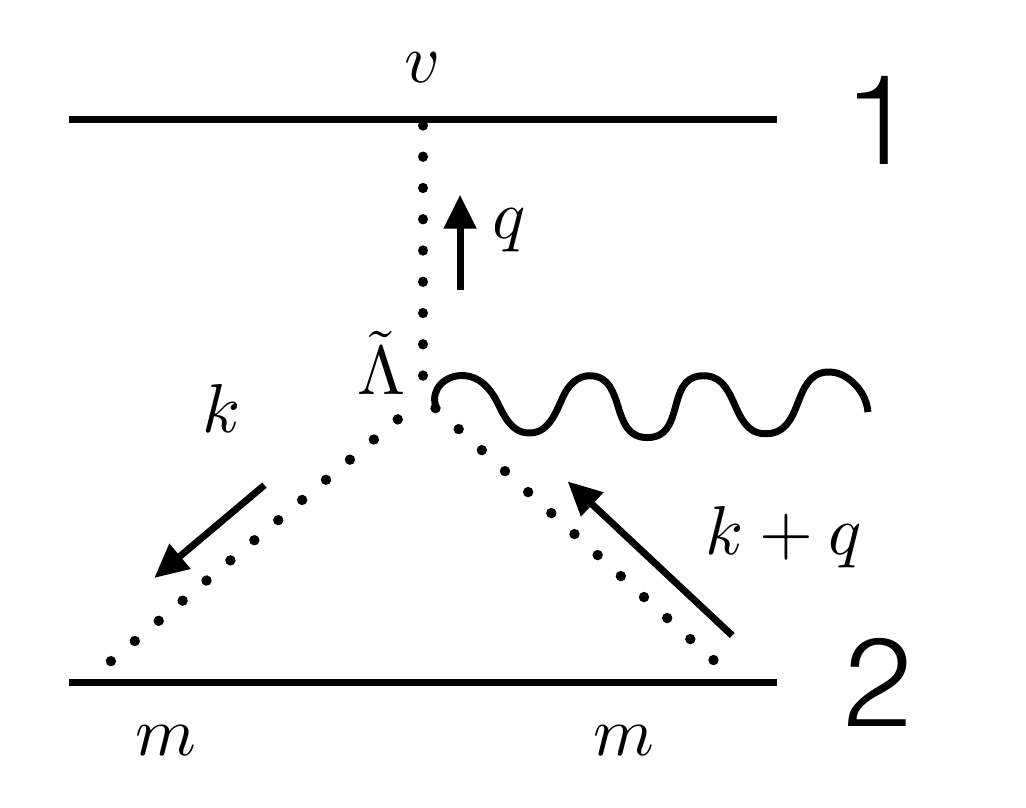}
   \caption{A diagram with the second ``topology" for the leading order radiative correction from the $\tilde{\mathcal{C     }}^2$ operator.}
   \label{RtildeR_radiative_2}
\end{figure}

Let us now compute the contribution from Fig. \ref{RtildeR_radiative_2} where the velocity vertex is isolated. Using our Feynman rules we have
\ba
\text{Figure \ref{RtildeR_radiative_2} } &=& 4i\frac{m_1 m_2^2}{\tilde \Lambda^6 \mpl^4}  \int dt \, \int_{\vec q}\int_{\vec k}  \frac{\epsilon_{nmr}(k+q)^r (k+q)^s q^s q^n \epsilon_{ijk}k^k k^l}{q^2 k^2 (\vec k+\vec q)^2 } e^{-i\vec q \cdot \vec {x}_{12}(t)}  \nonumber \\
&&\times  v_1(t)^m  \left(R^{ijl0}(\bar h)+2R^{ilj0}(\bar h)\right) \; .
\ea
Following almost identical manipulations as those for the previous diagram, we can first compute the loop integral over $\vec k$, which gives us
\ba
\text{Figure \ref{RtildeR_radiative_2} } &=& 2i\frac{m_1 m_2^2}{\tilde \Lambda^6 \mpl^4}  \int dt \, \int_{\vec q}  \left(\frac{-1}{64}\right)\frac{\epsilon_{nmr}\epsilon_{ijk}q^n T_3^{rkl}q^4}{q^2}
 e^{-i\vec q \cdot \vec {x}_{12}(t)}  \nonumber \\
&&\times  v_1(t)^m  \left(R^{ijl0}(\bar h)+2R^{ilj0}(\bar h)\right) \; .
\ea
As $\epsilon_{nmr}=-\epsilon_{rmn}$, we see that this diagram is exactly the same as the previous one where we have changed $v_1\leftrightarrow v_2$ and now has a minus sign. And so we obtain 
\ba
\text{Figure \ref{RtildeR_radiative_2} } &=& i\int dt\, \frac{m_1 m_2^2}{\tilde \Lambda^6 \mpl^4}\frac{1}{2\pi^2} \frac{1}{r^8}(-1)\left(6v_1(t)^i r^j r^l-v_1(t)^i r^2 \delta^{jl}  \right)  \times R^{ijl0}(\bar h)\; .
\ea

\subsubsection{Total radiative corrections}

Now, notice that when we take these two radiation diagrams together we get the nice structure (throwing out the trace term as it vanishes for the on-shell graviton):  
\ba
\text{Figure \ref{RtildeR_radiative_1} } +\text{Figure \ref{RtildeR_radiative_2} }= i\int dt\, \frac{m_1 m_2^2}{\tilde \Lambda^6 \mpl^4}\left(-\frac{3}{\pi^2}\right) \frac{1}{r^8}\Delta v_{12}^i r^j r^l \times R^{ijl0}(\bar h)
\ea
where $\vec{\Delta v}_{12}\equiv \vec v_1-\vec v_2$. Now we also need to compute the same diagrams but with the sources exchanged, i.e. $1\leftrightarrow2$. When we do so we  get a contribution that would exactly cancel that above (as $\vec{\Delta v}_{12}+\vec{\Delta v}_{21}=0$) were it not for the altered masses in the pre-factor. In summary, all of these diagrams combine to give: 
\ba
\text{Figure \ref{RtildeR_radiative_1} } +\text{Figure \ref{RtildeR_radiative_2} }+(1\leftrightarrow2)&=&i\int dt\, \frac{6m_1 m_2(m_1-m_2)}{\tilde \Lambda^6 \mpl^4} \frac{1}{r^8}\Delta v_{12}^i r^j r^l \times R^{ijl0}(\bar h)\nonumber\\
= i\int dt\,&& \frac{12}{\pi^6} \left(\frac{4 G^2 m_1 m_2 (m_1-m_2)}{r^2}\right)\left(\frac{2\pi}{\Lambda r}\right)^6\Delta v_{12}^i r^j r^l \times R^{ijl0}(\bar h) \nonumber\\
\;
\ea 
Notice that we write this as an effective ``magnetic'' quadrupole term, $\sim \Delta v_{12}^i r^j r^l$. Taking into account combinatorial coefficients, we have
\ba
S_{\tilde\Lambda,rad}=\int dt\, \frac{12}{\pi^6}\left(\frac{2\pi}{\tilde \Lambda r}\right)^6 \left(\frac{4G^2 m_1 m_2 (m_1-m_2)}{r^2}\right)  \Delta v_{12}^i r^j r^l \times R^{ijl0}(\bar h) \; ,
\ea 
The vanishing of the result for $m_1=m_2$ can be understood by noticing that the binary (and in particular the angular momentum distribution) in this limit becomes symmetric under a parity transformation around the origin. Therefore, the effective single-object action must be invariant under this same parity.

 So far, we have neglected another diagram for radiation where we generate an effective quadrupole. This diagram corresponds to contracting the Riemann of the radiation graviton with the velocity graviton-source vertex. Upon performing the integrations, we obtain an effective quadrupole term whose tensorial structure is of the form of the product of two angular momenta: $\epsilon_{ilm}v^l r^m\; \epsilon_{jpq}v^p r^q\sim L_i L_j$, minus its trace. This term is however subleading by one power of $v$ with respect to the current-quadrupole radiation for the emitted field, and therefore we neglect it here. However, one should keep in mind that this term would contribute comparably to the leading corrections if one were interested in the power emitted at the source, because this correction to the quadrupole would interfere with the newtonian quadrupole (unless the orbit is circular, in which case the interference vanishes).

\subsection{Corrections to current quadrupole: ${\mathcal{C}}\tilde{\mathcal{C}}$}

The structure of the leading diagram in this case is the same as in Fig.~\ref{R4_rad}, where we contract one of the Riemann tensors in $\tilde{\mathcal{C}}$ with the external graviton. Apart for combinatoric factors, the resulting diagram is identical to the one computed in Sec.~\ref{sec:radiativeonesec}, eq.~(\ref{eq:radiativeone}), with the replacement
\be
 R^{0i0j}(\bar h) \quad\to\quad -\epsilon^{iab} \left(\frac{R^{abj0}}{2}+ R^{ajb0}\right)\ \ \times \  \ \frac{2}{4} \ ,
\ee
where the factor of $2/4$ comes from the different combinatorics. We therefore obtain
\ba\nonumber
&&S_{\Lambda_-,rad}=-\int dt \,  \frac{21}{8\pi^6} \left(\frac{2\pi}{\Lambda_- r}\right)^6 \left(\frac{2G (m_1+m_2)}{r}\right)^2 \frac{m_1 m_2}{m_1+m_2} \left(r^ir^j-\frac{r^2 \delta^{ij}}{3} \right)\\ 
&&\qquad\qquad \times\ \epsilon^{iab} \left(\frac{R^{abj0}}{2}+ R^{ajb0}\right) \ .
\ea 
This can be interpreted as an effective current quadrupole with the tensor structure of $r_i r_j - r^2 \delta_{ij}/3$:
\be
J_{ij} = \frac{63}{8\pi^6} \left(\frac{2\pi}{\Lambda_- r}\right)^6 \left(\frac{2G (m_1+m_2)}{r}\right)^2 \frac{m_1 m_2}{m_1+m_2}\left(r^ir^j-\frac{r^2 \delta^{ij}}{3} \right)
\ee
Interestingly, due to the CP-odd nature of the  ${\mathcal{C}}\tilde{\mathcal{C}}$ term, we find there is a term with a tensor structure similar to the GR quadrupole that couples to the emitted graviton through an $\epsilon$-tensor, and therefore contributing as a $J_{ij}$ current quadruple (which normally, unlike here, contains an  $\epsilon$-tensor in its definition). In particular this means that the effective one body system violates parity.

\subsection{Summary}\label{sec:radiationsummary}

Combining the result of the calculation of the corrections to the quadrupole and current quadrupole of a binary system, the leading correction to the radiative coupling of the effective single object is given by  terms in the single-object effective action of the form
\ba\nonumber
\label{Lambda_rad_coupling}
S_{\Lambda,rad}=\int dt \,  \frac{21}{4\pi^6} \left(\frac{2\pi}{\Lambda r}\right)^6 \left(\frac{2G (m_1+m_2)}{r}\right)^2 \frac{m_1 m_2}{m_1+m_2} \left(r^ir^j-\frac{r^2 \delta^{ij}}{3} \right) \times R^{0i0j}(\bar h)\ ,\\ 
\ea
and
\ba
\label{Lambda_tilde_rad_coupling}
&&S_{\tilde\Lambda,rad}=\int dt\, \frac{12}{\pi^6}\left(\frac{2\pi}{\tilde \Lambda r}\right)^6 \left(\frac{4G^2 m_1 m_2 (m_1-m_2)}{r^2}\right)  \Delta v_{12}^i r^j r^l \times R^{ijl0}(\bar h) \; ,\\
\label{Lambda_minus_rad_coupling}\nonumber
&&S_{\tilde\Lambda_-,rad}=-\int dt\, \,  \frac{21}{8\pi^6} \left(\frac{2\pi}{\Lambda_- r}\right)^6 \left(\frac{2G (m_1+m_2)}{r}\right)^2 \frac{m_1 m_2}{m_1+m_2} \left(r^ir^j-\frac{r^2 \delta^{ij}}{3} \right)  \times \epsilon_{jkl} R^{kli0}(\bar h),\\  \; 
\ea
generated by the $\mathcal{C}^2/\Lambda^6$, $\tilde{\mathcal{C}}^2/\tilde\Lambda^6$ and $\tilde{\mathcal{C}}{\mathcal{C}}/\Lambda_-^6$ terms respectively. These expressions can be cast in a more familiar form by comparing to the structure of the leading gravitational multipole coupling (to on-shell gravitons) in the center of mass frame anticipated in~(\ref{eq:PNeffectiveaction}) (see for example~\cite{Goldberger:2009qd}):
\ba
S_{{\rm ext.\, obj.}}\supset \frac{1}{2}\int dt \, Q_{ij} R_{0i0j}-\frac{1}{3}\int dt \, J_{ij} \epsilon_{jkl}R^{kli0} + \dots \ ,
\ea
where $Q_{ij}$ and $J_{ij}$ are the mass and ``current'' quadrupole moments~\footnote{See also \cite{Maggiore:1900zz} for a discussion based on the a direct multipole expansion of the linearized Einstein equations.} given---to leading order---by the integrals
\ba
Q_{ij}&=&\int d^3x \; \left(x^i x^j-\frac{1}{3}\delta^{ij}x^2\right) T^{00}\quad \overset{\text{pp limit}}{\longrightarrow} \quad \sum_a m_a \left(x_a^i x_a^j-\frac{1}{3}\delta^{ij}x_a^2\right)\ , \\
J_{ij}&=&\int d^3x \; \left[x^{\{i}\epsilon^{j\}nm}- \frac{1}{3} \delta^{ij}x^{\{l}\epsilon^{l\}nm}\right]x^n T^{0m}  \quad \\ 
&&\qquad\overset{\text{pp limit}}{\longrightarrow} \quad \sum_a m_a \left[x_a^{\{i}\epsilon^{j\}nm}-\frac{1}{3} \delta^{ij}x_a^{\{l}\epsilon^{l\}nm}\right]x_a^n v_a^m\ ,
\ea
where $a^{\{i}b^{j\}}=\frac{1}{2} \left(a^i b^j+a^jb^i\right)$.
When we consider two point particles about their center of mass frame, we can write the quadrupole moments in a simplified form
\ba
Q_{ij}&=&\mu \left(r^i r^j -\frac{1}{3}\delta^{ij}r^2\right) \\
J_{ij}&=&\frac{\mu}{m_1+m_2} \left[r^{\{i}\epsilon^{j\}nm}-\frac{1}{3}\delta^{ij}r^{\{l}\epsilon^{l\}nm}\right] r^n (m_1v_2^m+m_2 v_1^m) \nonumber \\
&=& - \frac{\mu(m_1 -m_2)}{m_1+m_2}  \left[r^{\{i}\epsilon^{j\}nm}-\frac{1}{3}\delta^{ij}r^{\{l}\epsilon^{l\}nm}\right]r^n v_{12}^m 
\ea
where $\mu=m_1 m_2/(m_1+m_2)$ is the reduced mass. When we compare our results to these expressions we find that, for the terms in $\Lambda$ and $\tilde\Lambda$, the coupling is not only of a similar tensor structure (this is unsurprising as it really is just a general consequence of gauge invariance---see \cite{Goldberger:2009qd}) but it has the same structure as the leading PN case but with a modified coefficient. The term in $\Lambda_-$ has instead a different structure beyond the tensorial one. In other words, we can write our total radiative coupling as simply renormalized quadrupole and current quadrupole moments as follows
\ba
Q_{ij} &\rightarrow& \left(1+ \frac{21}{2\pi^6} \left(\frac{2\pi}{\Lambda r}\right)^6 \left(\frac{2G (m_1+m_2)}{r}\right)^2\right)Q^{(N)}_{ij} \\
J_{ij} &\rightarrow&\left(1-\frac{36}{\pi^6}\left(\frac{2\pi}{\tilde \Lambda r}\right)^6 \left(\frac{2G (m_1+m_2)}{r}\right)^2\right) J^{(N)}_{ij}\\ \nonumber 
&& +\frac{63}{8\pi^6} \left(\frac{2\pi}{\Lambda_- r}\right)^6 \left(\frac{2G (m_1+m_2)}{r}\right)^2 Q^{(N)}_{ij} \; ,
\ea
where $Q^{(N)}_{ij}$ and $J^{(N)}_{ij}$ are the Newtonian mass and current quadrupoles.

\section{Observable consequences for LIGO-VIRGO}\label{sec:observeligo}

In principle, we can fold in these corrections to the effective action and the radiative coupling to modify the dynamics of a compact binary during inspiral and deduce the observable consequences. In broad strokes, the effective potential changes the acceleration on each object which shifts the frequency of the emitted gravitational wave. Meanwhile, the corrected radiative coupling---as well as the shifted frequency itself---changes the amplitude of the emitted radiation and consequently the rate at which which power is emitted. Additionally, the effective potential also changes the energy as a function of the orbital parameters and so its modification effects the orbital decay.

To deduce the observable consequences in a completely accurate way for general orbits, one would use the results derived in the former sections and just numerically integrate until the PN expansion breaks down when $v/c \sim 1$. Exploring all of the parameter space (various mass ratios, ellipticity, spin orientations, etc.) goes beyond the scope of this first paper on the EFT. In the future, however, it would be very worthwhile to perform such an exploration and produce templates for the LIGO-VIRGO (and future gravitational wave observatories) pipeline.

For the purpose of this paper, we will restrict ourselves to a much simpler analysis, which is sufficient for us to illustrate the main observable effects. We consider (quasi) circular motion of two compact objects and treat the radiation reaction in an adiabatic manner, that is in the regime where $\omega_{\rm orb} \gg \dot r/r$, with $r$ being the orbital separation.

For a given orbital separation, the orbital frequency of the particles is given by the the full post-Newtonian equations of motion to some order, which we indicate as $\omega_{PN}(r, m_i, S_i)$. If we were to turn on the $\mathcal{C}^2/\Lambda^6$ term how would this frequency change?

Using (\ref{Lambda_pot}) to derive the acceleration on a single particle, we can compute (the leading order) change in the frequency as a function of $\omega_{PN}$ using the simple mechanics of circular motion as we already explained in footnote (\ref{footnote:omega}). We find that
\be
\frac{\Delta\omega_{\Lambda}}{\omega_{PN}}= -\frac{2304G^3 (m_1+m_2)(m_1^2+m_2^2)}{ \Lambda^6} \cdot \frac{1}{r^{11}(t)} \cdot \frac{1}{\omega_{ PN}^2}\ .
\ee
As we have computed the correction to the effective potential (and thus the equation of motion) to leading order in the PN-expansion it would be inconsistent to keep the full $\omega_{PN}$ in the full expression above. Consequently, we input the leading Newtonian contribution to the frequency $\omega_N=\sqrt{G (m_1+m_2)/r^3}$ yielding
\ba
&&\frac{\Delta\omega_{\Lambda}}{\omega_{N}}= -\frac{9}{\pi^6}\frac{4G^2 (m_1^2+m_2^2)}{ r^2}\left(\frac{2 \pi}{ \Lambda r}\right)^6
\ea
Of course, in order to be able to measure this effect, one should have computed the $\omega_{PN}$ to a sufficiently high order not to overshadow this effect.

Let us move on to the effect of the radiation. In the full post Newtonian treatment, the asymptotic strain tensor incident upon the detector is given by some function $h^{ij}_{PN}(t-R, \hat n)$ where $R$ is the distance to the source from the detector and $\hat n$ is the direction of propagation of the wave. We want to compute how,  in the quasi static approximation, this is altered by the presence of our new interactions. From the leading PN radiative coupling (i.e. the usual quadrupole formula \cite{Maggiore:1900zz}) we have that--in the usual TT gauge--
\be
\left[h_{ij}^{TT}(t,\vec x)\right]_{quad}=\frac{2 G}{R} \Lambda_{ij,kl}(\hat n) \ddot{Q}_{kl}(t-R)\ ,
\ee
where the $\Lambda$ tensor is given by
\be
\Lambda_{ij,kl}(\hat n)=P_{ik}P_{jl}-\frac{1}{2}P_{ij}P_{kl}\ ,
\ee
where $P_{ij}=\delta_{ij}-\hat n_i \hat n_j$. Restricting ourselves to a quasi-circular orbit, we take $\vec r=(r \cos (\omega t) , r\sin( \omega t), 0)$. We have then that the second time derivative of the quadrupole moment is given by 
\ba
\ddot Q_{ij}&=&-4 \omega^2 Q_{ij} = \frac{4}{3}\omega^2  \mu r^2 \delta_{ij}-4\omega^2 \mu r^2 \hat z_i \hat z_j \\
&=&-2 \mu r^2 \omega^2\left(
\begin{array}{ccc}
 \cos (2 \text{$\omega $t}) & \sin (2 \text{$\omega $t}) & 0 \\
 \sin (2 \text{$\omega $t}) & -\cos (2 \text{$\omega $t}) & 0 \\
 0 & 0 & 0 \\
\end{array}
\right) \; .
\ea
Note that the dominant frequency of the gravitational radiation is $2\omega$. As mentioned briefly above, one source of corrections from the EFT terms we are now considering is in the shift in frequency, another is in the ``rescaling'' of the quadrupole moment. When we put these effects together, we have to leading order that 
\ba
\left[\Delta h_{ij}^{TT}(t,\vec x)\right]_\Lambda&=& \left(\frac{2\pi}{\Lambda r}\right)^6 \frac{8 G^2}{\pi^6 r^2} \left( \frac{21(m_1+m_2)^2}{4}-9(m_1^2+m_2^2)\right) \nonumber\\
&& \times   \frac{2 G}{R} \Lambda_{ij,kl}(\hat n)  \ddot{Q}_{kl}(t-R)\ .
\ea
In terms of scaling, we can see directly that 
\be
\Delta h_\Lambda \sim h \times \left(\frac{\Delta \omega}{\omega} \right) \sim h \times \left(\frac{1}{\Lambda r} \right)^6  \left(\frac{G m}{r} \right)^2 \sim h \times \left(\frac{1}{\Lambda r} \right)^6  v^4 \; .
\ee
We can see that the effect is suppressed not only by $1/(\Lambda r)^6$, but also by four powers of~$v$. Therefore, for a given $\Lambda$ and $r$, in order to trust this correction, one needs to compute the ordinary PN waveform up to an order larger than $v^4$ by an amount  that can compensate for the $1/(\Lambda r)^6$ suppression.

Continuing on, we can compute the effects generated from the $\tilde{\mathcal{C}}^2/\tilde \Lambda^6$ term. Here the potential is not just a function of the radial distance, and so recovering the change in frequency is slightly more complicated than in the previous case. As a first step, let us compute the change in the equations of motion due to the effective potential. On particle~$1$, for instance, we have
\ba
[m_1 \Delta a_1]_j &=& \frac{\delta}{\delta x_1^j} \int dt \left(-V_{\tilde \Lambda}\right) \\
&=& \frac{216 G}{11\pi^6} \frac{4 G^2 m_2 m_1 }{r^2}\left(\frac{2\pi}{\tilde \Lambda r}\right)^6 \cdot \frac{1}{r^{3}}\times \left[ (m_1 S_1^i +m_2 S_2^i)\epsilon_{inm} \left(\frac{11 v_{12}^n  r_{12}^m r_{12}^j}{r^2}-2v_{12}^n \delta_m^j \right. \right. \nonumber \\
&&\left. \left. -\frac{11 r_{12}^n\delta_m^j (r_{12}\cdot v_{12})}{r^2}\right)- (m_1 \dot S_1^i +m_2 \dot S_2^i)\epsilon_{inj} r_{12}^n \right] \;,
\ea
and similarly for particle $2$. $\vec v_{12} = \vec v_{1} -\vec v_{2}$ is the relative velocity between the two particles. This expression looks a bit daunting, but there are a few simplifications that occur at the order we are working at. First of all, to leading order in the PN expansion $\dot S=0$ as the spin angular momentum is conserved. This means that at this order we may drop terms proportional to $\dot S$. When we work in the circular motion limit, $\vec v_{12}$ and $ \vec r_{12}$ are perpendicular, and if we again take $\vec r_{12}=(r \cos (\omega t) , r\sin( \omega t), 0)$ we can simplify the above force as 
\ba
[m_1 \Delta a_1]_j 
&=& \frac{216 G}{11\pi^6} \frac{4 G^2 m_2 m_1 }{r^2}\left(\frac{2\pi}{\tilde \Lambda r}\right)^6 \cdot \frac{1}{r^{3}} \times \Big[-11 v_{12} (m_1 S_1^z +m_2 S_2^z) \hat r_{12}^j  \nonumber \\
&&-2 \left( (m_1 \vec S_1 +m_2 \vec S_2)\times \vec v_{12}\right)^j\Big]\ ,
\ea
where $v_{12}$ is just the vector's magnitude. If the spin vectors are in arbitrary directions their components in the orbital plane serve to torque the orbit. To simplify the situation, let us just consider the objects' spin to be perpendicular to the orbital plane (this is also astrophysically the most likely scenario). In particular, let us take them to be in the same direction as the orbital angular momentum, that is, we take $S$ to have a positive value if it is pointed in the $+ \hat z$ direction. In this restricted scenario, we have that
\ba
[m_1 \Delta a_1]_j 
&=& -\frac{2808 G}{11\pi^6} \frac{4 G^2 m_2 m_1 }{r^2}\left(\frac{2\pi}{\tilde \Lambda r}\right)^6 \cdot \frac{ v_{12} (m_1 S_1 +m_2 S_2)}{r^{3}} \hat r_{12}^j \; .
\ea
We can see that the force is an attractive one for positive $(m_1 S_1 +m_2 S_2)$. We are now back to the form of a radially symmetric force, for which $\omega\propto\sqrt{ \frac{F_r}{m\, r}}$. Therefore computing the change in the orbital frequency we get
\be
\frac{\Delta \omega_{\tilde \Lambda}}{\omega_{PN}}=\frac{1404 }{11\pi^6} \left(\frac{2\pi}{\tilde \Lambda r}\right)^6 \cdot \frac{ 4 G^2 v_{12} (m_1 S_1 +m_2 S_2)}{r^3} \; .
\ee
This change in the orbital frequency changes the gravitational wave emission by shifting the frequency of the quadrupole motion. This contribution takes precisely the form computed above for the $\mathcal{C}^2/\Lambda^6$ term. It is given by
\ba\label{eq:tildeeffect}
\left[\Delta h_{ij}^{TT}(t,\vec x)\right]_{quad,\, \tilde\Lambda}&=& \frac{2808}{11\pi^6}\left(\frac{ 2\pi}{{\tilde \Lambda}r}\right)^6 \frac{4 G^2 v_{12} (m_1 S_1 +m_2 S_2)}{r^{3}} \nonumber \\
&& \times  \frac{2G}{R} \Lambda_{ij,kl}(\hat n) \ddot{Q}_{kl}(t-R) \; .
\ea
As before, we can estimate the size of this contribution. In the case of (nearly) equal binary compact objects of similar spin we have that
\be\label{eq:spinpotentialcoup}
\Delta h_{quad, \, \tilde \Lambda} \sim h \times \frac{\Delta \omega_{\tilde \Lambda}}{\omega_{PN}} \sim h \times \left(\frac{1}{\tilde \Lambda r} \right)^6  \left(\frac{G m}{r} \right)^2 \times v^{2+s}\sim h \times \left(\frac{1}{\tilde \Lambda r} \right)^6  v^{6+s}\ ,
\ee 
where we have used that the spin angular momentum scales as $S\sim L v^s$ where $ L \sim m r v$ is the orbital angular momentum and $s=1$ for maximally rotating compact objects and $s=4$ for co-rotating objects, non-rotating objects have $s=\infty$. As we can see
\be
\Delta h_{quad, \, \tilde \Lambda}  \sim \Delta h_{ \Lambda} \times \left(\frac{\tilde \Lambda}{\Lambda} \right)^6 v^{2+s}\ ,
\ee
which tells us the quadrupole effect from this operator is smaller than the one in $\mathcal{C}^2$ when~$\Lambda\sim\tilde \Lambda$.

For the $\tilde{\mathcal{C}}^2/\tilde \Lambda^6$ term there is also another way to change the gravitational waveform. We still have to consider the correction to the radiative coupling. The linearized gravitational wave given by the coupling to the current quadrupole is given by \cite{Maggiore:1900zz}
\be\label{eq:current_quadr_emission}
\left[h_{ij}^{TT}(t,\vec x)\right]_{curr \; quad}=\frac{1}{R}\frac{4 G}{3} \Lambda_{ij,kl}(\hat n) \hat n _m \left( \epsilon^{mkp}\ddot{J}^{pl}(t-R)+\epsilon^{mlp} \ddot{J}^{pk}(t-R)\right)\ .
\ee
Now, for circular orbits, $\vec r$ and $\vec v_{12}$ are always perpendicular, and their cross product is perpendicular to the orbital plane. In particular, by defining the direction  $\hat z$ such that $\epsilon^{jnm}r^n v_{12}^m=\omega r^2 \hat z ^j$, and 
  tensor $J^{pl}$ is given by  
\ba
J^{pl}&=& - \frac{\mu(m_1 -m_2)}{2(m_1+m_2)}r^2 \omega \left(r^p \hat z^l + r^l \hat z^p \right)\\
&=&- \frac{\mu(m_1 -m_2)}{2(m_1+m_2)}r^3 \omega\left(
\begin{array}{ccc}
 0 & 0 & \cos (\text{$\omega $t}) \\
 0 & 0 & \sin (\text{$\omega $t}) \\
 \cos (\text{$\omega $t}) & \sin (\text{$\omega $t}) & 0 \\
\end{array}
\right) \\
\Longrightarrow \quad \ddot{J}^{pl} &=&-w^2 J^{pl} \; .
\ea
Note that the frequency of the current quadrupole is just $\omega$, not $2\omega$ as in the mass quadrupole case. 

We can write the contribution to the gravitational wave coming from the ``renormalization'' of the current quadrupole moment,
\ba
\left[\Delta h_{ij}^{TT}(t,\vec x)\right]_{curr \; quad, \, \tilde{\Lambda}}&=&-\frac{36}{\pi^6}\left(\frac{2\pi}{\tilde \Lambda r}\right)^6 \frac{4G^2 (m_1+m_2)^2}{r^2}  \\ \nonumber
&&\times  \frac{4 G}{3 R}\Lambda_{ij,kl}(\hat n) \hat n _m \left( \epsilon^{mkp}\ddot{J}^{pl}(t-R)+\epsilon^{mlp} \ddot{J}^{pk}(t-R)\right) \; ,
\ea
where the second line is the second derivative of the leading current quadrupole moment from General Relativity. Dimensionally, this scales like
\be
\left[\Delta h_{ij}^{TT}(t,\vec x)\right]_{curr \; quad, \, \tilde{\Lambda}} \sim h\times \frac{\Delta J}{J} \times v \sim h \times \left(\frac{1}{\tilde \Lambda r} \right)^6  \left(\frac{G m}{ r} \right)^2 \times v \sim h\times \left(\frac{1}{\tilde \Lambda r} \right)^6 v^5\ ,
\ee
where $h$ above is the size of the leading quadrupole radiation.

As we can see
\be
\left[\Delta h_{ij}^{TT}(t,\vec x)\right]_{curr \; quad, \, \tilde{\Lambda}} \gg \left[\Delta h_{ij}^{TT}(t,\vec x)\right]_{quad, \, \tilde{\Lambda}} \; .
\ee
Notice that, if one is interested just in the emitter power, $\sim h^2$, the contribution from the current quadrupole has a polarization orthogonal to the one emitted from the quadrupole. We therefore have 
\be
[\Delta P]_{curr \; quad, \, \tilde{\Lambda}}  \sim \dot h^2 v^{6}\sim P_{N}\, v^6\; \sim  \frac{\left[\Delta P\right]_{quad, \, \tilde{\Lambda}}}{v^s}\gg \left[\Delta P\right]_{quad, \, \tilde{\Lambda}}\ .
\ee
and so the leading observable effect from the $\tilde{\mathcal{C}}^2/ \tilde{\Lambda}^6$ term for a binary inspiral comes dominantly from the corrected radiation coupling. 

We can perform a similar analysis for the $\mathcal{C}\tilde{\mathcal{C     }}$ operator. The leading effect for the amplitude of the emitted gravitational waves comes from the modified current quadrupole. Applying the formula~(\ref{eq:current_quadr_emission}) to our case, we have
\ba
\left[\Delta h_{ij}^{TT}(t,\vec x)\right]_{curr \; quad, \, {\Lambda}_-}&=&\frac{63}{8\pi^6}\left(\frac{2\pi}{ \Lambda_- r}\right)^6 \frac{4G^2 (m_1+m_2)^2}{r^2}  \\ \nonumber
&&\times  \frac{4 G}{3 R}\Lambda_{ij,kl}(\hat n) \hat n _m \left( \epsilon^{mkp}\ddot{Q}^{pl}(t-R)+\epsilon^{mlp} \ddot{Q}^{pk}(t-R)\right) \; ,
\ea
which, at parametric level, scales as
\be
\left[\Delta h_{ij}^{TT}(t,\vec x)\right]_{curr \; quad, \, {\Lambda_-}} \sim h \times \left(\frac{1}{ \Lambda_- r} \right)^6  \left(\frac{G m}{ r} \right)^2  \sim h\times \left(\frac{1}{ \Lambda_- r} \right)^6 v^4\ .
\ee
Notice however that if we were interested in the emitted power, the polarization of the total emitted gravity wave is orthogonal to the one associated to the mass quadrupole, as it comes from the current quadrupole. However, it is also true that, upon average in time and angle, the interference between the $\left[\Delta h_{ij}^{TT}(t,\vec x)\right]_{curr \; quad, \, {\Lambda}_-}$ and its Newtonian counterpart, $\left[\Delta h_{ij}^{TT}(t,\vec x)\right]_{curr \; quad, \, {N}}$,  vanishes. At subleading level, there is a correction to the matter quadrupole that scales as $\Delta Q/Q\sim v^5$. However, similarly to what happened for the current quadrupole, the form of the induced matter quadrupole will be such that it will not interfere with the newtonian one after time- and angle- integration.  Therefore, the leading contribution for the total emitted power comes from the correction to the potential, which is of order 
\be
V_{\Lambda_-}\sim \frac{G m^2}{r} \frac{1}{(\Lambda_- r)^6} \left(\frac{G m}{r}\right)^3 a\ ,
\ee
where $a=\sqrt{S_i S^i}/(G M^2)$ is the spin parameter in the Kerr metric of the faster spinning black hole. This implies $\Delta \omega/\omega\sim v^6 a$, and in turn we have
\be
[\Delta P]_{ quad, \, \Lambda_-}  \sim  P_{N}\, v^6\ a \sim P_{N}\, v^{5+s}\ . 
\ee
where we used that $a\sim v^{s-1}$.

Independently of the radiated power, there is a possibly interesting effect that one might consider, in relation to the evolution of the spins of a black hole binary systems. Depending on the sign of the coefficient of the $\mathcal{C}\tilde{\mathcal{C     }}$ operator, this might lead to enhancement or suppression of the component of the spin that is along the direction between the two black holes, an effect that can build up over many orbits and could induce very striking property for the spins of coalescing black holes, in contrast with the standard GR prediction.  A careful study of such effect requires taking into account spin-orbit and spin-spin couplings to the same order in $v$, which is beyond the scope of this paper.

A summary of the parametric dependence of the various contributions is given in Table~\ref{tab:vcount}. 
However, one should not naively disregard subleading contributions. For example,  one could consider the following effect. All the above arguments apply for the instantaneous wave detection. However, for real gravitational wave detectors, detection is a more complicated process that involves the analysis of many cycles of the source starting from some initial condition. Therefore, observationally it could be that we are particularly sensitive to effects that, although they are naively smaller in terms of the amplitude of the emitted gravitational wave, build up with time (such as, for example, the aforementioned effect on the spin alignment with the orbital plane), and in particular can affect the initial conditions at the observational window.  For example, for the lightest objects, such as Neutron star binaries, the sources can stay in the LIGO band for many cycles during their inspiral phase. Since studying these effects requires knowledge of the post-Newtonian evolution at the relevant order, which is not at our disposal here, we leave the study of these additional effects to future work.

This is also why we cannot give a sharp assessment of the implications of the first detections of LIGO-VIRGO on our EFT and in particular on the scales $\Lambda$'s.  The second event~\cite{Abbott:2016nmj} has probably too low a signal-to-noise ratio. In the first event~\cite{TheLIGOScientific:2016qqj}, the signal to noise is dominated by the highly relativistic phase, for which we do not have (yet) predictions. However, it is probably true that $\Lambda$'s such that $\Lambda r_s\sim 1$, where $r_s$ is the Schwartzschild radius of the final black hole, would give order one modifications to the signal. Therefore,  $\Lambda$'s such that $\Lambda \sim 1/r_s\sim 10^{-2} $km$^{-1} $ are probably excluded.  $\Lambda$'s such that $\Lambda \lesssim 1/r_s\sim 10^{-2} $km$^{-1}$ are probably also excluded, because the merger phase occurs beyond the regime of validity of the EFT, and so we expect order one corrections (though, as we stressed, this depends on the UV completion)~\footnote{These bounds appear to be confirmed by naively extrapolating the bounds on the PN parameters of~\cite{Yunes:2016jcc} where both events seem to contribute comparatively. In fact, even though our effect is 2PN, the shift in the phase produced scales as $v^{16}/(\Lambda r_s)^6$, so that one can rescale the bounds obtained on the 8th order PN parameter in~\cite{Yunes:2016jcc}. By very naively extrapolating the results of Fig.~4 of~\cite{Yunes:2016jcc} we find that indeed the bound is $\Lambda\gtrsim 10^{-1}/r_s$. However, one should not over-interpret this bound as one should have perturbative control of the theory in all the regions in $r$ where the signal-to-noise is relevant. Therefore, a dedicated analysis appears to us is still needed.}. A detailed study of the current bounds on $\Lambda$'s and forecast for constraints (or measurements!) from future observations is left to future work.

\begin{table}
\centering
\begin{tabular}{c|c|c|c}
\hline
& \multicolumn{2}{ |c| }{Parity Even}& Parity Odd \\
\hline
Operator            &$\mathcal{C}^2$ & $\tilde{\mathcal{C     }}^2$ & $\mathcal{C}\tilde{\mathcal{C     }}$ \\ \hline
$\Delta V/V$          & $v^4$   & $v^7 a$   	& $v^6 a$  \\
$\Delta \omega/\omega$        & $v^4$   & $v^7 a$ 		& $v^6 a$  \\
$\Delta Q_{ij}/Q_{ij}$     & $v^4$   & $-$ 				& $v^5$  \\
$\Delta J_{ij}/J_{ij} $    & $-$     & $v^4$		 	& $v^3$ \\
\hline
$\Delta P/P$& $v^4$   & $v^6$ 			& $v^6 a$  \\
\hline
$\Delta h/h $& $v^4$   & $v^5$ 			& $v^4$  \\
\hline
\end{tabular}
\caption{Summary of the relative contribution of the parity even and parity odd operators to various quantities in comparison to the leading contribution to the same quantity in General Relativity.  Only the dependence on $v$ and $a$, the spin parameter in the Kerr metric of the fastest spinning black hole, is shown (there is always, respectively, a factor of $1/(\Lambda r)^6$, $1/(\tilde \Lambda r)^6$ or $1/(\Lambda_- r)^6$  in each column, left implicit). $a$ is related to the spin vector of a black hole $S_i$, as $a=\sqrt{S_i S^i}/(G M^2)$. The~$`-'$ sign indicates that the contribution is subleading and was not estimated.}
\label{tab:vcount}
\end{table}

\subsubsection*{Finite size operators}

We add a final discussion. In this section, we have so far neglected the contribution of the finite size operators present in the point particle action in (\ref{eq:point}), such as  the one proportional to $d_2^{(1)}$ (see~\cite{Goldberger:2004jt} for a list of the leading finite-size operators). These operators do contribute to the effective potential and multipoles of the one-body effective action. An estimate gives an induced relative correction to the quadrupole and potential of order $\frac{d_2^{(1)}}{\mpl^2 r^5}$. The size of this contribution crucially depends on the coefficients, such as $d_2^{(1)}$, whose value is determined by the UV description of the system. For black holes in pure GR, all these coefficients are fixed in terms of the mass and spin of the black hole, and are parametrically of the form $d_2^{(1)}\sim m r_s^4$, where $r_s$ is the Schwarzschild radius. This makes this contribution scale as~$v^{10}$. 

\begin{figure}[!ht]
  \centering
      \includegraphics[trim = 5cm 8cm 3cm 9cm, clip, width=0.3\textwidth]{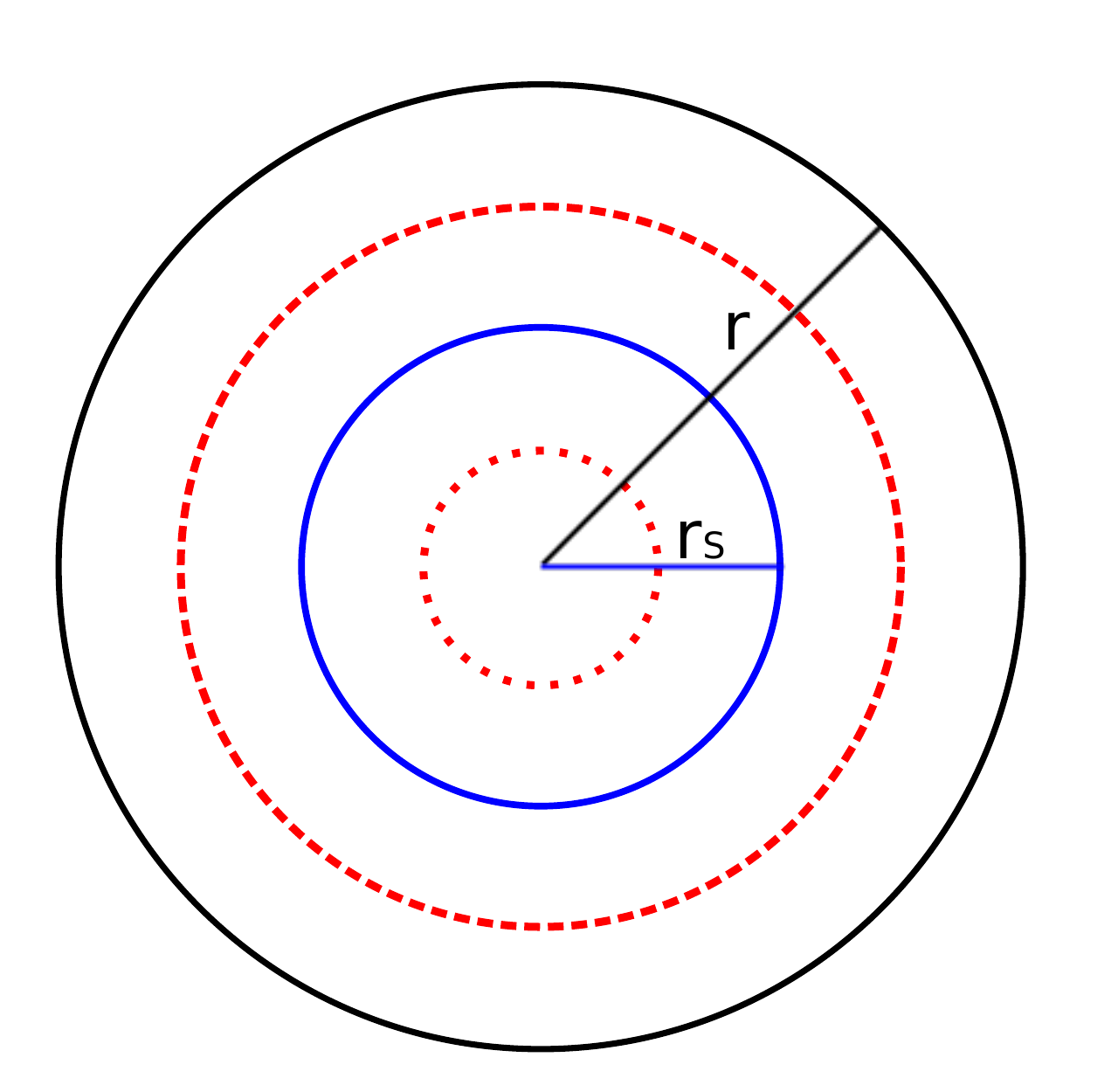}      
  \caption{A schematic view of the scales in our theory. The blue solid circle shows the Schwarzschild radius of a black hole ($r_S \sim G M$), and the black solid circle shows the orbit of the second black hole with radius $r$. The dashed and dotted red circles are two possible choices of the scale $\Lambda$ for a given black hole: $1/\Lambda\gtrless r_S$, that lead to different qualitative contributions, and that are discussed in the text.      \label{fig:scaleLambda21111}
}
\end{figure}

In the case of black holes in our EFT, we should consider two limits (see Fig.~\ref{fig:scaleLambda21111}). If $\Lambda\gtrsim 1/r_s$, then we can compute these terms within the validity of the EFT, and find that  $d_2^{(1)}\sim m r_s^4 (\Lambda r_s)^{-6}$, so that the induced $\Delta h/h\sim v^{10} (\Lambda r_s)^{-6}$. In this regime, this effect is leading with respect to the one we computed. However, this is also the regime where the post-Newtonian result is very small, and furthermore exactly the regime where the perturbative numerical simulation discussed in sec.~\ref{sec:numerical} can be performed all the way to the merger (as the effect of our operators is perturbative up to the horizon). So we neglect to compute this contribution in this regime.

When instead $\Lambda\lesssim 1/r_s$, the black hole solution cannot be derived within the regime of our EFT, and therefore the calculation of the finite-size coefficients, such as $d_2^{(1)}$,  depends on the unknown UV completion. In our setup, in order to avoid constraints from small-scale experiments,  it is crucial to assume the `softness' of the UV completion (see Sec.~\ref{sec:action}). Since 
at $r\sim 1/\Lambda\gtrsim r_s$ we are still in the post-Newtonian regime, the gravitational field is still weak, and therefore our softness assumption implies that at distances shorter than $\Lambda$, effects suppressed by powers of $1/\Lambda$ disappear. This implies that the induced finite-size terms scale as $d_2^{(1)}\sim m r_s^4$, without additional powers of $1/\Lambda$. In this case, the effect from the finite-size operators is dominant over the ones we compute only for large distances $r\gtrsim 1/(\Lambda^2 r_s)$. Therefore, the results presented in this section give the leading effect in the parametric window
\be
r_s\lesssim \frac{1}{\Lambda}\lesssim r\lesssim \frac{1}{\Lambda^2 r_s}\ ,
\ee
which is the most interesting region, as it is where the effect is the largest. Similar conclusions apply for the operators suppressed by $\tilde\Lambda$ and $\Lambda_-$.

If we now pass to neutron stars, we can see that a similar discussion holds. We can argue that the induced modifications to the ordinary contact terms from our EFT are even smaller than in the case of the black hole: since the system is never completely relativistic, our UV-softness assumption suggests that there is a further suppression. Therefore, these corrections are contributing to an order smaller than $v^{10}$ and therefore we can neglect them as in the case of the black holes.

\section{Constraints from other experiments\label{sec:otherexp}}

Experimental measurements of our effective action can be classified into two categories depending on the strength of the measured gravitational potential, weak field or strong field. We will first summarize current measurements in weak gravity systems by discussing their current precision and the one required to probe our effective operators. Secondly, we will discuss the only known strong gravity system before the gravitational wave events at LIGO: the X-ray binaries. 

\subsection{Weak gravity systems}

Weak gravity systems are systems where the gravitational potential is much smaller than unity in natural units. They can contain strongly relativistic objects like neutron stars and black holes as long as the gravitational field experienced by the test mass is weak. Phenomenologically, weak gravity systems correspond to the situation where the distance $r$ is so large that $ v^2 \sim G M/r$ is small. In these systems, in order for measurements to constrain the higher dimensional operators, one needs experimental measurements with precision at least of order $v^4$ in the case of $\mathcal{C}^2$ terms, at least $v^6$ in the case of $\tilde{\mathcal{C     }}^2$ terms, and at least $v^6a$ in the case of $\tilde{\mathcal{C     }}{\mathcal{C     }}$ (see Table~\ref{tab:vcount}). The most relevant of the current experimental tests of gravity in weak gravity systems are weak equivalence principle tests ~\cite{Wagner:2012ui,Smith:1999cr,Dimopoulos:2006nk} and indirect measurement of gravitational wave through orbital decay~\cite{Taylor:1982zz}. In the following, we will summarize the current measurements.

Earth based experiments, in particular the E$\ddot{o}$t-Wash experiment, constrain violations of the weak equivalence principle through measurement of the differential acceleration of berylium and titanium in the earth gravitational field to a precision of $(a_{\rm Be} - a_{\rm Ti})/a = (0.3 \pm 1.8) \times 10^{-13}$~\cite{Wagner:2012ui,Smith:1999cr}. However, as we described, within our assumptions the violation of the weak equivalence principle from our effective Lagrangians is negligibly small.

Lunar laser ranging accurately measures the distance between the earth and moon~\cite{Turyshev:2006gm}. The system has a typical $v\sim 10^{-5}$. Because of our UV-softness assumption that we discussed at length in Sec.~\ref{sec:action}, the effect of our operators is never larger than the one obtained at $r\sim 1/\Lambda$. Therefore the effect is at most of order $v^4 \sim 10^{-22}$. This means that it is negligibly small, independently of the value of $\Lambda$.    

Neutron star binaries are binary star systems where two neutron stars orbit around a common center of mass. Two famous systems are the Hulse-Taylor pulsar and double pulsar PSR J0737-3039. The closest of these binary systems seen up to date have orbital period of a few hours. The period decay rate of the binary neutron star systems are measured to $10^{-6}$ precision. The correction due to the higher dimensional operators are at most  at the level of $v^4 \sim 10^{-13}$, much beyond current precision. Similar measurement of the orbital decay rate can be done with another type of compact binary objects, the low mass X-ray binaries (LMXB), for example A0620-00~\cite{McClintock} and XTE J1118+480~\cite{Orosz:2004ac}. However, the orbital decay in these systems is not yet measured to high precision, and therefore, with a companion orbital period of a few hours, these systems do not place constraint on our theory unless the precision of measurement reaches $v^4\sim (GM/T)^{4/3}$, with $T$ being the orbital period. For the strongest system, this reaches $\sim 10^{-11}$.
 
Short distance modifications to the gravitational force are measured to distances as small as a few micron, much smaller than the $1/\Lambda$'s that are of interest for the LIGO-like experiments. However, as we argued in the introduction and  in Sec.~\ref{sec:action}, since these experiments are performed with very light sources~\cite{Hoskins:1985tn,Kapner:2006si,Geraci:2008hb} (for a review, see~\cite{Adelberger:2009zz}), our `UV-softness' assumptions implies that the effect of our extension to GR is negligibly small in these cases.  

To conclude, current measurements of weak gravity system do not constrain the parameter space of our theory due to the smallness of gravitational binding energy  ($\sim G M m/r$) compared with the energy of the lighter object. In particular, the higher dimensional operators violate the strong equivalence principle in a very special way: gravitational fields have additional couplings among themselves. In weak gravity systems, the gravitational energy is a very small component of the total energy density that gravitates and therefore one needs to perform very precise experiments in order to measure these effects.

\subsection{Strong gravity system: X-ray binaries}

X-ray binaries are binary systems that consist of a massive star and a compact object: a neutron star or a black hole. In this section, we will discuss a type of X-ray binary where the emission from the accretion of the massive star onto a black hole can be observed to good precision~\cite{Reynolds:2013qqa}. The emission profile of the accretion disk can be measured with continuum fitting and X-ray relativistic reflection methods. These measurement of the emission profile of the accretion disk can be used to determine the innermost stable orbit, and provide an alternative measurement of the mass and the spin of the compact object, compared to the measurement of the orbital period of the binary. 

The measurements of X-ray binaries, especially the measurement of GRS 1915+105 and Cygnus X-1 can be used to put constraints on deviations from the Kerr metric of a black hole~\cite{Bambi:2014nta,Bambi:2014oca}. For example, the mass of the host black hole of Cygnus X-1 can be determined by measuring the orbital period of the massive star up to sub-leading corrections due to spin-orbit couplings. In this case, the sub-leading terms due to spin and our higher-dimensional operators are both velocity suppressed since the massive star is at a location where gravitational field is already weak, and are much beyond current sensitivity of orbital period measurements, which are at percent level.

The spin of the black hole and new corrections from  our higher dimensional operators can furthermore be in principle determined by measuring the innermost orbit of the same (or similar) Kerr black hole through measurement of the X-ray emission from the accretion disk~\cite{Reynolds:2013qqa}. The $\tilde{\mathcal{C     }}^2$ term corrects the Kerr-metric in a way that is proportional to the spin of the black hole because $\tilde{\mathcal{C     }}$ vanishes for  the Schwarzschild metric, while the $\mathcal{C}^2$ term corrects the metric even in absence of spin.

Focussing first on the ${\mathcal{C}}^2$ operator,  the Newtonian potential and the potential in eq.~(\ref{Lambda_pot}) can be reorganized, in the limit $m_2\ll m_1$, in the following way
\begin{align}
V_\Lambda = \frac{G m_1 m_2}{r}\left(1 + \frac{2}{\pi^6}\left(\frac{2 G m_1}{r}\right)^2 \left(\frac{2 \pi}{\Lambda r}\right)^6 \right)\ .
\end{align} 
It is clear that there is a sizable correction for $\Lambda r\sim1$, which could well be probed by X-ray observations. However, the region $\Lambda r\sim 1$ is exactly where our EFT is supposed to break down. Our EFT is indeed an expansion in $\Lambda r\gg1$, and predictions for $\Lambda r\sim 1$ strongly depend on the UV completion. Of course, given the strong dependence on $\Lambda r$, the effect quickly becomes very small as we make $\Lambda r\gg 1$. Therefore, overall, it appears to us that these are potentially very interesting probes, if the associated observations are able to control the statistics and the potential systematic effects associated to the astrophysical matter present in the environment, so that they can set limits also when $\Lambda r$ is safely larger than one (for recent reviews, see~\cite{Reynolds:2013qqa,McClintock:2013vwa}). Computing the modifications to the metric due to our extension of gravity is rather straightforward, by solving the perturbative extended Einstein equations~(\ref{eq_temp_source}). However, it is clear that to compute observable quantities for these systems, much more astrophysical ingredients are needed, that we leave for future work.

Similarly, we can show that the $\tilde{\mathcal{C     }}^2$ term generates corrections to the leading order newtonian potential depending on the spin of the black hole. The size of correction is~$\mathcal{O} (10^{-2})$ for a maximally rotating black hole with $\frac{2\pi}{\Lambda r} \sim 1$. 
With current measurement, the observational effect of the above operators in an X-ray binary system is quite degenerate with the spin of the black hole. Therefore, unless the corrections due to the higher dimensional operator is such that the best fit dimensionless black hole spin parameter is measured to be larger than unity ($a/m > 1$), such an effect is experimentally in practice indistinguishable.

Finally, let us mention the constraint coming from a system where we usually observe the strong field regime of GR, which is cosmology. Here, the most powerful bound comes from BBN, which however constraints our scales $\Lambda$'s to be shorter than just about a thousand kilometers. This is clearly a subleading constraint. Of course one might wonder what happens in our set up to the early universe cosmology, for example to inflation. The constraint from BBN leaves some room in energy for inflation to happen within the validity of our EFT. However, it could also be that inflation happens at higher energies. In this case, the answer will depend on the UV completion, which is not at our disposal. In particular, if the UV completion will follow the ``softness" assumptions we made in this paper, then it might be easier to expect that inflation will happen in the usual way. Similar considerations apply to other strong field phenomena that might happen in the universe. Cleary, it would be interesting to study these aspects further.

\section{Conclusion\label{sec:conclusions}}

The recent discovery of gravitational waves from black hole merger by the LIGO-VIRGO collaboration opens up a new observational window on the universe and the laws that determine it. One of the characteristics associated to these observations is that black holes probe the strong field regime of gravity, where the deviation of the metric from Minkowski is order one. Only in the cosmological setting this regime of gravity has been probed in a comparably precise way, but, as we have described in the paper, cosmology is not yet directly sensitive to very high values of  the curvature.

We have therefore constructed the most general effective field theory for a modification of General Relativity that satisfies the following constraints. The first is that it is testable with gravitational wave observations. This means that the effect of new physics should not be largely suppressed with respect to the curvature scale of the probed compact objects. This forces our higher dimension operators to be suppressed by a scale of order of a few inverse km. Second, this modification of GR must not be ruled out by already existing experiments. This forces our theory not to alter the coupling to matter. We discuss that this choice is stable under radiative corrections. Furthermore, this same requirement imposes a strong condition on the UV completion of our EFT. We need to assume that the physical effect of our extension to GR {\it saturates} at the scale when our EFT breaks down. We argue that this is both essential for the viability of the  EFT, but also a rather common phenomenon in field theory. Third, it must not violate any  widely accepted principle of physics, such as Lorentz invariance, unitarity and locality. In particular this implies that our theory does not allow for superluminal propagation of signals, which in turns, forces the leading operators in our EFT to take the form of four contracted  Riemann tensors with some restrictions on the relative sign and size of the coefficients. Fourth and last, but not least, we restrict our EFT not to have any additional light degrees of freedom beyond the two helicity-two states of the usual graviton. While this is a limitation (that be rather easily fixed), as experiments like LIGO-VIRGO can probe also theories with additional  degrees of freedom with masses smaller than a few inverse km, our EFT represents the most general extension to GR in the UV without additional degree of freedom. Therefore, by testing our EFT, one is guaranteed to investigate a vast class of physically consistent theories all at once.  

After setting up the EFT formalism, we have calculated the effect of the leading higher dimensional operators in the inspiral phase of a black hole merger. We have done this by adapting the EFT for extended objects, which was formulated for GR, to our theory.  The resulting effects, which amount to a shift in the phase, amplitude and polarization of the emitted waves, though small corrections compared to the leading gravitational wave emissions, can in principle be extracted from future gravitational wave events. For this reason, it will be worthwhile to compute the actual modification to the waveform that our EFT produces in the inspiral phase.

The merger phase, with $v\approx 1$, probes regions where our higher dimensional operators have the largest effect on $h$.
It is very important to calculate their impact in this phase. We have argued and highlighted a procedure according to which currently available numerical codes should be modifiable to compute these effects. \vspace{0.15cm}

Of course, detecting a deviation from GR in the form of our EFT would represent a revolution in physics, though quite unexpeted.

%%%%%%%%%%%%%%%%%%%%%%%%%%%%%%%%%%%%%%%%%%%%%%%%%%%%%%%%%%%%%%%%%%%%%%%%%%%%%%%%%%%%%%%%%%%%%%%%%%%%%%%%%%%%%%%%%%%%%%%%%%%%%%%%%%%%%%%%%%%%%%%%%%%%%%%%%%%%%%%%%%%%%%%%%%%%%%%%%%%%%%%%%%%%%%%%%%%%%%%%%%%%%%%%%%%%%%%%%%%%%%%%%%%%%%%%%%%%%%%%%%%%%%%%%%%%%%%%%%%%%%%%

%%%%%%%%%%%%
%
%
 %           Acknowledgements
 %
 %
 %%%%%%%%%%%%%%%%

\subsubsection*{Acknowledgments}
We thank S.~Dimopoulos, W.~East, G.~Giudice, P. Graham, M~Okounkova, M.~Mirbabayi, L.~Shao, L.~Stein M.~Zaldarriaga and S.~Zhiboedov, for interesting conversations and comments on the draft. We thank Frans Pretorius for carefully reading and commenting a preliminary draft. SE and LS are partially supported by DOE Early Career Award DE-FG02-12ER41854. JH and VG are partially supported by NSF grant  PHYS-1316699.

\section*{Appendices}
\appendix

\section{Review and Feynman Rules\label{app:A}}

In this appendix, we will briefly review the NR-EFT framework and set up some notations (for a review, see~\cite{Goldberger:2007hy,Porto:2016pyg}). Since we are dealing with the post-Newtonian regime, we can focus on distances much larger than  Schwarzschild radius of the astronomical black holes, and therefore we can expand the Riemann tensor around flat space to leading order in terms of the graviton $h$. The gravitons can be decomposed into the potential gravitons $H$ and the long wavelength radiation field $\bar{h}$
\begin{equation}
h_{\mu \nu} = H_{\mu \nu} + \bar{h}_{\mu\nu}.
\end{equation}
The potential gravitons $H$ are gravitions with typical energy and momentum $(p^0 \sim v/r,{\bf p}\sim 1/r)$, while for gravitons that are emitted by the system, the typical energy and momentum carried by these gravitons is $(p^0 \sim v/r,{\bf p}\sim v/r)$ since they are on shell (i.e. gravitational waves satisfy the relativistic dispersion relation~$p^0=p$).
	
The dynamics of the two-body system can be described by an effective action $S_{\rm eff}$ consisting of two parts~\cite{Goldberger:2004jt}, $S_{pp}$ which describes the interaction between the black holes, which we treat as a point particle, and the graviton, while $S_{\rm eff}$ describes the interactions between the gravitons. The lowest order vertices between black holes and gravitons used in this paper can be found from the two-body system $S_{pp}$ (after canonical normalization)~\cite{Goldberger:2004jt,Porto:2005ac}:
\begin{equation}\label{eq:pointverteces}
S_{pp} \supset \int { d} t \left(-\frac{m H_{00}}{2\mpl} -\frac{mH_{0i}v^i}{\mpl} -\frac{\d_iH_{0j} S^{ij}}{2\mpl}\right)\ .
\end{equation}
where the relation between $\mpl$ and $G$ is $G = \frac{1}{32 \pi \mpl^2}$. $m$ is the mass of the black hole and $S_{ij}$ is the associated spin in the center of mass frame. We also define the parameter $S_i \equiv \epsilon_{ijk} S^{jk}$ (related to the parameter $a$ of the Kerr metric as $a=\frac{\sqrt{S_i S^i}}{G M^2}$). Once inserted in the graph, the vertex must be multiplied by $i$. The graviton propagator and the self-interactions can be found from $S_{\rm eff}$
\begin{equation}
S_{\rm eff} \supset\int { d}^4 x \sqrt{-g} 2 \mpl^2 \left(-R + \frac{\mathcal{C}^2}{\Lambda^6} + \frac{\tilde{\mathcal{C     }}^2}{\tilde{\Lambda}^6}+ \frac{\tilde{\mathcal{C     }}{\mathcal{C     }}}{\tilde{\Lambda}_-^6}+\ldots\right)\ .
\end{equation} 
The propagator can be derived from the Einstein-Hilbert action, in the harmonic gauge, to be (after canonical normalization), for momenta $q$ and $k$ both pointing outwards,
\be
(-i) \frac{P_{\mu \nu ; \alpha \beta}}{k^2}\delta(t_1-t_2)(2\pi)^3\delta^{(3)}(\vec q+\vec k)
\ee
where $P_{\mu \nu ; \alpha \beta}=\frac{1}{2}[\eta_{\mu \alpha} \eta_{\nu \beta}+\eta_{\mu \beta} \eta_{\nu \alpha}-\frac{2}{d-2}\eta_{\mu \nu} \eta_{\alpha \beta} ]$ with $d$ the spacetime dimension. The propagator is found with the gauge fixing term

\begin{equation}
S_{\rm GF} =\int { d}^4 x \sqrt{-g}\mpl^2 \Gamma_{\mu} \Gamma^{\mu} \ ,
\end{equation} 
where $\Gamma_{\mu} = \nabla_{\alpha} H^{\alpha}_{\mu} - \frac{1}{2} \nabla_{\mu} H^{\alpha}_{\alpha}$, where  $\nabla_\mu$ is the covariant derivative with respect to the background metric $\bar h$. Notice the instantaneous propagation for the Newtonian potential. Interaction vertices from Einstein-Hilbert action will not be needed to calculate the leading order effect from our higher-dimensional operators, and we postpone the derivation of the interaction vertices from our higher-dimensional operators to appendix~\ref{App:new_quartics}.

As we mentioned, we will be interested in computing the correction to the multipoles of the effective single body, and, in order to compute their time dependence, we will also need the potential between the two-bodies. The potential can be computed as the following. First, compute the Feynmann diagram in Fourier space for the scattering of two bodies. For our quartic vertices, this is a function of four three-momenta, $\vec k_i$. Then, compute the Fourier transform to go to real space, by multiplying by four factors of $e^{-i\vec k_i^{(a)}\cdot\vec x_a(t)}$, where $a=1,2$, and the $\vec k_i^{(a)}$ corresponds to the momenta going to the body at position $\vec x_a$. After accounting for momentum conservation, these exponential factors will be summed into $e^{ i \vec q \cdot \vec x_{12}(t)}$, where $q$ is the momentum that propagates through the diagram, and $x_{12} = x_1 - x_2$ is the distance between the two bodies in a binary.
 This gives the correction to the potential. The correction to multipoles is done in an identical way apart for the inclusion in the diagram of an on-shell graviton $\bar h$.

In order to calculate corrections from our higher-dimensional operators, we also need both time and spacial derivatives acting on the gravitons. For potential graviton, each spacial derivative $\d_i$ on an incoming graviton with momentum $\vec k$ translates into a $-ik_i$, while the time derivatives must be dealt with by taking a time derivative of the appropriate propagator. This yields a time derivative on the time $\delta$-function which can be integrated by parts to get $\vec v \cdot \vec k$. The outgoing gravitons studied in this paper will only come from higher dimensional operators, which will be discussed also in appendix~\ref{App:new_quartics}.

In the end, we will need to integrate over all the internal momenta $\vec k$ with $\int { d}^3 k /(2\pi)^3$. These loop integrals will be discussed in more detail in appendix~\ref{app:loops}. This set of Feynman rules corresponds to computing $i\, S_{\rm ext.\, obj.}$~\footnote{In computing these diagrams, one needs to properly account for combinatorial factors. To determine these, it is useful to remember that computing our diagrams correspond to performing the following path integral in a perturbative way:
\be
e^{i S_{\rm ext.\, obj.}}=\int {\cal{D}}H_{\mu\nu} \; e^{i S_{\rm eff}+ i S_{pp}}\ .
\ee
By Taylor expanding the vertices entering in each diagram, and computing all the possible contractions, the combinatoric factors are obtained. 
}.

\begin{figure}[!ht]
  \centering
      \includegraphics[trim = 6cm 18.5cm 6cm 3cm, clip, width=0.4\textwidth]{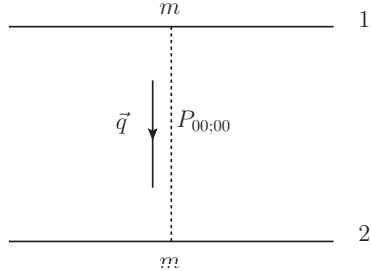}
   \caption{The diagram that gives the newtonian potential in the NR-EFT language. The two horizontal solid lines (labeled 1 and 2) represent the world line of the two black holes with masses $m_1$ and $m_2$, respectively. The dotted line between the two black holes represents a potential graviton $H$, with $3$-momentum $\vec q$. For the newtonian potential, only the $P_{00;00}$ component of the potential graviton propagator contributes.}
   \label{fig:newton}
\end{figure}

To give a definite example, we will calculate using this formalism the newtonian potential between two compact objects(see~\cite{Goldberger:2004jt,Goldberger:2007hy} for more examples). The newtonian potential can be found with the diagram in Fig.~\ref{fig:newton}. Here, the two horizontal solid lines (labeled 1 and 2) represent the world line of the two black holes with masses $m_1$ and $m_2$, respectively. The dotted vertical line between the two black holes represents a virtual potential graviton $H$, with $3$-momentum $\vec q$. The vertex labeled $m$ in the figure should be replaced by $-im_1/2\mpl$ or $-im_2/2\mpl$ depending on the black hole the vertex is attached to. We therefore have
\ba\nonumber
&&\text{Figure \ref{fig:newton}} = \\
&=& \int dt_1 \int dt_2   \int_{\vec k} \int_{\vec q}  \left[\frac{-i P_{\mu\nu;\alpha\beta}(2\pi)^3 \delta^{(3)}(\vec k+\vec q)}{k^2} \right]\nonumber\\
&& \left[-i\delta_{\alpha}^0 \delta_{\beta}^{0}\frac{m_1}{2\mpl} e^{-i \vec k\cdot\vec x_1}\right]\left[-i\delta_{\mu}^0 \delta_{\nu}^{0}\frac{m_2}{2\mpl}e^{-i \vec q\cdot\vec x_2}\right]  \delta(t_1-t_2) \nonumber \\
&=& i \frac{m_1 m_2}{(2\mpl)^2} \int dt \, \int_{\vec q} \frac{1}{2 { q^2}} e^{i {\vec q} \cdot {\vec x_{12}}}  = i \frac{m_1 m_2}{ 8 \mpl^2} \int dt \, \int \frac{d^{3-\epsilon} q}{(2\pi)^{3-\epsilon}} \frac{1}{{q^2}} e^{i {\vec q} \cdot {\vec x_{12}}}\nonumber \\ 
&=& i \int dt \frac{m_1 m_2}{ 32\pi \mpl^2 x_{12} } = i \int dt \frac{G m_1 m_2}{x_{12} },
\ea
where 
\be
\int_{\vec k}=\int \frac{d^3k}{(2\pi)^3}\ ,
\ee
 and 
where ${\vec x_{12}} = {\vec x_{1}} - {\vec x_{2}}$ is the distance between the black holes in the binary.  Since the action $S$ contains $-\int dt V$, and our Feynman rules compute $i S_{\rm ext. obj.}$, we read the familiar Newtonian gravitational potential $V=-\frac{G m_1 m_2}{x_{12}} $.

%%%%%%%%%%%%%%%%%%%%%%%%%%%%%%%%%%%%%%%%%%%%%%%%%%%%%%%%%%%%%%%%%%%%%%%%%%%%%%%%%%%%%%%%%%%%%%%%

\section{Expansion of $R_{\mu\nu\rho\sigma}$ and its contractions\label{App:new_quartics}}

Our purpose is to identify the quartic vertices associated to our operators. There will be two kinds of vertices that we will need: one where there are four potential gravitons, $\sim H^4$, which is relevant to compute the force between the objects, and one with three potential gravitons and one on-shell graviton, $\sim H^3\bar h$, required for the coupling to radiation. Since our operators are products of two scalar operators, ${\mathcal C}$ and $\tilde{\mathcal{C     }}$, it is enough to find the quadratic vertices from $\mathcal{C}$ and $\tilde{\mathcal{C     }}$ of the form $H H$  and $H\bar h$.

To linear order in the fluctuations about Minkowski 
\be
R_{\alpha \beta \delta \gamma} \rightarrow \delta R_{\alpha \beta \delta \gamma}=\d_\alpha \d_{[\delta }h_{\gamma ] \beta}-\d_\beta \d_{[\delta }h_{\gamma ] \alpha}
\ee
where $[x y]=\frac{1}{2}(xy-yx)$.

\subsection{${\mathcal{C}}$}
Using this, we can compute to leading order (in the PN expansion) $\delta R (H) \delta R (h)$ which will be relevant for both the potential and the multipole corrections as we can take $h\rightarrow \bar h$, when computing radiation, or $h\rightarrow H$, when computing the potential. In general,
\ba
\delta R_{\alpha \beta \gamma \delta} (h) \delta R^{\alpha \beta \gamma \delta} (h)&=& 2\d_\alpha \d_\gamma h_{\delta \beta} \delta R^{\alpha \beta \gamma \delta}(h) \\
\rightarrow \quad \delta R_{\alpha \beta \gamma \delta} (H) \delta R^{\alpha \beta \gamma \delta} (h)&=& 2\d_i \d_j H_{\delta \beta} \delta R^{i \beta j \delta}(h) \quad \text{to leading order} \, .
\ea

In all of our our diagrams, for $\mathcal{C}^2$, the insertion of the leading graviton-source vertex, $-mH_{00}/2\mpl$, gives a non-vanishing contribution. Therefore, one $H$ in the vertex $\mathcal{C}^2$ will be contracted with $H_{00}$ in the gravity-source vertex (through the propagator of momentum ${\bf k}$), which, in the standard schematic (and maybe potentially confusing) notation, we denote as $H\to H_{00}$.  Consequently, for the $H\bar h$ contribution, to leading order we may write, using $P_{\delta \beta \, ; 00}=\frac{1}{2}(2\eta_{\delta 0} \eta_{\beta 0}{ -}\eta_{\delta \beta})$:
\ba
\delta R (H\rightarrow H_{00}) \delta R (\bar h) &\rightarrow& (ik)_i (ik)_j\left(2 \delta R^{i0j0}(\bar h)+\delta R^{i\beta j}\,_\beta(\bar h) \right) \nonumber\\ 
&\rightarrow&2 (ik)_i (ik)_j \delta R^{i0j0}(\bar h) \label{RR_H00_hbar}\, ,
\ea
where in the last line we used the fact that $\bar h$ is one-shell and consequently all traces of the Riemann are zero.

Similarly,  for the $HH$ contribution, we have
\ba
\delta R (H\rightarrow H_{00}) \delta R (H\rightarrow H_{00}) &\rightarrow& \frac{1}{2}(ik)_i (ik)_j (iq)_i (iq)_j+\frac{1}{4}k^2 q^2  \ .
\label{RR_H00_H00} 
\ea
We will see that the term in $k^2 q^2$ never contributes to our diagrams.

\subsection{$\tilde{\mathcal{C}}$}

For $\tilde{\mathcal{C}}$, the leading order structure is given by contracting one leg with the lowest order graviton-source vertex, $-mH_{00}/2\mpl$, and one other with the next-to-leading graviton-source vertex, $-mv^iH_{0i}/\mpl$, or, if the spin is relevant, alternatively with the leading graviton-spin vertex, $-\frac{\d_iH_{0j}S^{ij}}{2\mpl}$.

In particular, for the $HH$ contribution, we have 
\be
\delta R(h\rightarrow H_{00}(\vec q)) \delta \tilde R (h\rightarrow  H_{0m}(\vec k)) =\epsilon_{nmi} (iq^i)(iq^j)(ik^j)(ik^n)
\ee
where $k$ is the momentum going into the $v$ vertex (or $S$ vertex) and $q$ is the one going into the $m$ vertex. Additionally, we have set our notation where $\epsilon_{0nmi}=\epsilon_{nmi}$.

Similarly, for the $H\bar h$ contribution, at leading order, we have: 
\ba
\delta  R (H\rightarrow H_{00}) \delta \tilde R (\bar h) &\rightarrow& - 2\epsilon_{ijk}\,^0(ik)_k (ik)_l \delta R^{ijl0}(\bar h)+ \epsilon_{\mu \nu k \gamma}(ik)_k (ik)_l \delta R^{\mu \nu l \gamma}(\bar h)  \nonumber\\
  &\rightarrow& \epsilon_{ijk}(ik)_k (ik)_l \left(\delta R^{ijl0}(\bar h)+2 \delta R^{ilj0}(\bar h) \right) \, .
\ea
Note that this can not be simplified by the first Bianchi identity.

%%%%%%%%%%%%%%%%%%%%%%%%%%%%%%%%%%%%%%%%
%%%%%%%%%%%%%%%%%%%%%%%%%%%%%%%%%%%%%%%%

\section{Loop integrals\label{app:loops}}

The kind of loop integrals that we will be interested in will be of the form
\be
\label{integral_to_solve}
\int_{\vec k} \frac{k^{i_1} \ldots k^{i_N}}{k^2[(k+q)^2]^M}
\ee
where $\int_{\vec k} \equiv \int \frac{d^Dk}{(2\pi)^D}$. We will regularize these integrals using dimensional regularization and will set $D=3-\epsilon$. A useful reference for dimensional regularization is~\cite{Collins:1984xc}. 

\subsection{Tensor structure}

The first point to note about  (\ref{integral_to_solve}) is that any trace of two indices in the numerator will yield a vanishing result. This can be seen by noting that the integral will then have the form of
\be
\int_{\vec k} \frac{k^{i_1} \ldots k^{i_{N-2}}}{[(k+q)^2]^M}\quad \rightarrow\quad \int_{\vec p} \frac{(p-q)^{i_1} \ldots (p-q)^{i_{N-2}}}{[p^2]^M}\ ,
\ee
where we have shifted the integration variable. Now, expanding the numerator we get a sum of integrals of the form
\be
\int_{\vec p}\frac{p^{i_1} \ldots p^{i_{2m}}}{[p^2]^M}\quad \propto\quad \left(\delta^{i_{1} i_{2}} \ldots \delta^{i_{2m-1} i_{2m}} + \text{perm} \right)  \int_{\vec p} \left(p^2\right)^{m-M}
\ee
which is, by definition, purely a power law integrand in any dimension, and so it vanishes (see~\cite{Collins:1984xc}, eq. 4.3.1a).
This means that, structurally, the result of integration over $k$ in~(\ref{integral_to_solve}) must be proportional to a {symmetric} combination of Kronecker $\delta$'s and $\vec q$'s such that it is zero under contraction of any two indices. A quick analysis indicates that there is one such unique term for a given $N$. For our purposes we will need only $N=2,3$. For reference these tensor structures are the following, up to $N=6$: 
\ba
T_2^{ij} &=&\frac{1}{(D-1)}[D \hat q \hat q-\delta ]^{ij}\\
T_3^{ijk}&=&\frac{1}{(D-1)}[(D+2) \hat q \hat q \hat q -(\hat q \delta+\hat q \delta+\hat q \delta) ]^{ijk}\\
T_4^{ijkl}&=&\frac{1}{(D+1)(D-1)}[(D+4)(D+2) \hat q \hat q \hat q \hat q -(D+2)(\hat q \hat q \delta+\text{$5$ perm}) \nonumber \\&& +(\delta \delta +\delta \delta +\delta \delta) ]^{ijkl}\\
T_5^{ijklm}&=&\frac{1}{(D+1)(D-1)}[(D+6)(D+4) \hat q \hat q \hat q \hat q \hat q -(D+4)(\hat q \hat q \hat q \delta+\text{$9$ perm}) \nonumber \\&& +(\hat q \delta \delta +\text{$14$ perm}) ]^{ijklm}\\
T_6^{ijklmn}&=&\frac{1}{(D+3)(D+1)(D-1)}[(D+8)(D+6)(D+4) \hat q \hat q \hat q \hat q \hat q \hat q \nonumber \\&& -(D+6)(D+4)(\hat q \hat q \hat q \hat q \delta+\text{$14$ perm})  +(D+4)(\hat q \hat q \delta \delta +\text{$44$ perm}) \nonumber \\&& -(\delta \delta \delta+\text{$14$ perm})]^{ijklmn} \; .
\ea
We have set the normalization by requiring that $\hat q^{i_1} \ldots \hat q^{i_N} T^{i_1 \ldots i_{N}}_N=1$. The pattern above seems generalizable to larger $N$ but they are not necessary for our purposes here.

It should also be noted that, obviously,
\be
\label{identity}
\hat q^ n\cdot T_{M+1}^{ni_1 \cdots i_M}=T_{M}^{i_1 \cdots i_M} \; 
\ee
as the contracted tensor $\hat q \cdot T_{M+1}$ is a properly-normalized, symmetric and traceless tensor with $M$ indece.

\subsection{Pre-factor}

Structurally, we have that 
\be \label{general_structure}
\int_{\vec k} \frac{k^{i_1} \ldots k^{i_N}}{k^2[(k+q)^2]^M} =C_{N}[M,D](q)\; T^{i_1 \ldots i_{N}}_N (\hat q) \; .
\ee
Our goal is now to find $C_N[M,D](q)$. The $q$ dependence follows from dimensional analysis but the exact numerical pre-factor requires more work. Let us do that now.

To solve an integral of the form (\ref{integral_to_solve}) in the usual way we first use the technique of Feynman parameters to simplify the denominator. We can write (\ref{integral_to_solve}) in the form:
\be
\int_{\vec k} \frac{k^{i_1} \ldots k^{i_N}}{k^2[(k+q)^2]^M}=\int_0^1dx \, M (1-x)^{M-1}\int_{\vec l} \frac{(l-(1-x)q)^{i_1} \ldots(l-(1-x)q)^{i_N} }{(l^2+\Delta)^{M+1}}
\ee
with $\Delta=x(1-x)q^2$. When we multiply the numerator together and collect the terms, they will come with various powers of $l$ and $q$. A term with an odd power of $l$ vanishes by symmetry and we therefore see that the sum of the terms of the form $ (l)^{2m}(q)^p$ will schematically give
\be
\sum (l)^{2m}(q)^p \rightarrow \sum(\delta)^m (q)^p\propto T_{N=2m+p}(\hat q) \; ,
\ee
that is, they will conspire to form the tensor structure of $T_{N=2m+p}$. Consequently, we need only to compute one such term to determine the overall normalization. The simplest way to do so is to look at the $(\hat q)^N$ term. Isolating this term, we have
\be
\int_0^1dx \, (-1)^N M (1-x)^{M-1+N}\int_{\vec l} \frac{q^N}{(l^2+\Delta)^{M+1}}\; .
\ee
Using the general formula
\be
\int \frac{d^D l}{(2\pi)^D} \frac{(l^2)^a}{(l^2+\Delta)^b}=\frac{\Gamma(b-a-D/2)\Gamma(a+D/2)}{(4\pi)^{D/2}\Gamma(b)\Gamma(D/2)}\times \Delta^{D/2+a-b}\ ,
\ee
we have
\be
\int_0^1dx \, (-1)^N M x^{D/2-M-1}(1-x)^{N+D/2-2} q^{N}(q^2)^{D/2-M-1}\frac{\Gamma(M+1-D/2)}{(4\pi)^{D/2}\Gamma(M+1)} \; .
\ee
Now using the fact that
\be
\int _0^1 dx\, x^a (1-x)^b=\frac{\Gamma(a+1)\Gamma(b+1)}{\Gamma(2+a+b)}\ ,
\ee
we have that the part of (\ref{integral_to_solve}) proportional to all $\hat q$'s is given by 
\be
(-1)^N M q^{N}(q^2)^{D/2-M-1} \times \frac{\Gamma(M+1-D/2)}{(4\pi)^{D/2}\Gamma(M+1)} \cdot\frac{\Gamma(D/2-M)\Gamma(N+D/2-1)}{\Gamma(N+D-M-1)} \; .
\ee
Comparing this to (\ref{general_structure}) and noting that we can write the coefficient in front of the all $\hat q$'s term as 
\be
\frac{\Gamma(D/2+N-1)\Gamma(D/2-1/2)}{\Gamma(D/2+N/2-1)\Gamma(D/2+N/2-1/2)}\ ,
\ee
we can see immediately that 
\ba
\label{master_prefactor}
C_N[M,D](q)&=&(-1)^N M q^{N+D-2M-2} \times \frac{\Gamma(M+1-D/2)}{(4\pi)^{D/2}\Gamma(M+1)}  \\
&& \times \frac{\Gamma(D/2-M)}{\Gamma(N+D-M-1)} \cdot \frac{\Gamma(D/2+N/2-1/2)\Gamma(D/2+N/2-1)}{\Gamma(D/2-1/2)}\ , \nonumber
\ea
and thus we have computed the integral for arbitrary $N$ and $M$ and $D$.  There is another interesting relation between these $C_N[M,D](q)$'s. The identity 
\be
2qC_N[M,D](q)=C_{N-1}[M-1,D](q)-q^2C_{N-1}[M,D](q)
\ee
happens to be true. This identity and that of (\ref{identity}) are directly related.

Equation (\ref{master_prefactor}) is complicated, we will only need it for a couple of particular values. One of the only real values that we will need for leading order computations are those with $M=1$ and $D\rightarrow3$. As we can see, there are no poles we have to watch out for, since one-loop integrals are finite in odd dimensions and in our case we do not need to keep track of the order $\epsilon$ contributions, and we can take $D=3$ immediately in the above formulas. As it turns out, they simplify a great deal and we have that
\be
C_N(q)[M=1,D=3]=(-1)^N\frac{1}{8}\cdot \frac{q^{N-1}}{2^N} \; ,
\ee
or, collecting everything together, we have that
\be
\label{final_loop_int_form}
\left . \int_{\vec k} \frac{k^{i_1} \ldots k^{i_N}}{k^2[(k+q)^2]}\right|_{D\rightarrow3} =(-1)^N\frac{1}{8}\cdot \frac{q^{N-1}}{2^N} T^{i_1 \ldots i_{N}}_N (\hat q)
\ee
which is often all we will need to proceed.

We will also use infinite parts of integrals with $M=-3/2$ that appear in our non-factorized two-loop diagrams, and that read:
\ba
C_2(q)[-3/2,3-\epsilon]&=&\frac{q^4}{315 \pi ^2 \epsilon}+\frac{q^4 \left(\gamma -3 -2\log (q)+ \log(\pi )+4  \log (2)+2 \psi ^{(0)}\left(\frac{11}{2}\right)\right)}{630 \pi
   ^2}\nonumber\\
   &+&\mathcal{O}\left(\epsilon^1\right) \\ \label{eq:maseq1}
C_3(q)[-3/2,3-\epsilon]&=&-\frac{q^4}{1155 \pi ^2 \epsilon}+\frac{q^4 \left(-3 \gamma +11+6 \log (q)-3\log (\pi )-6 \log (4)-6  \psi
   ^{(0)}\left(\frac{13}{2}\right)\right)}{6930 \pi ^2}\nonumber\\
   &+&\mathcal{O}\left(\epsilon^1\right), \label{eq:maseq2}
\ea
where $\psi^{(0)}(x)$ is polygamma function.

\subsection{$q$ integrals}

Throughout our computations the final result always has some integral over the three-momentum exchange. These integrals take the following form
\be\label{eq:trensotial}
\int \frac{d^{3-\epsilon}q}{(2\pi)^{3-\epsilon}}q^{i_1} \dots q^{i_n}q^me^{i\vec q \cdot \vec r} =  (-i \d_{i_1})\dots(-i \d_{i_n}) \int \frac{d^{3-\epsilon}q}{(2\pi)^{3-\epsilon}}q^m e^{i\vec q \cdot \vec r} \; .
\ee
The latter integral can be easily done in a general dimension: 
\ba
\int \frac{d^{3-\epsilon}q}{(2\pi)^{3-\epsilon}}q^m e^{i\vec q \cdot \vec r}&=&\frac{S[1-\epsilon]}{(2\pi)^{3-\epsilon}}\int dq \,  q^{m+2-\epsilon }\int_{-1}^1du \,  e^{i u q r} \nonumber \\
&=&\frac{S[1-\epsilon]}{(2\pi)^{3-\epsilon}}\frac{2}{r}\int dq \,  q^{m+1-\epsilon }\sin(q r)  \nonumber \\
&=& \frac{2 S[1-\epsilon]}{(2\pi)^{3-\epsilon}}\frac{\Gamma(m+2-\epsilon)}{r^{m+3-\epsilon}}\cos\left(\frac{m+1-\epsilon}{2}\pi\right) \; ,
\label{qInt}
\ea
where $S(n)$ is the area of the sphere in $n$ dimensions: $S(n-1)=2\pi^{n/2}/\Gamma(n/2)$.
Notice that because of the cosine, this vanishes in the $\epsilon \rightarrow 0$ limit for even non-negative~$m$.

\section{Gauge theory example\label{app:example}}

In this section, we will show a gauge theory example where features similar to our UV modification to General Relativity show up in the low energy effective theory. However, unlike the gravity theory, the gauge theory example can be easily UV completed into a theory that is valid all the way up to a scale that resembles the planck scale in General Relativity. In the following, we will discuss one example of such a theory starting with the effective action at the lowest energy. The low energy effective theory can be schematically written down as:

\begin{align}
\mathcal{L} \supset  f \left(\frac{\partial^2}{(g_2 v)^2}\right) \;\frac{1}{v^2}\left(\phi^{\dagger} \partial \phi \right)^2 + \frac{\alpha_1 \phi}{v} F_{\mu \nu} F^{\mu \nu}+m_{\phi}^2 \phi^2 + \cdots \label{eq:lowenergy}
\end{align}
where $F_{\mu\nu}$ represents the field strength of a massless $U(1)$ gauge boson with fine structure constant $\alpha_1$, and $\phi$ is a scalar with mass $m_{\phi} < g_2 v$. The function $f(\partial^2/(g_2 v)^2)$ is a polynomial function that stems from integration out UV degrees of freedom with mass $g_2 v$ (and it is implied that derivatives are distributed in all possible ways). Such an effective theory is valid within the energy interval $m_{\phi} < E < g_2 v$ where the scalar $\phi$ mediates a long range interaction between gauge bosons. The derivative expansion in this low energy effective theory is similar to the expansion of the higher-dimensional operators in the schematic form of 
\begin{equation}
M_{\rm pl}^2 R_{\mu\nu\rho\sigma} \;f \left(\frac{\nabla_\mu}{\Lambda}\right)\; g \left(\frac{R_{\mu\nu\rho\sigma}}{\Lambda^2}\right) \ ,
\end{equation}
in our theory of UV extensions to General Relativity, where $f$ and $g$ are two different polynormial functions. The mapping between the fields and scales in the gauge theory example and in the UV extension to General Relativity is displayed in Table~\ref{tab:mapping}. The expansion $h_{\mu\nu}/M_{\rm pl}$ is mapped to the expansion of $\phi/v$ while the expansion $\nabla/\Lambda$ is mapped to the expansion of $\partial/(g_2 v)$. 

\begin{table}
\centering
\begin{tabular}{c|c|c|c|c|c}
GR extension & Matter & Graviton & $M_{\rm pl}$ & $\Lambda$ & $\frac{\mathcal{C}^2}{\Lambda^6},\; \frac{\tilde{\mathcal{C     }}^2}{\tilde\Lambda^6},\; \frac{\tilde{\mathcal{C     }} {\mathcal{C     }}}{\Lambda_-^6} $\\[0.2cm] \hline\vspace{0.4cm}
Gauge theory & $A_{\mu}$ & $\phi$ & $v$ & $g_2 v$ & $\frac{\phi\d\phi}{v} \frac{\d^2}{(g_2 v)^2} \frac{\phi\d\phi}{v}$\\
\end{tabular}
\caption{The mapping between the GR extension studied in this paper and the toy gauge theory example.}\label{tab:mapping}
\end{table}

In the gauge theory example, as one can see, the theory hits a cutoff when the momentum $p \sim g_2 v$, at which point the derivative expansion is not valid any more. This is similar to the case of UV extensions of General Relativity when the momentum $p \sim \Lambda$. However, such cutoff does not signal a strong coupling scale. In fact, one can resolve the cutoff by introducing a single new state $A'_{\mu}$, with masses at $g_2 v$~(\footnote{We would of course be very happy to be able to perform this step for gravity.}). The effective theory that is valid at energies larger than $g_2 v$ is

\begin{align}
\mathcal{L} \supset &-\frac{1}{4} F_{\mu \nu} F^{\mu \nu} - \frac{1}{4} F'_{\mu \nu} F'^{\mu \nu} - (\partial \phi)^2 \nonumber\\
&+ g_2^2 v^2 A_{\mu}^{\prime 2} + (\lambda v^2) \phi^2 + \frac{\alpha_1 \phi}{v} F_{\mu \nu} F^{\mu \nu} + (g_2 A'_{\mu} \phi )^2 + (g_2 A'_{\mu} \phi)(\partial^{\mu}\phi) + \cdots\ .
\label{eq:mediumenergy}
\end{align}

\begin{figure}[!ht]
  \centering
      \includegraphics[trim = 6cm 20cm 6cm 3cm, clip, width=0.32\textwidth]{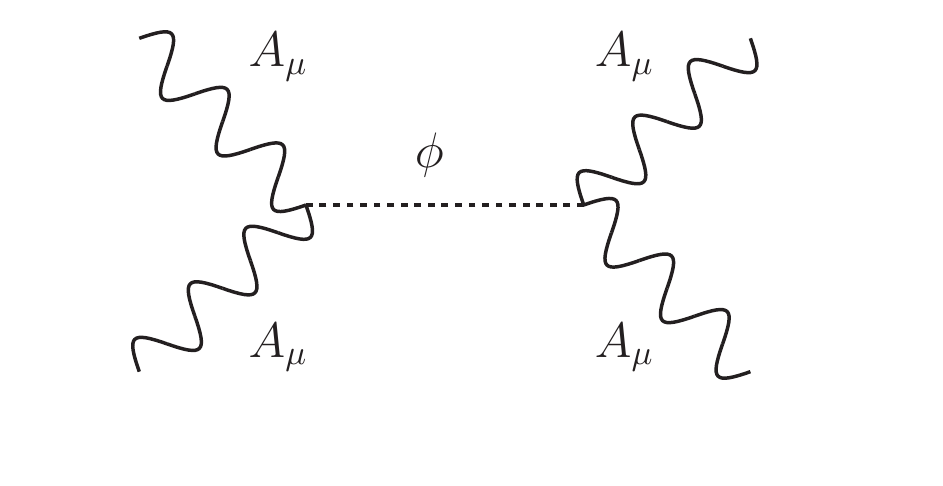}
            \includegraphics[trim = 4cm 16cm 3cm 3cm, clip, width=0.32\textwidth]{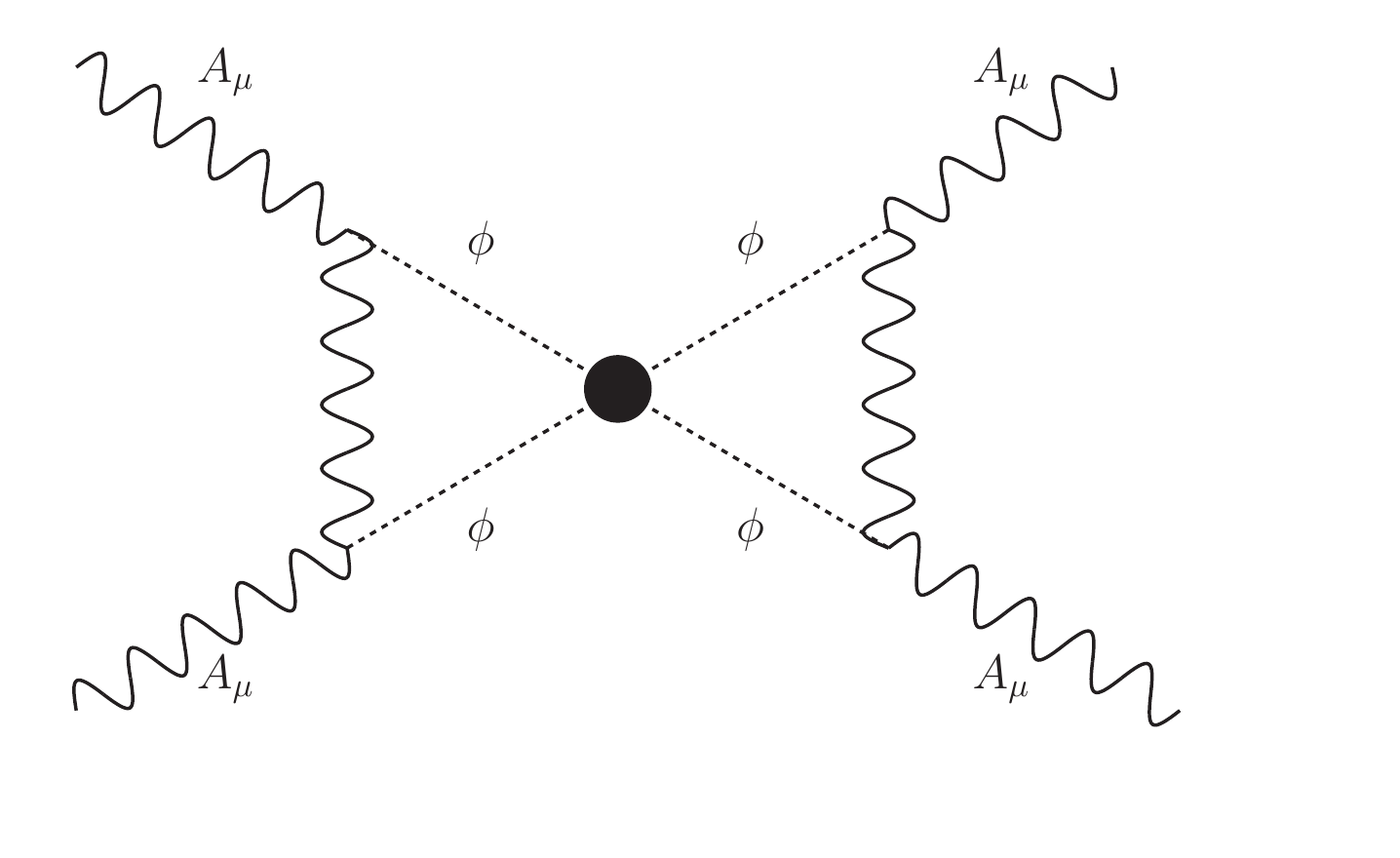}
      \includegraphics[trim = 6cm 18cm 3cm 3cm, clip, width=0.32\textwidth]{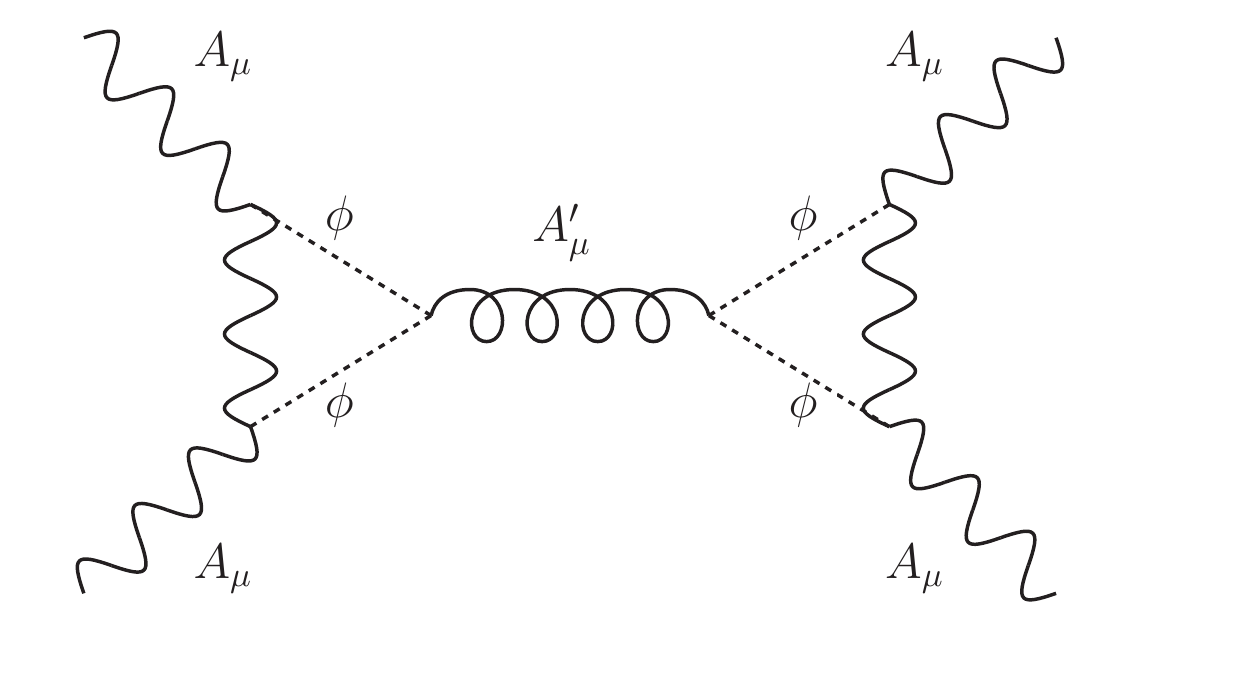}
  \caption{(Left) The interaction between photons mediated by the scalar $\phi$ which is similar to the gravitational interaction between matter in General Relativity. (Middle) The interaction mediated by the scalar $\phi$ with the effective operator similar to the $\mathcal{C}^2$, $\tilde{\mathcal{C}}^2$, ${\mathcal{C}}\tilde{\mathcal{C}}$ interaction studied in this paper, which has a UV cutoff at $g_2 v$. (Right) The interaction between photons mediated by the scalar $\phi$ and the gauge boson $A'_{\mu}$ as a UV completion of the diagram in the middle. }\label{fig:treeloop}
\end{figure}

The interaction mediated by the new state $A'_{\mu}$ can never become stronger than the leading order interaction mediated by $\phi$ (both shown in Fig.~\ref{fig:treeloop}) though the scattering amplitude in the low energy effective theory where $A'_{\mu}$ is integrated out blows up. The scattering of the scalar $\phi$ in the low energy effective theory in~(\ref{eq:lowenergy}) increases with energy as $E^2/v^2 (1 + (E/g_2 v)^2 + \cdots)$, while the same scattering mediated by $A_\mu'$ in the UV completion in~(\ref{eq:mediumenergy}) is finite and  ${\cal O}(g_2^2)$ at tree level and suppressed by additional powers of $g_2$ at loop level. Therefore this UV completion provides an example of what we call ``UV softness".  A similar UV behavior was assumed  to be true, and it was actually essential in order for not being already ruled out, in the case of our UV extension to General Relativity, though a UV completion is not available. The small $E/v$ limit corresponds to the limit where the test mass (BH mass) goes to zero, though due to difference in derivative structure, $E/v \rightarrow 0$ should be mapped to the combination $Gm/r \rightarrow 0$ instead of $m/M_{\rm pl} \rightarrow 0$.

Readers familiar with particle physics will notice that the gauge theory example can be further UV completed into a theory that is very similar to the Standard Model, where the higher dimensional coupling between the gauge boson $A_{\mu}$ and the scalar $\phi$ is UV completed by adding new fermions at the scale $v$:

\begin{align}
\mathcal{L} \supset &-\frac{1}{4} F_{\mu \nu} F^{\mu \nu} - \frac{1}{4} F'_{\mu \nu} F'^{\mu \nu} \nonumber\\
&+ L^{\dagger} \slashed{D} L + E^{\dagger}  \slashed{D}' E + y \Phi L E^c + conj. \nonumber\\
&+ \left(D'_{\mu} \Phi\right)^2 - \lambda \left(\Phi^{\dagger}\Phi -v^2\right)^2 
\end{align}
where $D'_{\mu} = \partial_{\mu} + i g_2 A'_{\mu}$ and $D_{\mu} = \partial_{\mu} + i g_1 A_{\mu} + i g_2 A'_{\mu}$. The fermions $L$ and $E$ are coupled with the scalar $\Phi$ with $O(1)$ yukawa coupling $y$. At energies much below the scale $v$, the fermions can be integrated out and the low energy spectrum includes the gauge boson $A_{\mu}$, the massive gauge boson $A'_{\mu}$ and the `higgs' $\phi$ with effective action shown in~(\ref{eq:mediumenergy}).~\footnote{Similarly to the case of the Standard model, for $\mathcal{O}(1)$ yukawa coupling, a natural theory would require tuning to get a small mass of $\phi$ and therefore $v$. However, such a tuning is only required so as to get a large ``planck scale'' of the theory. Moreover, by chosing $y$ and $g_2$ carefully in a way that $\lambda \ll g_2^2 < y^2 \ll \lambda^{1/2} \ll 1$, one can find a parameter range where everything we discussed above is true in a natural theory, given that supersymmetry or some other symmetries exist to cutoff the gauge boson loop at a scale $g_2 v \ll \Lambda \lesssim \sqrt{\lambda} v/g_2$.}

With this gauge theory example, we hope to convince the readers that derivative expansion breakdown in an effective theory does not always signal a new strong scale or the appearance of new states that couple to all the states in the effective theory. In fact, we present a new theory where the new gauge boson $A'_{\mu}$ cannot couple to the massless $A_{\mu}$ directly at leading order. In the language of the UV extension to General Relativity, this means that we do {\it not} necessarily need to introduce new states at the scale $\Lambda$ that couple directly to SM matter but just the graviton. Therefore, though we do not have a UV completion to our higher-dimensional operators beyond the scale $\Lambda$, we can imagine UV completions where similar behaviors may show up.

\subsection{Classical object}

We argued in the text that even though our gravitational EFT gives negligible corrections to small scale experiments, it nevertheless can give reliable and sizable corrections for experiments involving strong gravitational fields. We are now going to show that the same phenomenon can happen for our gauge theory EFT example.

We therefore study a classical object in this gauge theory example which has properties similar to  the strong gravity objects like black holes. A simple classical object that has this feature is a solenoid. For a solenoid with magnetic field $B$ and radius $R$, the scalar field $\phi$ develops a profile outside the solenoid
\begin{equation}
\phi(r) \sim \frac{\alpha_1 B^2 R^2}{v}\log[r/R]\ .
\end{equation}

In the context of the EFT for modifications of gravity, we have that around black holes, the value of gravitational potential $h\sim 1$, but this does not guarantees the testability of the theory unless it also happens that the curvature scale, $R^{\mu\nu\rho\sigma}$, is of order $\Lambda^2$. Similarly in the case of this gauge EFT, we have that when the solenoid has a large enough $B$ field, $B^2 \sim v/(R^2\alpha_1)$ (or $|A_{\mu}| \sim B R \sim v$ for $\alpha_1 \sim 1$) then around the solenoid $\phi\sim v$. This is analogous to forming a black hole. Around this solution, the effect of the higher derivative operators perturbs the $\phi$ solution as \begin{equation}
\frac{\delta \phi}{\phi} \sim \left(\frac{R}{r}\right)^{m} \left(\frac{1}{g_2 v r}\right)^{n} \left(\frac{\phi}{v}\right)^2\ .
\end{equation}
In order for this correction to be sizable, we need $R$ sufficiently small ($R\lesssim 1/(g_2 v)$). Such small solenoids are the analogue of the black holes where the curvature length is so small that the $1/\Lambda$ operators are important.

The operation of keeping the solenoid size fixed while decreasing the magnetic field decreases $|A_{\mu}|$ and leads to systems where $\phi \ll v$ (weak $\phi$ systems). Such an operation is similar to decreading the mass of a stellar object while keeping the size fixed. The stellar object will cease to be a black hole and become a weak gravity system in General Relativity. In the limit where the magnetic field inside the solenoid is taken to zero,  the correction to the force between the two solenoids vanishes. The zero magnetic field limit, in this case, corresponds to the scattering of two photons shown in Fig.~\ref{fig:treeloop}, where it is evident that all the effects are small and finite when $g_2 \rightarrow 0$ and obeys perturbation theory in $\alpha_{1,2}$ and $E/v$ as long as the center of mass energy is smaller than $v$, the cutoff of the intermediate theory (\ref{eq:mediumenergy}). Notice importantly that this true even in the energy interval above the cutoff of the lowest EFT: $g_2v<E<v$.

Another interesting operation is to keep the combination $B R$ fixed ($|A_{\mu}|$ fixed) while decreasing $R$. This corresponds to the case where we probe smaller and smaller black holes. Two possibilities remain in this case. If the solenoid is probed with charged particles near the solenoid surface, one expect to see $O(1)$ corrections when the size of the solenoid is smaller than $1/g_2 v$. Such is the reason why merger of smaller black holes probe higher cutoff scale $\Lambda$. However, if one probes the system at distances $r \gg R$, that is, when one measure the force between two solenoid that is very far apart, the corrections decrease as $R/r$ decreases and we do not expect to see $O(1)$ deviations and precise measurement of the system is needed to place constraint on this model~\footnote{One difference between the two cases is that unlike in General Relativity and its extension where the coefficient of the series expansion of $h_{\mu\nu}/M_{\rm pl}$ is determined by the symmetry structure of Einstein-Hilbert action, in our gauge theory example, one can write down an infinite power of series of $\phi/v$ whose coefficient is arbitrary. However, just like the leading operator in~(\ref{eq:lowenergy}), $(\phi^{\rm \dag}\partial \phi)^2/v^2$, the coefficients of operators in the $\phi/v$ expansion can be measured with large solenoid to, in principle, arbitrary precision. The coefficients of these type of corrections will be the same for both large and small solenoid, and can be subtracted when measuring deviation due to operators that are sensitive to the scale $g_2 v$. Similarly, in our UV extension to General Relativity, we expect the corrections we propose can be distinguished from the higher order corrections in $h_{\mu\nu}/\mpl$.}. For the same reason, the neutron stars or black holes, though intrinsically systems with rather strong gravitational potential near the surface, do not significantly constrain our model when the distance between them is much greater than the corresponding Schwarzschild radius (this corresponds to the limit of small velocity $v$). 

To conclude, in this section, we presented a gauge theory model which exhibits similar behaviors as our EFT for UV modifications of General Relativity. However, unlike the UV extension of General Relativity, this gauge theory model can be easily UV completed and one can understand the behavior of the theory at all energies. In particular, scattering between the `matter fields' in the gauge theory is small and finite even at energies higher than $g_2 v$ as long as it does not exceed the cutoff scale $v$. The apparent singular point at $g_2 v$ can be resolved by adding one weakly coupled state with mass at $g_2 v$. In this sense, this offers a field theory example of what we call "UV-softness", and also of how the theory might be not testable in weak field experiments, but only in strong field ones.

\bibliography{reference}
\bibliographystyle{jhep}

\end{document}